\documentclass{aa}
\usepackage{colortbl}
\usepackage{graphicx}
\usepackage{txfonts}
\usepackage{rotating}
\begin{document}

\title{The X-ray emission of $\gamma$~Cassiopeiae during the 2020-2021 disc eruption\thanks{Based on observations with {\it XMM-Newton}, and on optical spectra collected with the TIGRE telescope (La Luz, Mexico) and at Observatoire de Haute Provence (France).}}
\author{G.\ Rauw\inst{1} \and Y.\ Naz\'e\inst{1}\fnmsep\thanks{Senior Research Associate FRS-FNRS (Belgium)} \and C.\ Motch\inst{2} \and M.A.\ Smith\inst{3} \and J.\ Guarro Fl\'o\inst{4} \and R.\ Lopes de Oliveira\inst{5,6,7}}

\offprints{G.\ Rauw}
\mail{rauw@astro.ulg.ac.be}
\institute{Space sciences, Technologies and Astrophysics Research (STAR) Institute, Universit\'e de Li\`ege, All\'ee du 6 Ao\^ut, 19c, B\^at B5c, 4000 Li\`ege, Belgium \and Universit\'e de Strasbourg, CNRS, Observatoire astronomique de Strasbourg, UMR 7550, F-67000 Strasbourg, France \and National Optical Astronomical Observatory, 950 N.\ Cherry Ave., Tucson, AZ, USA \and Balmes, 2, 08784 Piera, Barcelona, Spain \and Departamento de F\'{\i}sica, Universidade Federal de Sergipe, Av. Marechal Rondon, S/N, 49100-000 S\~{a}o Crist\'{o}v\~{a}o, SE, Brazil \and Departamento de Astronomia, Instituto de Astronomia, Geof\'{i}sica e Ci\^{e}ncias Atmosf\'ericas, Universidade de S\~{a}o Paulo, R. do Mat\~{a}o 1226, Cidade Universit\'aria, 05508-090, S\~{a}o Paulo, SP, Brazil \and Observat\'{o}rio Nacional, Rua Gal.\ Jos\'e Cristino 77, 20921-400,Rio de Janeiro, RJ, Brazil}
\date{Received date / Accepted date}
\abstract{$\gamma$~Cas is known for its unusually hard and intense X-ray emission. This emission could trace accretion by a compact companion, wind interaction with a hot sub-dwarf companion, or magnetic interaction between the star and its Be decretion disc.}{These various scenarios should lead to diverse dependences of the hard X-ray emission on disc density. To test these scenarios, we collected X-ray observations of $\gamma$~Cas during an episode of enhanced disc activity that took place around January 2021.}{We investigate the variations in the disc properties using time series of dedicated optical spectroscopy and existing broadband photometry. Equivalent widths and peak velocity separations are measured for a number of prominent emission lines. Epoch-dependent Doppler maps of the H$\alpha$, H$\beta,$ and He\,{\sc i} $\lambda$~5876 emission lines are built to characterise the emission regions in velocity space. We analyse four {\it XMM-Newton} observations obtained between January 2021 and January 2022 at key phases of the episode of enhanced disc activity. Archival {\it XMM-Newton}, {\it Chandra}, MAXI, and RXTE-ASM data are also used to study the long-term correlation between optical and X-ray emission.}{Optical spectroscopy unveils a clear increase in the radial extent of the emission regions during the episode of enhanced disc activity, whilst no increase in the $V$-band flux is recorded. Our Doppler maps do not reveal any stable feature in the disc resulting from the putative action of the companion on the outer parts of the Be disc. Whilst the hard X-ray emission is found to display the usual level and type of variability, no specific increase in the hard emission is observed in relation to the enhanced disc activity. However, at two occasions, including at the maximum disc activity, the soft X-ray emission of $\gamma$~Cas is strongly attenuated, suggesting more efficient obscuration by material from a large flaring Be disc. In addition, there is a strong correlation between the long-term variations in the X-ray flux and the optical variations in the $V$-band photometry.}{The observed behaviour of $\gamma$~Cas suggests no direct link between the properties of the outer regions of the Be disc and the hard X-ray emission, but it favours a link between the level of X-ray emission and the properties of the inner part of the Be disc. These results thus disfavour an accretion or colliding wind scenario.}
\authorrunning{G. Rauw et al.}
\titlerunning{$\gamma$~Cas}
\maketitle

\section{Introduction \label{intro}}
Over 150 years ago, \citet{Secchi} first reported the presence of bright Balmer emission lines in the optical spectrum of $\gamma$~Cas (HD~5394, B0.5\,IVe). The star thus became the prototype of classical Be stars, which are defined as main-sequence or giant B stars that display, or have displayed, emission in their Balmer lines \citep[for a review, see][]{Riv13}. These emissions arise from a viscous circumstellar decretion disc. For the nearest and brightest Be stars, such as $\gamma$~Cas, discs have indeed been resolved in interferometry \citep{Qui97,Tyc06,Stee12} and shown to be Keplerian \citep{Stee12}. Over the last century, the H\,{\sc i} emission lines of $\gamma$~Cas have displayed significant variability, with the Balmer lines (including H$\alpha$) episodically going into absorption in 1944 and undergoing brief shell phases in 1935--1936 and 1939--1940 \citep[e.g.][]{Cow68,How79,Doa83}. Variations in the radial velocity (RV) of the H$\alpha$ line with a semi-amplitude near 4\,km\,s$^{-1}$ and a period of 203.5\,days revealed that $\gamma$~Cas is a binary system \citep{Har00,Miro02,Nem12,Smi12} with a nearly circular orbit ($e \leq 0.03$). The nature of the unseen secondary star remains unknown: its likely mass (0.8 -- 1.0\,M$_{\odot}$) is compatible with a late-type main-sequence star, a hot helium star sub-dwarf (sdO), or a white dwarf. 

Over the last few decades, the X-ray emission of $\gamma$~Cas has triggered a lot of interest (see \citealt{Smi16} for a review).  Indeed, this star is now the prototype of a subcategory of early Be stars that exhibit unusually bright and hard thermal X-ray emission \citep{LMH06,Naz18,Naz20a}. These objects have X-ray luminosities in the range $\log{L_{\rm X}} \sim 31.6$ to $33.2$, with $\log{(L_{\rm X}/L_{\rm bol})} \sim -6.2$ to $-4$, intermediate between normal OB stars and high mass X-ray binaries (HMXBs). The plasma temperatures are $kT \geq 5$\,keV, with $kT \sim 12$\,keV for $\gamma$~Cas itself\footnote{For comparison, normal OB stars have $\log{(L_{\rm X}/L_{\rm bol})} \simeq -7$ and display thermal X-ray emission with $kT$ usually below $1$\,keV \citep[e.g.][]{Naz09}. The lowest luminosity HMXBs containing a Be star, the so-called persistent Be HMXBs, have $\log{L_{\rm X}} > 34$. Their X-ray spectrum is best described by an absorbed power-law model, and they display pulsations at the spin period of the accreting neutron star \citep[e.g.][]{Rei11,LaP21}.}. Currently, the class of $\gamma$~Cas stars features 25 confirmed members and two additional candidates \citep{Naz20a}. Yet, the origin of this X-ray emission remains controversial.

Following its discovery as a moderately bright and hard X-ray source with the Small Astronomy Satellite (SAS-3) in December 1975 \citep{Jer76,Mas76,Bra77}, the X-ray emission of $\gamma$~Cas was investigated with an impressive number of satellites that progressively unveiled the complexity of the phenomenon. Unlike typical Be HMXBs, no X-ray pulsations were found despite intensive searches, notably with the {\it Uhuru}, SAS-3 \citep{Pet82}, {\it Einstein} \citep{Whi82}, X-ray Observatory Satellite \citep[EXOSAT;][]{Par93}, and {\it Ginga} \citep{Hor94} satellites. Instead, \citet{Par93}, using EXOSAT data, and \citet{Smi98a}, using Rossi X-ray Timing Explorer (RXTE) observations, showed that the X-ray light curve is actually dominated by irregular positive fluctuations, so-called shots \citep{Smi98a}, on timescales ranging from  $\sim 4$\,s to $\sim 10$\,min. This chaotic variability results in a red noise-dominated power spectrum. The shots are not associated with a significant change in the hardness ratio and come on top of a basal X-ray flux that varies on longer timescales \citep{Smi98a,Rob00}. Besides these rapid shots, the X-ray flux of $\gamma$~Cas also exhibits undulations on timescales of several hours \citep[e.g.][]{Rob02}. These undulations have been interpreted as quasi-periodic oscillations in the accreting white dwarf scenario \citep{Haberl} or as the passage of translucent clouds in forced corotation with the Be star in the magnetic interaction scenario \citep{Smi12}. A number of multi-wavelength studies have unveiled the existence of correlations between the variability of $\gamma$~Cas in the X-ray, UV, and optical domains \citep{Smi98a,Smi98b,Smi99,Rob00,Rob02,Smi06,Mot15,Smi19b} on timescales of hours, days, months, and years, without any obvious delay between the variations.

The presence of Fe\,{\sc xxv} and Fe\,{\sc xxvi} lines at 6.7 and 6.97\,keV, first unveiled by {\it Tenma} observations \citep{Mur86}, clearly demonstrates the thermal nature of the X-ray emission. The X-ray emission is dominated at all energies by a hot ($kT \sim 12$ -- 14\,keV) thermal plasma component that was detected with {\it Suzaku} and with the International Gamma-Ray Astrophysics Laboratory (INTEGRAL) up to energies of 100\,keV without evidence of any power-law component at very high X-ray energies \citep{Shr15}. {\it Chandra} and {\it XMM-Newton} observations revealed the presence of additional fainter thermal components with $kT$ near 0.1, 0.5, and (possibly) 3\,keV that mostly contribute to the low-energy part of the spectrum \citep{Smi04,LSM10,Smi12}.\\

The fact that the X-ray properties of $\gamma$~Cas stars differ significantly from those of Be HMXBs \citep{Rei11} led to a lively debate about the origin of their emission. The three groups of proposed scenarios involve accretion onto a compact companion (a neutron star or a white dwarf), magnetic interactions between the Be star and its disc, or the collision of the wind of a helium star companion with the Be disc.

At first, it was suggested that the X-ray emission results from accretion onto a neutron star companion \citep{Whi82}. Yet, the comparatively low level of the X-ray emission, the absence of pulsations, and the thermal nature of the emission are arguments against such a scenario. \citet{Pos17} revived the neutron star companion scenario, suggesting that $\gamma$~Cas stars might consist of a Be star orbited by a fast spinning magnetised neutron star. Direct accretion would be impeded by the propeller mechanism, leading to the formation of a hot shell of material around the neutron star magnetosphere that would emit thermal X-rays and not produce pulses. \citet{Smi17} presented a number of objections against this propeller scenario. They notably pointed out that this scenario does not account for the observed direct correlations between optical/UV and X-ray variations. Furthermore, they showed that, from an evolutionary point of view, such a propeller phase would be too short to account for the growing number of $\gamma$~Cas stars identified over recent years. To avoid the difficulties of the accreting neutron star scenario, \citet{Mur86} and later \citet{Haberl}, \citet{Kub98}, \citet{Apa02}, \citet{Ham16}, and \citet{Tsu18} proposed that the X-ray emission instead arises from an accreting white dwarf companion. A problem with this scenario is the large mass-loss rate from the Be star required to power the accretion process, which exceeds the wind mass-loss rate of $\gamma$~Cas by several orders of magnitude. \citet{Shr15} further noted the absence of variability at energies above 20\,keV in the INTEGRAL data of $\gamma$~Cas collected over nine years, whereas accreting white dwarf systems do exhibit variability at these energies. 

\citet{Lan20} proposed that the hard X-ray emission arises from the collision of the wind of a hot helium star companion with the Be disc and/or wind. Indeed, many Be stars were probably spun up through mass transfer in a close binary system, leaving the former primary star (i.e.\ the mass donor) as a hot stripped-off helium star sub-dwarf (sdO). \cite{Wan18} report the results of a search for such sdO companions among Be stars in the far UV. No signature of an sdO was found in the case of $\gamma$~Cas, though. Whilst this does not rule out the possibility of the secondary being a helium star, it sets an upper limit on the luminosity (and hence the wind strength) of such a stripped companion, thereby challenging the \citet{Lan20} scenario. Moreover, in view of the wind parameters estimated for presumably single sdOs \citep{JH10,Groh,Vink}, it seems unlikely that a wind collision could account for the very high temperature of the dominant plasma component. Another issue comes from the very short timescale of the shots, which implies plasma electron densities of at least $10^{14}$\,cm$^{-3}$ \citep{Smi98a}, significantly higher than those expected at the interface between the sdO wind and the outer Be disc.

The correlations between the X-ray, UV, and optical variations, and the difficulties of the accretion models in explaining the observed properties of $\gamma$~Cas, led to the elaboration of an alternative magnetic interaction scenario \citep{Smi98a,Smi99,Mot15,Smi19}. In this scenario, small-scale magnetic fields emerging from the Be star become entangled with a toroidal magnetic field produced in the Be disc. Due to the different rotation rates of the two systems of magnetic field lines, the lines stretch and sever, leading to reconnection events. These events accelerate electron beams onto the Be star's surface, where they produce the hard X-ray flux \citep[][and references therein]{Smi19}. One issue with this scenario is the absence of direct detections of stellar magnetic fields in Be stars (\citealt{Wad16} indicate that 50\% of their sample of 85 classical Be stars should have a dipole field strength of less than 50\,G). However, current spectropolarimetry techniques can only detect large-scale magnetic fields, and the presence of such fields was shown to be inconsistent with the existence of a Keplerian circumstellar disc \citep{udD18}. The stellar magnetic fields required for the star-disc interaction scenario should instead be localised fields generated in a thin subsurface convective layer powered by the iron opacity peak \citep{Can09}. \citet{Can11} quantify the strength of the ensuing localised magnetic fields that could emerge at the surface via magnetic buoyancy as a function of the star's position in the Hertzsprung-Russell diagram. For a star similar to $\gamma$~Cas, \citet{Can11} predict a minimum surface magnetic field strength of 10 -- 20\,G. Such putative localised fields are beyond the reach of current spectropolarimetric facilities. Indeed, \citet{Koc13} investigated the detectability of small-scale magnetic fields by means of high-resolution spectropolarimetry. They showed that the expected signature of such fields decreases with increasing $v\,\sin{i}$. With current spectropolarimetric observations, small-scale magnetic fields of about 1\,kG could be detected in the spectrum of a star with $v\,\sin{i} \simeq 200$\,km\,s$^{-1}$. For a very fast rotator, such as $\gamma$~Cas, the detection limit would obviously be higher than 1\,kG.

The various scenarios described above all imply that the properties of the X-ray emission should depend somehow on the properties of the Be disc. Indeed, in the accretion scenarios, although the Be disc is most probably truncated well within the Roche lobe of the Be star \citep{Oka01}, the radiation field of the Be star erodes the disc \citep{Kee16}, and the ensuing wind Roche lobe overflow certainly depends on the amount of material present in the disc. An increase in the Be star mass-loss rate should lead to an increase in the X-ray luminosity in a Be plus neutron star binary system \citep[with $L_{\rm X} \propto \dot{M}^\beta$, where $\beta \simeq 1$ or $5$ depending on the wind properties;][]{Wat89}. Likewise, the properties of the collision between the wind of a putative sdO companion and the Be disc obviously depend on the disc density. Finally, in the case of a complete disc dissipation, the magnetic interaction scenario would lose one of its key ingredients, and the hard X-ray emission is expected to vanish. The different scenarios imply different time delays between the variations in the disc properties and the X-ray response \citep{Mot15}: whilst the response should be almost instantaneous for the magnetic interaction scenario, significantly longer delays are expected for the accretion scenarios. Besides these temporal differences, the various scenarios are sensitive to the properties of different parts of the disc: whereas the magnetic field scenario most strongly depends on the conditions in the inner disc regions, both the accretion and the wind--disc collision scenarios are expected to most strongly react to changes in the density of the outer disc. 

To investigate the link between disc strength and X-ray emission, we initiated an optical spectroscopic monitoring of a sample of $\gamma$~Cas stars to uncover disc dissipation or outburst events that could be used as triggers for X-ray observations. Supporting evidence for a connection between large changes in the disc properties and the hard X-ray emission was obtained in the case of the O9:npe star HD~45\,314, the earliest spectral-type $\gamma$~Cas star known to date. Starting from October 2014, the Oe emission lines of this star underwent important morphology and intensity changes, including one episode, in early 2016, where the optical emission lines nearly vanished, suggesting a partial dissipation of the disc. An {\it XMM-Newton} observation taken during this episode revealed an X-ray emission a factor of 10 fainter and significantly softer than that observed during the 2012 observation that led to the discovery of the $\gamma$~Cas behaviour of this star \citep{Rau18}. At odds with this, only moderate changes in the X-ray emission were observed for three other $\gamma$~Cas stars whose H$\alpha$ emission strength underwent strong variations. In the case of the B1\,Ve star $\pi$~Aqr, \citet{Naz19} and \citet{Naz22} reported only a moderate increase in the X-ray flux, by a factor of 1.5, between two different emission states, with the H$\alpha$   equivalent width (EW) changing from $-1.7$\,\AA\ to $-23$\,\AA. Similarly, HD~119\,682 (B0\,Ve) and V~767\,Cen (B2\,Ve) were both found to display a hard X-ray emission that was essentially unaffected by the large changes in their H$\alpha$ emission strength over recent years, including one episode in July 2020 when the line emission of HD~119\,682 had (nearly) entirely disappeared \citep{Naz22}.

In the present paper we report the results of our campaign devoted to $\gamma$~Cas and, more specifically, the enhanced emission event it underwent over the 2020-2021 visibility season. Our observational material is described in Sect.\,\ref{obs}, and our analysis of the optical and X-ray data is provided in Sect.\,\ref{analyse}. The issue of long-term variations in optical and X-ray spectra is discussed in Sect.\,\ref{longtermvar}. Finally, Sect.\,\ref{discus} summarises our results and discusses their implications.  
\begin{table*}
  \caption{Journal of the X-ray observations of $\gamma$~Cas analysed here.\label{journalX}}
  \begin{center}
  \begin{tabular}{l c c c c c c c}
    \hline
    \multicolumn{8}{l}{{\it Chandra}-HETG} \\
    \hline
    \multicolumn{1}{c}{Obs.} &  & HJD-2450000 & Duration &  & & & $\phi_{\rm Bin}$ \\
         &       &  & (ks)    &                 \\
    \hline
    Aug.\ 01  & & 2132.202 & 52.5 & & & & 0.25 \\
    \hline
    \multicolumn{8}{l}{\it XMM-Newton} \\
    \hline
    \multicolumn{1}{c}{Obs.} & Rev. & HJD-2450000 & Duration & \multicolumn{3}{c}{EPIC modes} & $\phi_{\rm Bin}$ \\
         & &  &           (ks)    & MOS1 & MOS2 & pn & \\
    \hline
    Feb.\ 04  &  762 & 3041.662 & 65.2 & T & T & T & 0.72 \\
    Jul.\ 10a & 1937 & 5384.901 & 17.5 & S & S & S & 0.23 \\
    Jul.\ 10b & 1945 & 5401.741 & 15.7 & S & S & S & 0.31 \\
    Aug.\ 10a & 1950 & 5411.182 & 17.5 & S & S & S & 0.36 \\
    Aug.\ 10b & 1959 & 5428.685 & 22.4 & S & S & S & 0.44 \\
    Jul.\ 14  & 2678 & 6863.268 & 32.5 & S & S & S & 0.49 \\
    Jan.\ 21  & 3859 & 9219.190 & 16.3 & S & T & T & 0.07 \\
    Feb.\ 21  & 3881 & 9263.018 &  5.2 & S & T & T & 0.28 \\
    Jul.\ 21  & 3960 & 9420.564 &  6.4 & S & T & T & 0.06 \\
    Jan.\ 22  & 4049 & 9598.080 &  4.5 & S & T & T & 0.93 \\ 
    \hline
  \end{tabular}
  \end{center}
  \tablefoot{The date quoted in Col. 3 and the duration given in Col. 4 correspond respectively to the middle and the total duration of the usable exposure of the EPIC-pn instrument (i.e.\ after discarding intervals of soft proton background flares). The EPIC modes are labelled as S for small window mode and T for timing mode. The last column quotes the orbital phase of the observation using binary orbital solution number 5 of \citet{Nem12}, with $P = 203.523$\,d and $T_0 =$ 2\,452\,081.89 corresponding to the epoch of minimum RV of the Be star.}
\end{table*}
\section{Observations \label{obs}}
\subsection{Optical spectroscopy}
In 2019 we started monitoring the optical spectrum of $\gamma$~Cas with the 1.2~m Telescopio Internacional de Guanajuato, Rob\'otico-Espectrosc\'opico \citep[TIGRE, formerly known as the Hamburg Robotic Telescope,][]{Hempelmann,Schmitt} at La Luz Observatory near Guanajuato (Mexico). The telescope is operated in a fully robotic way, and spectra are taken with the refurbished Heidelberg Extended Range Optical Spectrograph (HEROS) echelle spectrograph \citep{Kaufer2,Schmitt}, which offers a spectral resolving power of 20\,000 over the full optical range from 3760 -- 8700\,\AA, with a small gap between 5660 and 5780\,\AA. At first, we obtained typically one observation every three weeks (weather permitting). By the end of 2020, we noted a strong gradual increase in the strength of the H\,{\sc i} Balmer and He\,{\sc i} emission lines. As a result, we increased the monitoring cadence of our campaign to one spectrum every four days. Individual observations consisted of the combination of six integrations of $20$\,s duration each to achieve a S/N of typically about 150 in the continuum whilst simultaneously avoiding saturation of the detector in the strong emission lines. The data were reduced with the HEROS reduction pipeline \citep{Mittag,Schmitt}.

In October 2021, $\gamma$~Cas was monitored during five consecutive nights with the Aur\'elie spectrograph \citep{Gillet} at the 1.52\,m telescope of the Observatoire de Haute Provence (OHP) in France. The spectrograph was equipped with a 600\,l\,mm$^{-1}$ grating, blazed at 5000\,\AA, providing a resolving power of about 10\,500 over the wavelength domain from 6415\,\AA\ to 6845\,\AA. The detector was an Andor Newton CCD with $2048 \times 512$ pixels of 13.5\,$\mu$m squared. Typical integration times of individual exposures ranged between 10\,s and 2\,min depending on the atmospheric conditions. Between 75 and 100 exposures were taken per night and were combined to provide a nightly mean spectrum. The mean S/N of these nightly means, evaluated in the continuum between 6736 and 6745\,\AA, was 1200. The data reduction was performed using version 17FEBpl\,1.2 of the {\sc midas} software developed at the European Southern Observatory.

These data were complemented by dedicated amateur spectra taken by co-author Joan Guarro Fl\'o either from Piera or remotely from Santa Maria de Montmagastrell (T\`arrega) in Catalonia. The equipment used on both sites was a 40.6\,cm Schmidt-Cassegrain with a focal reducer to f/6.5 along with a self-made echelle spectrograph. The spectrograph provided a resolving power of 8500 over the range from 3750 to 9500\,\AA\ with an ASI2600MM CCD camera. The data were reduced using the ISIS-V6.1.1 software. 
Additional archival amateur spectra of $\gamma$~Cas taken during the visibility gap of the star with TIGRE or obtained contemporaneously with older X-ray observations were extracted from the Be Star Spectra (BeSS) database\footnote{\tt http://basebe.obspm.fr} \citep{Nei11}.

We used the {\tt telluric} tool within {\sc iraf} along with the atlas of telluric lines of \citet{Hinkle} to remove the telluric absorptions in the spectral regions around the He\,{\sc i} $\lambda$\,5876 and H$\alpha$ lines. To achieve as homogeneous a normalisation as possible, the spectra were continuum normalised using the MIDAS software adopting best-fit spline functions adjusted to the same set of carefully chosen continuum windows for all spectra.

\begin{figure*}
  \resizebox{15cm}{!}{\includegraphics[angle=0]{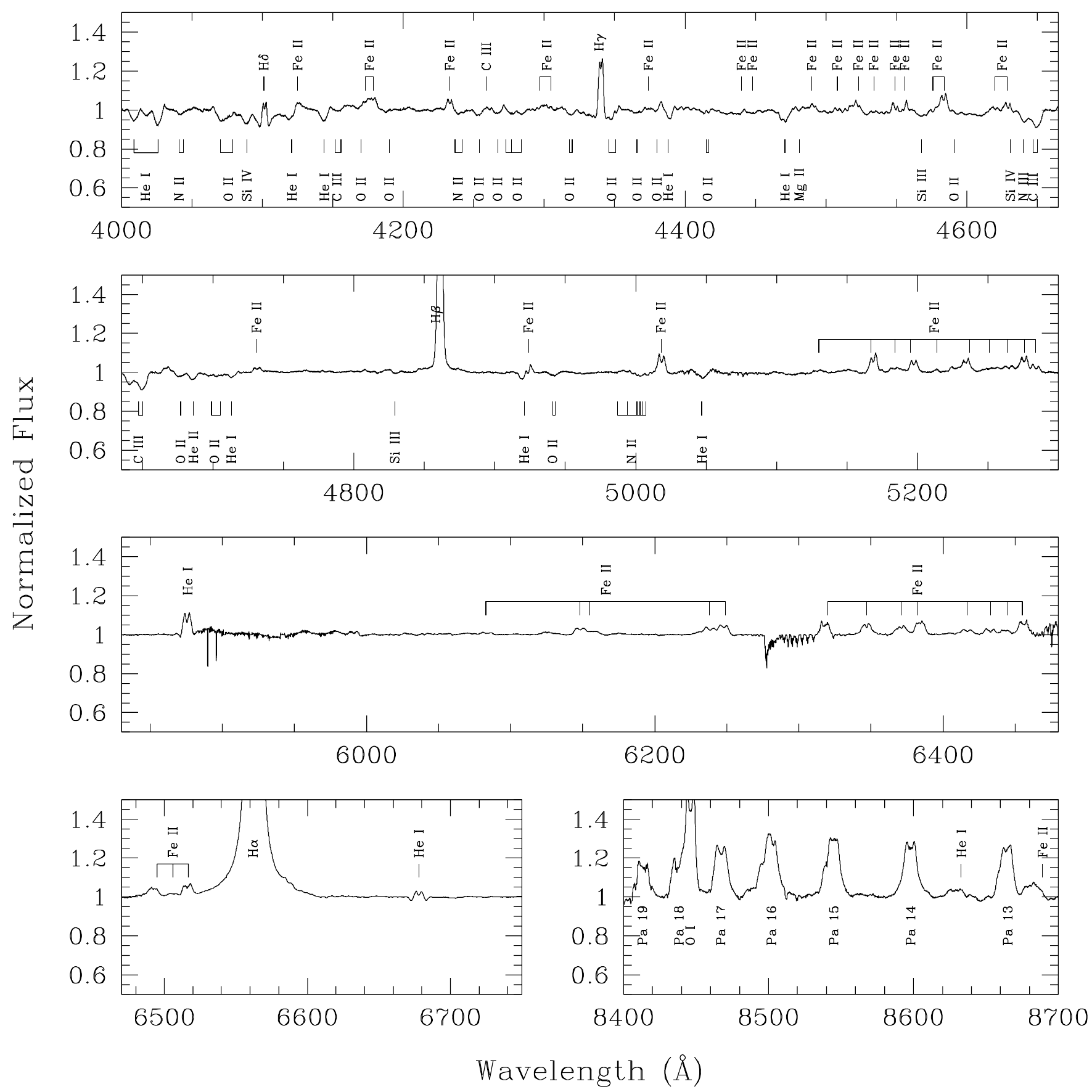}}
  \caption{Mean HEROS spectrum of $\gamma$~Cas. The telluric lines around the He\,{\sc i} $\lambda$~5876 and H$\alpha$ lines have been removed (see the main text). \label{meanHEROS}}
\end{figure*}

\subsection{X-ray observations}
During its recent high-emission state, $\gamma$~Cas was observed four times with {\it XMM-Newton} \citep{Jansen} in the context of our target of opportunity programme to follow the evolution of the X-ray spectrum during such an event. The European Photon Imaging Camera (EPIC) instruments were used to acquire CCD-resolution spectra. The EPIC-MOS1 camera \citep{MOS} was operated in small window mode, whilst the EPIC-MOS2 and EPIC-pn instruments \citep{pn} were operated in timing mode to further limit the impact of photon pileup. All three EPIC cameras used the thick filter to prevent optical and UV photons from reaching the detectors. Simultaneously with the EPIC data, {\it XMM-Newton} collected high-resolution X-ray spectra of $\gamma$~Cas with the Reflection Grating Spectrometer \citep[RGS;][]{RGS}.

The data were processed with the Science Analysis System (SAS) software version 18.0.0 and using the current calibration files available in January 2021. The {\tt epatplot} SAS command was used to check for the presence of pileup in the MOS1 data. No obvious indication of severe pileup was found. Yet, the MOS1 count rate during the observation significantly exceeded the pileup limit for small window mode. As a result, the MOS1 spectra were extracted over an annulus of inner and outer radii of 10 and 40\,arcsec. In this way, the core of the point spread function where pileup should be more severe was excluded from the spectral extraction.

Whilst our first and third observations (revolutions\ 3859 and 3960) were free of any significant soft proton background flares, the second and fourth observations (revolutions\ 3881 and 4049) were strongly impacted by such events. We thus filtered our data against high background (i.e.\ discarding time intervals with count rates exceeding 0.25\,ct\,s$^{-1}$ and  0.5\,ct\,s$^{-1}$ at energies above 10\,keV for the EPIC-MOS and EPIC-pn cameras, respectively) and kept only the clean part of these exposures. \\ 

We further retrieved all archival (2004 -- 2014) {\it XMM-Newton} observations of $\gamma$~Cas \citep{LSM10,Smi12,Smi19b} and reprocessed them with SAS version 18.0.0 in the same way as our new data. In addition, {\it Chandra} High-Energy Transmission Grating (HETG) data \citep{Smi04} were extracted from the {\it Chandra} Transmission Grating Data Archive and Catalog \citep[TGCat;][]{Hue11}.
The properties of the X-ray observations considered in this paper are summarised in Table\,\ref{journalX}.

To investigate the long-term variability of the X-ray flux and its correlation with optical photometry, we also considered data collected with the Monitor of All-sky X-ray Image (MAXI) installed since August 2009 on the Japanese Experiment Module Exposed Facility of the International Space Station \citep[ISS;][]{Mat09}. The MAXI data up to mid-2021 were downloaded from their public archive\footnote{MAXI on-demand processing, http://maxi.riken.jp/mxondem/}. We examined only the data in the 2 -- 6\,keV energy band, as data in the 6 -- 20\,keV band, even after background subtraction, suffer from a $\sim 70$\,d modulation of the background linked to the precession of the ISS orbit \citep{Mot15}.

\section{Data analysis \label{analyse}}
\subsection{Optical spectroscopy}
The optical spectrum of $\gamma$~Cas (Fig.\,\ref{meanHEROS}) is dominated by the H$\alpha$, H$\beta$ and H$\gamma$ Balmer emission lines, whereas the near-IR part exhibits emissions from the Paschen series of hydrogen. Upon closer inspection of the spectrum, one can also find several double-peaked He\,{\sc i} emissions (notably He\,{\sc i} $\lambda$~5876 and $\lambda$~6678) as well as numerous Fe\,{\sc ii} emissions, many of which also display a double-peaked morphology \citep{Cow68,SmiBal06}.

Before analysing the variability of the optical spectrum of $\gamma$~Cas, it is important to recall that the various ingredients of this spectrum arise from physically distinct regions \citep{Stee98}. Indeed, the $V$-band continuum comes mostly from the stellar photosphere and from reflection on the innermost part (within a radius of a few R$_*$) of the decretion disc. The He\,{\sc i} $\lambda$~5876 and $\lambda$~6678 lines consist of a mix of photospheric absorptions and emissions forming over a smaller part of the decretion disc than the $V$ band. The H\,{\sc i} Paschen and Fe\,{\sc ii} emission lines likely arise from essentially the same regions of the disc as the $V$-band flux. Finally, the Balmer emissions probe much larger parts of the disc, with the size of the emission region significantly increasing from H$\gamma$ to H$\alpha$, reaching out to radii of $\sim 10\,R_*$ for the latter.
\begin{figure*}
  \resizebox{16cm}{!}{\includegraphics[angle=0]{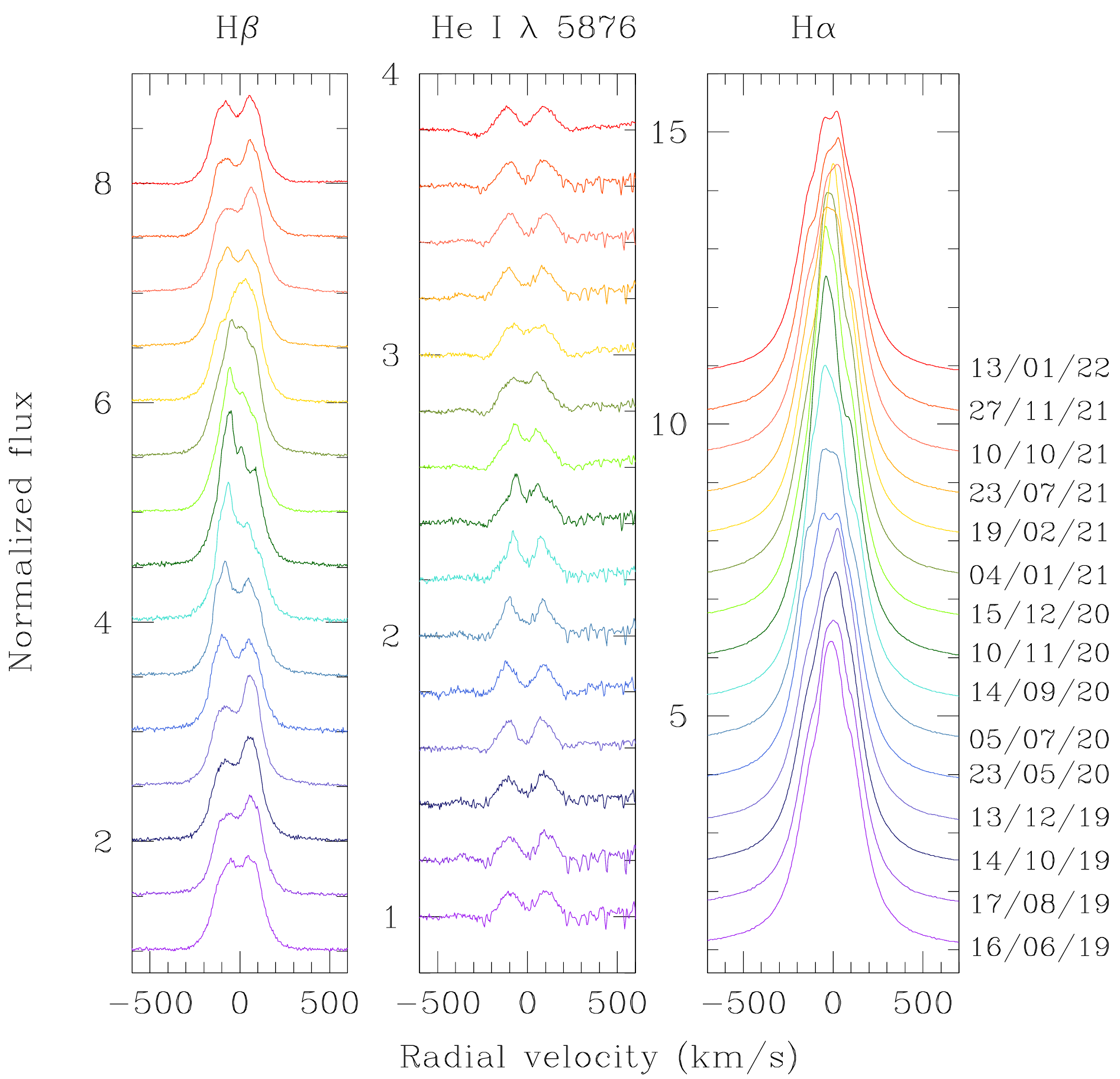}}
  \caption{Variations in the strongest emission lines in the HEROS spectrum of $\gamma$~Cas over the last two years. Subsequent spectra are shifted upwards to enhance the visibility. The labels on the right yield the date of the observation in the format dd/mm/yy.\label{montage}}
\end{figure*}

\begin{figure*}
  \begin{minipage}{9cm}
    \resizebox{9cm}{!}{\includegraphics[angle=0]{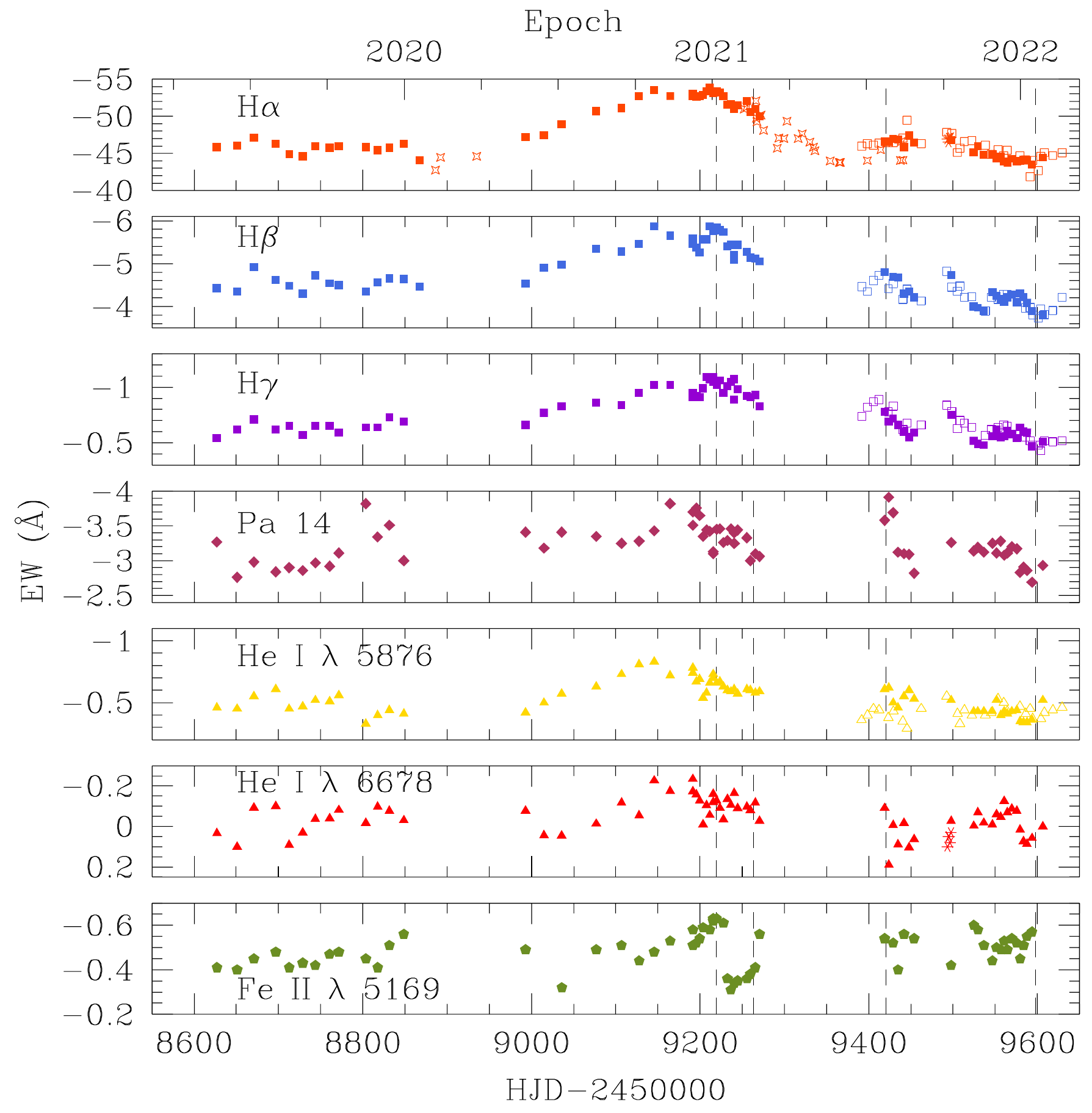}}
  \end{minipage}
  \hfill
  \begin{minipage}{9cm}
    \resizebox{9cm}{!}{\includegraphics[angle=0]{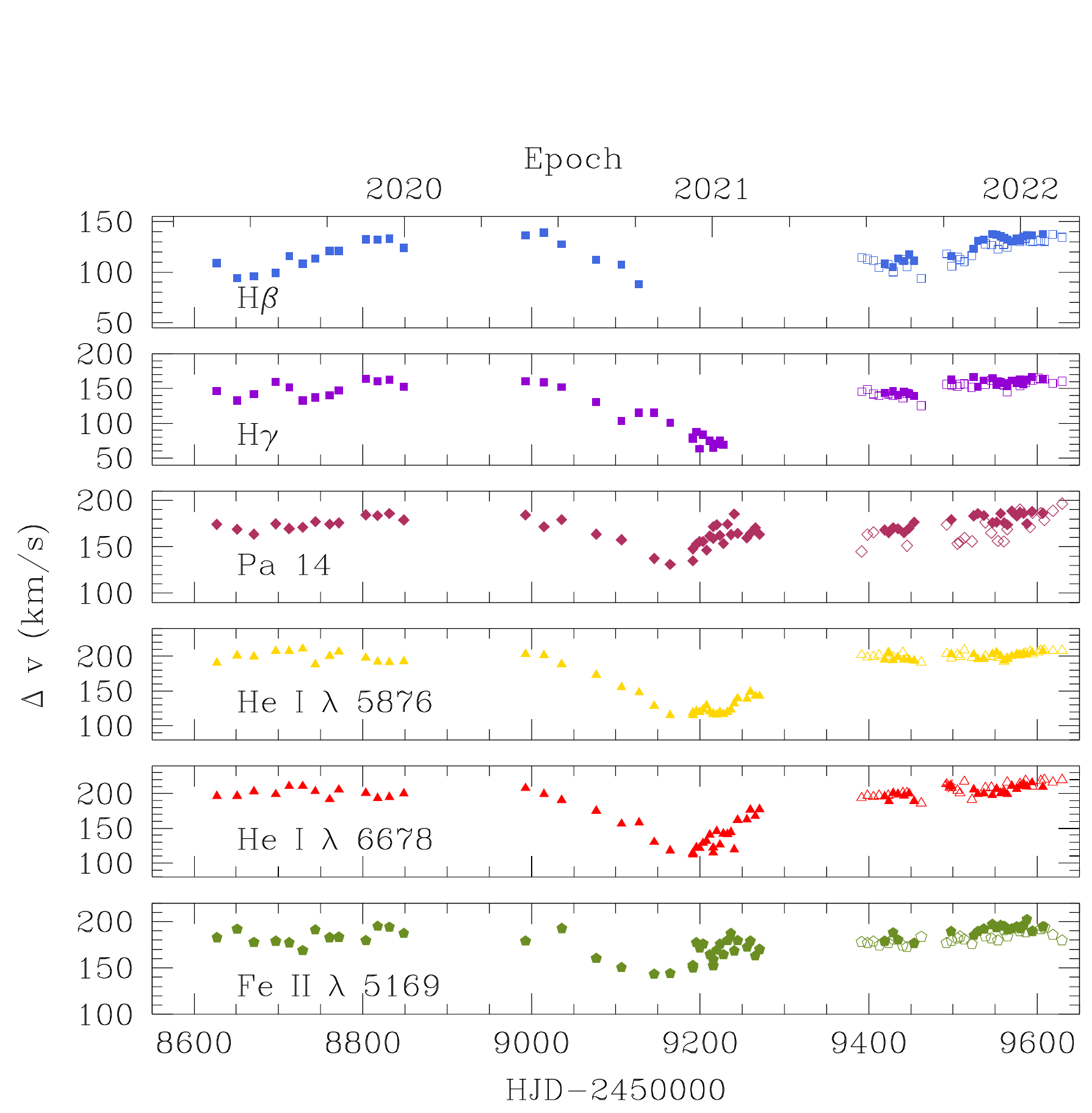}}
  \end{minipage}
%  \hfill
%  \begin{minipage}{6cm}
%    \resizebox{6.25cm}{!}{\includegraphics[angle=0]{Fig3c.pdf}}
%  \end{minipage}
    \caption{Results from the optical monitoring of $\gamma$~Cas. The left panel illustrates the EWs of various prominent lines as a function of time, whilst the right panel yields the RV separation of the double-peaked profiles. The dashed vertical lines in the EW plots correspond to the dates of the new {\it XMM-Newton} observations. Filled and open symbols identify TIGRE data and spectra collected by co-author JGF, respectively. Asterisks (H$\alpha$ and He\,{\sc i} $\lambda$\,6678 panels) correspond to Aur\'elie data, and four-point stars (H$\alpha$ panel) refer to BeSS archival spectra.\label{monit}}
\end{figure*}

\subsubsection{Emission line variability \label{lpv}}
\begin{figure}
    \resizebox{9cm}{!}{\includegraphics[angle=0]{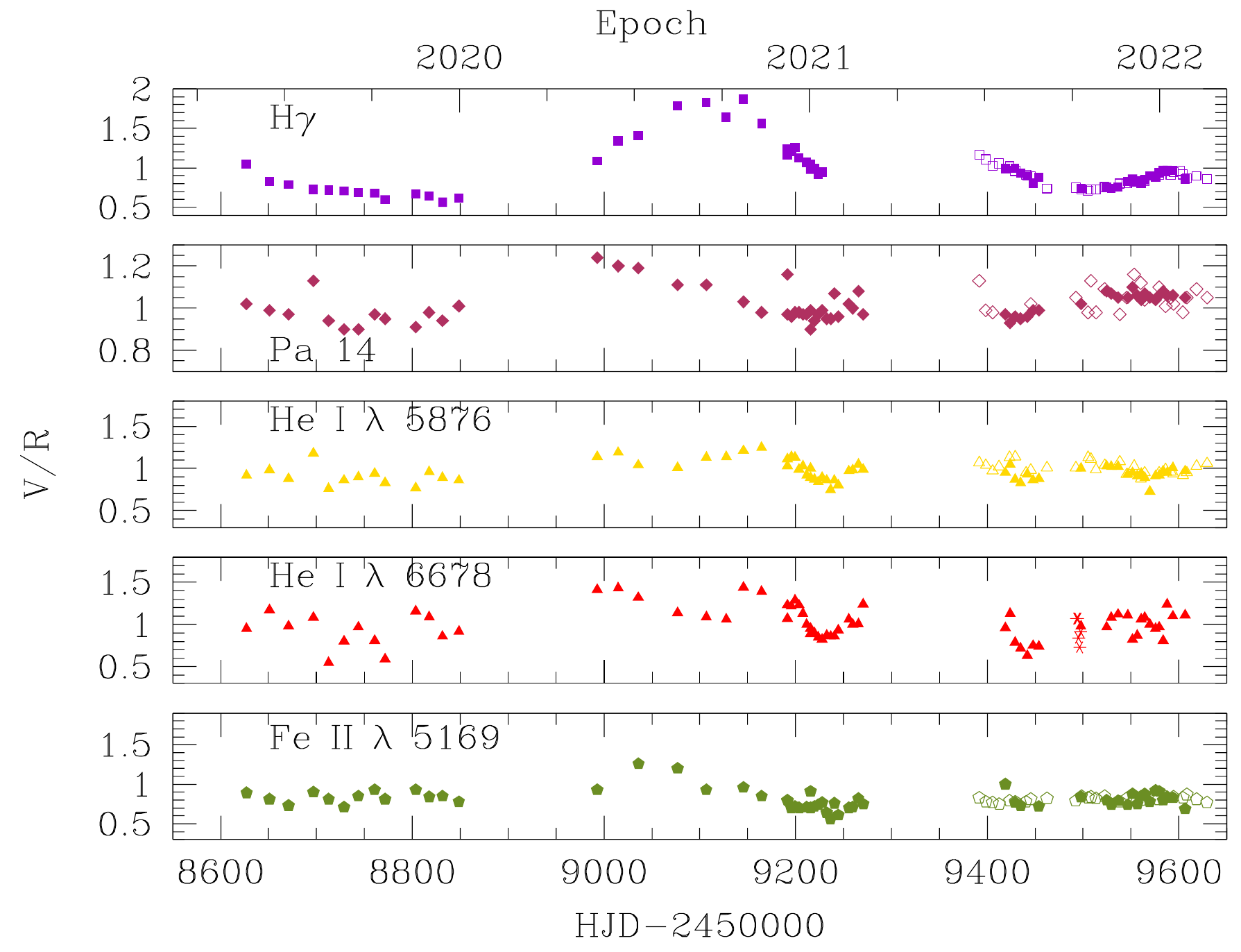}}
    \caption{V/R ratios of prominent lines in the optical spectrum of  $\gamma$~Cas as a function of time. The symbols have the same meaning as in Fig.\,\ref{monit}.\label{VR}}
\end{figure}

Figures\,\ref{montage}, \ref{monit}, and \ref{VR} illustrate the variations in the strongest emission lines over the duration of our HEROS/TIGRE campaign. These figures show that the emission gradually strengthened from June 2019 until January 2021. Following the maximum in December 2020 - January 2021, the strength of the emission lines declined again \citep[see Fig.\,\ref{monit}, and][]{Pol21} and reached a plateau in June-July 2021, followed by a slower decline.

The H$\beta$ line started with a double-peaked morphology and was undergoing V/R variations (see Fig.\,\ref{montage}). From December 2020 on, the line morphology became more complex with a dominant blue peak and two additional weaker sub-peaks. In the following months, the visibility of the sub-peaks decreased; by the end of February 2021, the profile displayed a single, relatively broad peak. In June 2021, the H$\beta$ line had recovered its double-peaked morphology, and, in the following months, the V/R variations were again clearly seen.   

In the vast majority of our spectra, the H$\alpha$ emission displayed a single emission peak on top of a somewhat broader emission component (Fig.\,\ref{montage}). The only exceptions were the spectra taken in May 2020 and January 2022. Both epochs correspond to transitions of the V/R ratio of the H$\beta$ profile switching from $< 1$ to $> 1$. Finally, the He\,{\sc i} $\lambda$~5876 line always appeared as a double-peaked feature, although the separation between the peaks clearly changed with time (see below). From July 2021 on, profile variations of He\,{\sc i} $\lambda$~5876 were essentially absent, whilst there were still important variations in the H$\alpha$ and H$\beta$ line profiles.

We measured the EWs of H$\gamma$, H$\beta$, Fe\,{\sc ii} $\lambda$~5018, Fe\,{\sc ii} $\lambda$~5169, He\,{\sc i} $\lambda$~5876, H$\alpha$, He\,{\sc i} $\lambda$~6678, and of the relatively well-isolated H\,{\sc i} Pa\,14 line\footnote{These EWs were evaluated by integrating the normalised spectra between 4336.5 and 4343.5\AA, 4850 and 4874\,\AA, 5011.5 and 5024\,\AA, 5163 and 5173.5\,\AA, 5865 and 5880\,\AA, 6530 and 6600\,\AA, 6665 and 6690\,\AA, and 8577 and 8617\,\AA for the H$\gamma$, H$\beta$, Fe\,{\sc ii} $\lambda$~5018, Fe\,{\sc ii} $\lambda$~5169, He\,{\sc i} $\lambda$~5876, H$\alpha$, He\,{\sc i} $\lambda$~6678, and Pa\,14 lines, respectively.}. For those lines that exhibited a double-peaked morphology, we also measured the RV separation between the peaks ($\Delta\,v$) as well as the V/R intensity ratios. The results are shown in Fig.\,\ref{monit}.

The gradual increase in the emission strengths noted in Fig.\,\ref{montage} is clearly seen in the EWs of H$\gamma$, H$\beta$, H$\alpha$, and He\,{\sc i} $\lambda$~5876 displayed in Fig.\,\ref{monit}. This high-emission event culminated in December 2020 - January 2021 as independently reported by \citet{Pol21}. Between 1973 and 2002, EW(H$\alpha$) varied between $\simeq -20$\,\AA\ and $-50$\,\AA\ with irregular cycles on timescales of about 10 years \citep{Miro02}. More recent data suggest a longer duration for the cycle. Indeed, \citet{Pol14} reported on a brief episode of EW(H$\alpha$) $\simeq -50$\,\AA\ in 1995 that was followed by a decay to $\simeq -25$\,\AA\ by November 2001. Since then, the H$\alpha$ emission strength gradually increased until early 2021. Because the EWs were not always measured in the same way, it is difficult to exactly compare our numbers with these literature values. Yet, our values agree very well with those of \citet{Pol14}, and since the strongest values reported in the literature agree with our peak measurements, it seems likely that the 2020--2021 event corresponds indeed to the H$\alpha$ emission strength culminating near its maximum value for this star. On top of these long-term trends, we note the presence of much lower amplitude oscillations of EW(H$\alpha$) that were previously noted by \citet{Hor94}. Following a decline phase that lasted from February 2021 until about May 2021, the value of EW(H$\alpha$) stabilised at about $-46$\,\AA\ in July-August 2021, with a shallow trend towards a slow decrease in line strength during the second half of the year 2021.

A similar overall behaviour is seen for the other H\,{\sc i} Balmer lines and for the He\,{\sc i} $\lambda$~5876 line in Fig.\,\ref{monit}. The 2020--2021 event is much less obvious in the EW data of the Pa\,14 and He\,{\sc i} $\lambda$~6678 lines. These lines are most likely formed in the inner parts of the disc \citep{Stee98}, and the relative impact of a change in disc radius on the EW will thus be more modest compared to H$\alpha$, which forms over a much wider region. Moreover, for the He\,{\sc i} $\lambda$~6678 line, we observe a rather large dispersion of the EW data, due to fluctuations occurring on short timescales \citep[see also][]{Pol21}. These rapid variations might result from non-radial pulsations and/or migrating sub-features that were previously observed in high-cadence time series of this line \citep{Smi95}. Alternatively, they could reflect ongoing mass injection from the star into the disc, which would mostly affect the conditions in the innermost part of the disc where the He\,{\sc i} $\lambda$~6678 line is formed \citep{Pol21}. Yet, even in those lines, we can find a clear signature of a change in the circumstellar envelope of $\gamma$~Cas during the high-emission event. Indeed, the RV separation of the emission peaks significantly decreased during the event.

For an optically thin Keplerian disc located in the stellar equatorial plane, the velocity separation of the emission peaks reflects to first order the radius of the line emitting region in the disc \citep[e.g.][]{HV95,Cat13,Zam19}:
\begin{equation}
  \Delta v = 2\,v\,\sin{i}\,\left(\frac{R_*}{R_{\rm disc}}\right)^{1/2}\,\cdot
  \label{eqZam}
\end{equation}
A $\Delta v$ decrease is seen for the H$\gamma$ and H$\beta$ lines (Fig.\,\ref{monit}), but our time series of $\Delta v$ measurements for those lines are truncated because of the disappearance of their double-peaked morphology around the time of maximum emission strength\footnote{Even the H$\gamma$ line displayed a single-peak morphology between January 18 and (at least) February 25, 2021.}. The He\,{\sc i} $\lambda$~5876, He\,{\sc i} $\lambda$~6678 and Pa\,14 lines maintained a double-peaked morphology throughout our entire campaign and we observe for all of them a clear drop in $\Delta v$ at the time of the high-emission event. In June-July 2021, the $\Delta v$ values stabilised close to their pre-eruption level. During the high-emission event, the radii of the emitting regions inferred from Eq.\,\ref{eqZam} apparently increased between a factor of 1.5 (for the Fe\,{\sc ii} line) to a factor of 3 (for the He\,{\sc i} lines) and even a factor of 5 (for H$\gamma$). We shall come back to the implications of Eq.\ref{eqZam} in Sect.\,\ref{trunc}.

An interesting result is that the He\,{\sc i}, Pa\,14 and Fe\,{\sc ii} lines reached their minimum $\Delta v$ about 50 -- 60 days before the Balmer lines displayed theirs (H$\gamma$) and reached their maximum emission strength (H$\gamma$, H$\beta$ and H$\alpha$). This delay reflects the difference in the formation region of these lines for a density enhancement propagating outwards across the disc.
  
Finally, from Fig.\,\ref{VR}, we note that the V/R ratio of H$\gamma$ (as long as the line displayed two emission peaks) underwent variations that are most likely a manifestation of one-armed oscillations precessing in the disc \citep{Ber99}. Indeed, such variations have been reported for $\gamma$~Cas in the literature, although the duration of the associated cycles changed from epoch to epoch \citep[][and references therein]{Tel93,Miro02}. For instance, between 1970 and 1990, the V/R variations in the H$\beta$ line had a typical timescale of $5 \pm 1$\,yr with a trend for the cycle duration to increase with time \citep{Tel93,Hor94}. Our data suggest apparently shorter timescales (of order 600 and 500\,days for H$\gamma$ and Pa\,14). The variations in the V/R ratio in the Pa\,14 line are not in phase with those of H$\gamma$, and the same holds for the rather low-level variations in the V/R of the He\,{\sc i} lines. This situation most likely again stems from different formation regions of these lines. \citet{PolGua14} collected V/R measurements of the He\,{\sc i} $\lambda$\,6678 line between 2009 and 2014. They inferred a period near 280\,days for these variations. Whilst we find that the V/R of the He\,{\sc i} lines apparently varies on shorter timescales than for the H\,{\sc i} lines, we cannot establish a clear cyclic behaviour from our data. The V/R variations in the Fe\,{\sc ii} lines mimic those of the Pa\,14 line, that is, they are shifted with respect to those of H$\gamma$ or of the He\,{\sc i} lines. The V/R data of the different lines clearly show that there is no explicit persistent correlation of their variations with the binary orbital period (203.5\,d; see also Sect.\,\ref{tomo} below). 

\begin{figure}
  \resizebox{9cm}{!}{\includegraphics[angle=0]{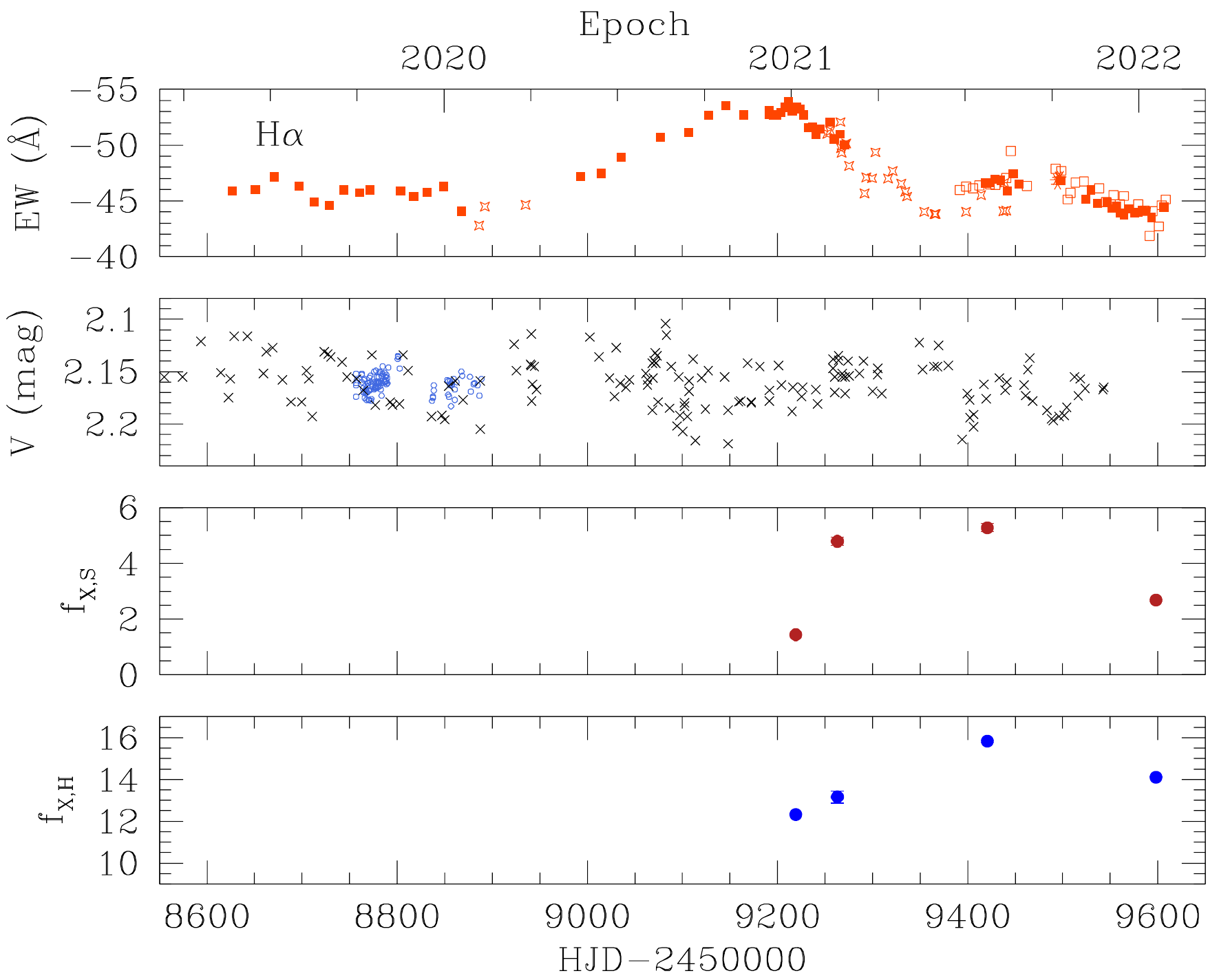}}
  \caption{Variations in EW(H$\alpha$) (top panel), in the APT and AAVSO $V$-band photometry (second panel), and in the soft (third panel) and hard X-ray fluxes (bottom panel) of $\gamma$~Cas during the 2021 disc eruption. In the second panel, APT differential magnitudes from \citet{Smi21} were shifted by 6.49\,mag and are shown as blue circles, and AAVSO photometry by observer Wolfgang Vollmann is shown as black crosses. The X-ray fluxes are in units of $10^{-11}$\,erg\,cm$^{-2}$\,s$^{-1}$.\label{AAVSOlc}}
\end{figure}

For EW(H$\alpha$) between $-27.0$ and $-36.6$\,\AA, \citet{Pol14} found a positive correlation between the H$\alpha$ emission strength and the $V$-band brightness (see also Sect.\,\ref{longtermvar}). Unfortunately, the intensive photometric monitoring with the Automated Photometric Telescope \citep[APT;][]{Rob02,Smi06,Hen12,Smi21} does not cover the events analysed here. We thus retrieved $V$-band photometry collected by observer Wolfgang Vollmann from the American Association of Variable Star Observers\footnote{https://www.aavso.org} \citep[AAVSO;][]{Kaf16} archive (see Fig.\,\ref{AAVSOlc}). The $V$-band light curve does not show any conspicuous variability at the time of the 2020--2021 high-emission event. Hence, the positive correlation between H$\alpha$ emission and $V$-band brightness does not apply to the most recent epochs. If anything, the star became slightly fainter when EW(H$\alpha$) reached its maximum, though the change in $V$-band magnitude is most probably not significant. 

\subsubsection{Radial velocities\label{radvel}}
\begin{figure}
  \resizebox{9cm}{!}{\includegraphics[angle=0]{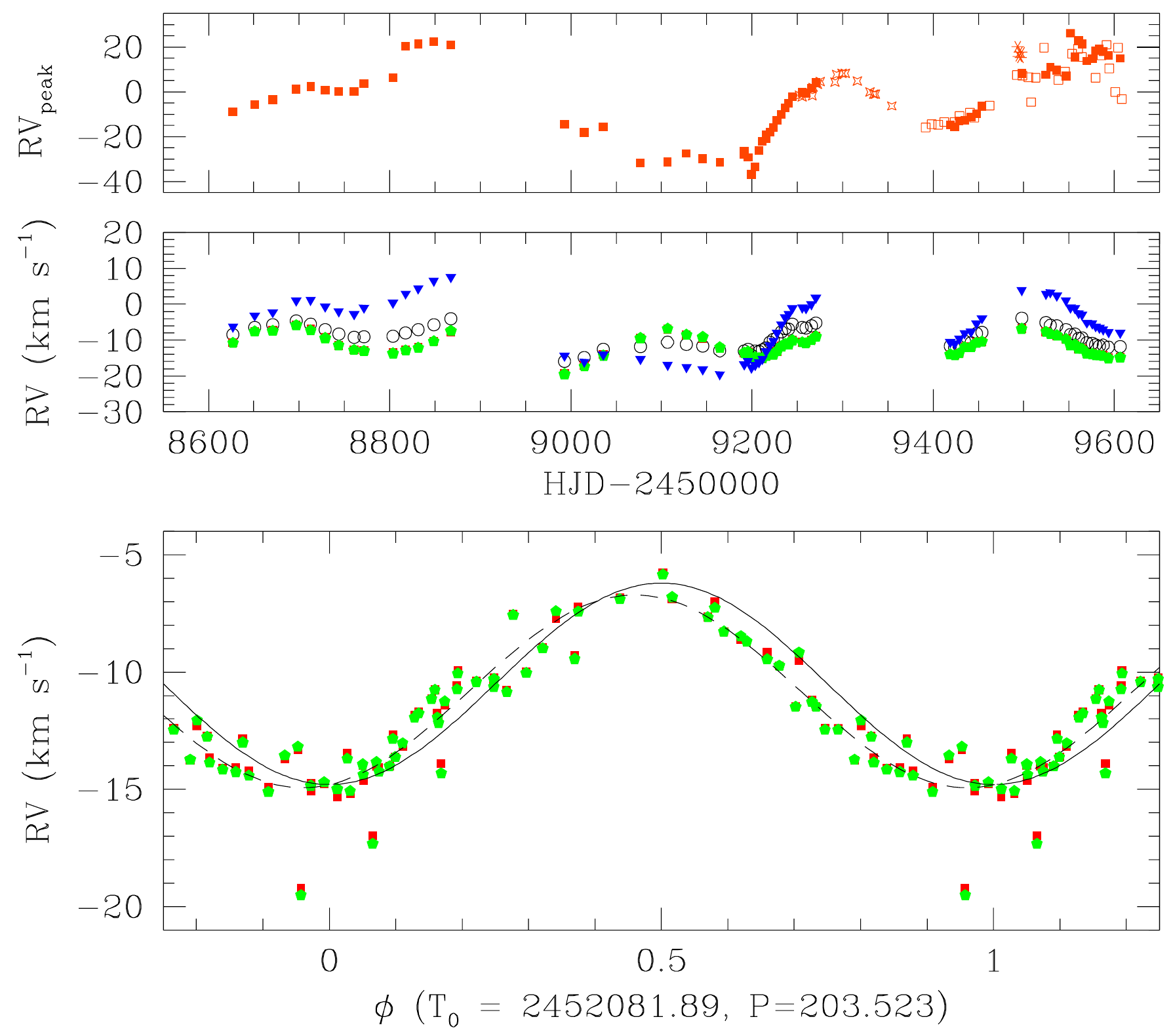}}
  \caption{RVs of the H$\alpha$ line of $\gamma$~Cas during the 2020 -- 2022 monitoring campaign. The top panel illustrates the RVs of the H$\alpha$ emission peak. The middle panel illustrates the overall RVs as a function of time. The different symbols stand for the methods used to determine the overall RVs: open dots represent first-order moment; red squares the mirror profile; green pentagons the correlation with a two-Gaussian function; and blue triangles the correlation with the February 19, 2021, spectrum. The bottom panel illustrates the RVs from the mirror profile and the correlation with two-Gaussian function methods as a function of orbital phase computed with the ephemerides of solution number 5 of \citet{Nem12}. The solid line in the bottom panel yields the RV curve of \citet{Nem12} with a systemic velocity set to  $-10.5$\,km\,s$^{-1}$, whilst the dashed line corresponds to the best-fit orbital solution obtained from our RVs (see the main text and Table\,\ref{RVsol}). The symbols have the same meaning as in the middle panel. 
\label{RVs}}
\end{figure}

For the H$\alpha$ line, we measured the RV of the emission peak (RV$_{\rm peak}$) by fitting a Gaussian to the top part of the emission (see the top panel of Fig.\,\ref{RVs}). Moreover, for the TIGRE spectra, overall H$\alpha$ RVs were determined by four different methods \citep[for details, see][]{Naz19,Naz21}: the first-order moment of the entire line profile, a comparison between the blue wing and the mirrored red wing\footnote{The `wings' refer here to those parts of the profile with intensities between $\sim 20$\% and $\sim 60$\% of the line's peak intensity above the continuum. We thus avoid the high-velocity wings where electron scattering affects the profile. Moreover, the importance of electron scattering is probably reduced as the disc is likely truncated, and does not extend to very low densities.} obtained by reversing the velocities, correlation of the line profile to a two-Gaussian function (bisector derivation at half width), and correlation of the observed profiles against the profile obtained on February 19, 2021. The results of all these measurements are displayed in the middle and bottom panels of Fig.\,\ref{RVs}.

Since the H$\alpha$ line most of the time displays a single-peaked morphology\footnote{\citet{Nem12} indicate that this was also the case between 1993 and 2010, that is, during epochs when the H$\alpha$ emission was significantly weaker. This morphology is thus not specific to the line strength reaching an extreme value, although our Fig.\,\ref{Bess} reveals a double-peaked morphology in August 2001.}, we cannot measure its V/R ratio or its $\Delta v$ value. Nevertheless, for this line, the V/R variations appear through shifts of RV$_{\rm peak}$. Comparing RV$_{\rm peak}$ of the H$\alpha$ emission (Fig.\,\ref{RVs}) to the V/R variations seen in the H$\gamma$ line (Fig.\,\ref{VR}), we note a rather good correspondence between both quantities: the H$\alpha$ RV$_{\rm peak}$ is negative when the H$\gamma$ violet peak dominates over the red peak (V/R $> 1.0$) and vice versa.

The overall RVs of the H$\alpha$ line are shown in Fig.\,\ref{RVs}. As one can see, the V/R variations that cause the changes in RV$_{\rm peak}$ (see the top panel of Fig.\,\ref{RVs}) affect the values of the RVs determined via the cross-correlation method and, to a lesser extent, via the first order moment. Indeed, these methods probe the full line profile, and can thus be affected by changes in its core. By contrast, the mirror method and the two-Gaussian correlation method both focus on the line wings. Hence, their results are much less sensitive to the long-term V/R trends. Therefore, these two methods provide a more sensitive diagnostic of the orbital motion of the Be star. We used these RVs as input to the Li\`ege Orbital Solution Package \citep[LOSP;][]{San13} code to compute an orbital solution for $\gamma$~Cas. We set the orbital period to 203.523\,d and tested both eccentric and circular orbits. Allowing for an elliptical orbit leads to a non-significant eccentricity of $e = 0.056 \pm 0.050$, and does not improve the fit compared to a circular orbit. The best-fit LOSP circular orbital solution for the RVs determined via the mirror method is displayed by the dashed curve in Fig.\,\ref{RVs} and summarised in Table\,\ref{RVsol}. Both methods of RV determination yield results that are in excellent agreement with each other and agree remarkably well with the orbital solutions of \citet{Nem12} and \citet{Smi12}, which relied on RVs inferred with similar methods. The \citet{Nem12} and \citet{Smi12} orbital solutions were obtained from data collected at epochs when the H$\alpha$ emission was significantly weaker than during our campaign, indicating that there is an excellent long-term coherence of the RV curve, at least at epochs when the line is in emission. This situation contrasts with the case of $\pi$~Aqr, where more extreme changes in emission strength were recorded. Indeed, for this star, the amplitude of the RV variations when H$\alpha$ was in strong emission was found to be half the amplitude measured when the line was in absorption \citep{Naz19}.

We observe a small shift by $\sim 0.04$ in phase (see Fig.\,\ref{RVs}) between our new orbital solution and the solution number 5 of \citet{Nem12}. This shift could indicate that the orbital period of 203.523\,days is slightly overestimated by about 0.2\,days, hence suggesting a value closer to that of orbital solutions number 1 or 6 of \citet{Nem12}.   
\begin{table}
  \caption{Orbital solution of H$\alpha$ RVs from the mirror method.\label{RVsol}}
  \begin{center}
  \begin{tabular}{l c}
    \hline\hline
    Element & LOSP solution \\
    \hline
    $P_{\rm orb}$ & 203.523 (adopted) \\
    $e$ & 0.0 (adopted) \\
    $\gamma$ (km\,s$^{-1}$) & $-10.8 \pm 0.1$ \\
    $K_{\rm Be}$ (km\,s$^{-1}$) & $4.1 \pm 0.3$ \\
    $T_0$ (HJD-2450000) & $9197.5 \pm 1.4$ \\
    $f(m)$ (M$_{\odot}$) & $0.00146 \pm 0.00031$ \\
    rms (km\,s$^{-1}$) & 1.02 \\
    \hline
  \end{tabular}
  \end{center}
  \tablefoot{$\gamma$ is the apparent systemic velocity, $K_{\rm Be}$ stands for the RV amplitude of the Be star, and $T_0$ is the time of minimum RV of the Be star. $f(m)$ is the mass function.}
\end{table}

Assuming an orbital inclination of $42^{\circ}$ \citep{Stee12} and a mass of the Be star of 16\,M$_{\odot}$, the mass function $f(m)$ implies a mass of the companion near 1.1\,M$_{\odot}$ and an orbital radius of 375\,R$_{\odot} \simeq 47.5\,R_*$.  

\begin{figure*}
  \begin{minipage}{6cm}
    \resizebox{6cm}{!}{\includegraphics[angle=0]{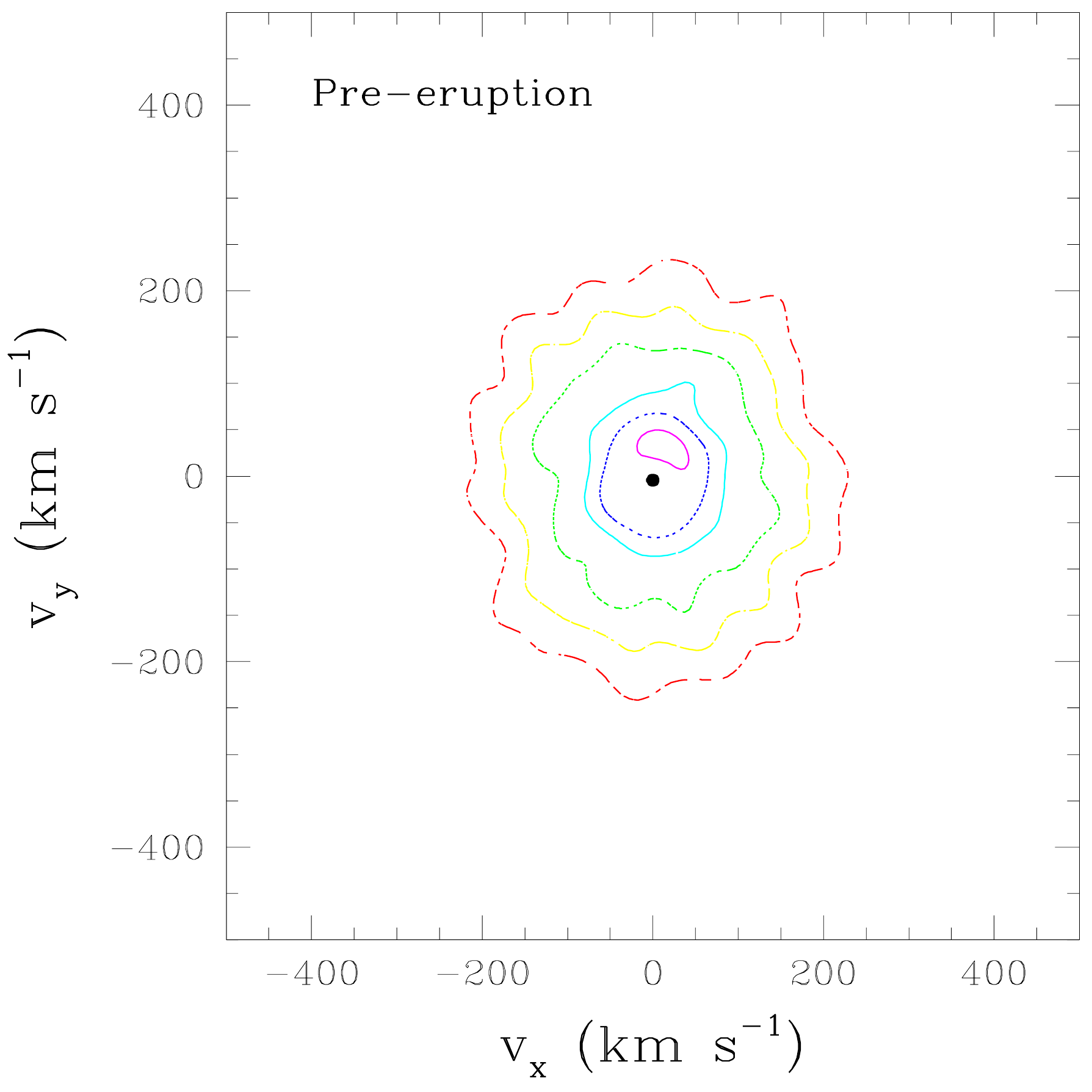}}
    \hspace*{1.15cm}
    \resizebox{4.8cm}{!}{\includegraphics[angle=0]{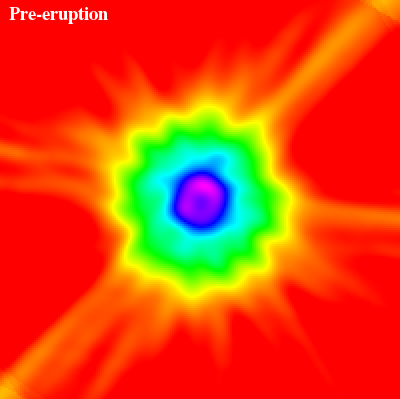}}
  \end{minipage}
  \hfill
  \begin{minipage}{6cm}
    \resizebox{6cm}{!}{\includegraphics[angle=0]{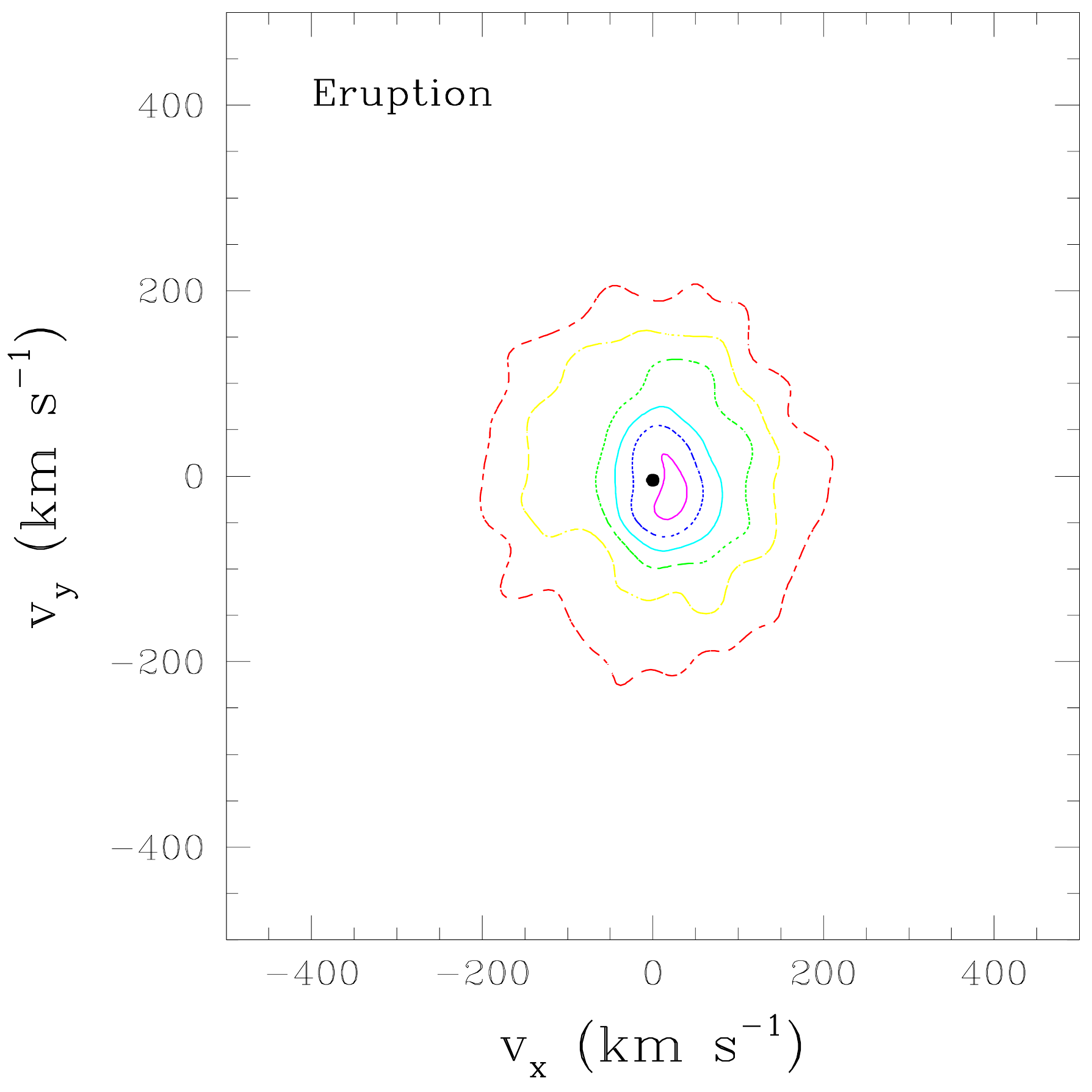}}
    \hspace*{1.15cm}
    \resizebox{4.8cm}{!}{\includegraphics[angle=0]{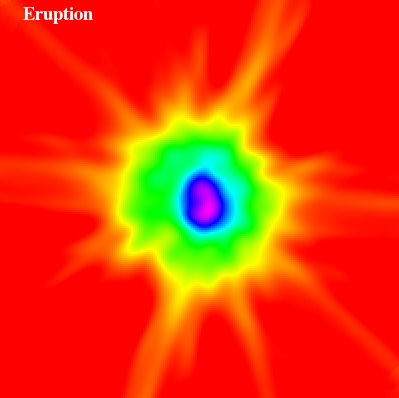}}
  \end{minipage}
  \hfill
  \begin{minipage}{6cm}
    \resizebox{6cm}{!}{\includegraphics[angle=0]{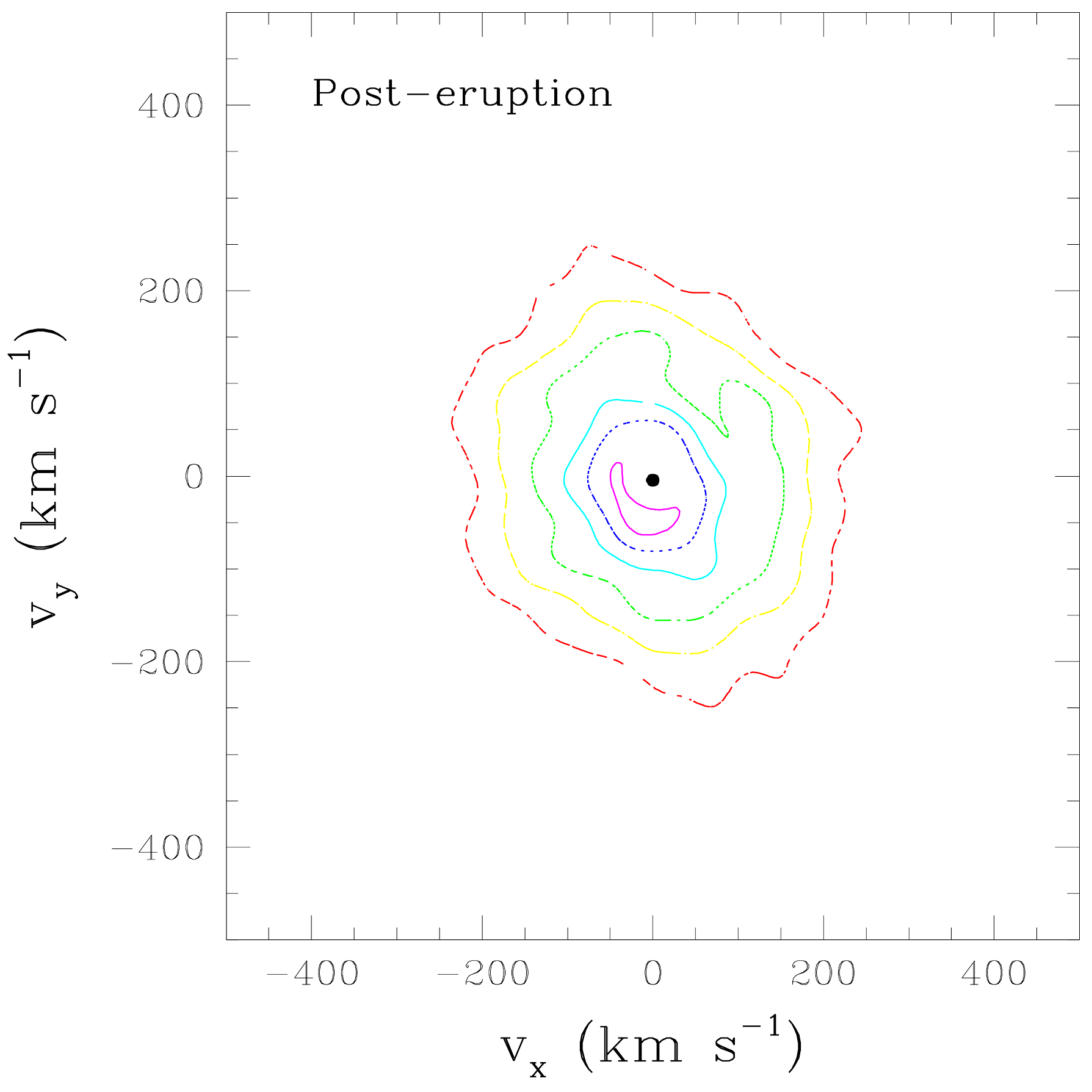}}
    \hspace*{1.15cm}
    \resizebox{4.8cm}{!}{\includegraphics[angle=0]{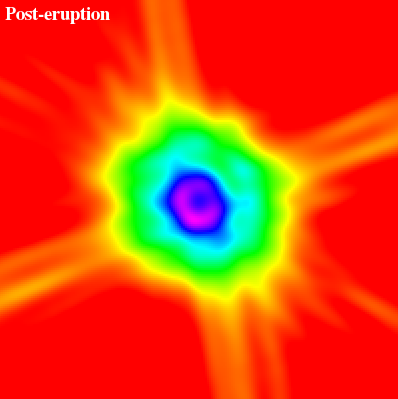}}
  \end{minipage}
  \caption{Doppler maps of the H$\alpha$ emission. From left to right the different maps correspond to the pre-eruption, eruption, and post-eruption epochs. For each epoch, the top panel illustrates the map with contours corresponding to 95\% (magenta), 80\% (blue), 65\% (cyan), 50\% (green), 35\% (yellow), and 20\% (red) of the value of the highest peak in the map. The black dot corresponds to the orbital motion of the Be star with $K_{\rm Be} = 4.3$\,km\,s$^{-1}$. The bottom panel illustrates the same map as a colour plot.\label{HaDoppmap}}
\end{figure*}
\subsubsection{Doppler tomography \label{tomo}}
The near circular orbit of $\gamma$~Cas allows us to apply our Doppler tomography algorithm \citep{Rau02,Naz19b} to map the Be disc in velocity space. Our method is based on the Fourier-filtered back-projection technique \citep{Hor91}. In the rotating frame of reference of a binary system with a circular orbit, the RV, as seen by an external observer, of a stationary structure is given by
\begin{eqnarray}
  v(\phi) & = & V_x\,\sin{i}\,\sin(2\,\pi\,\phi) + V_y\,\sin{i}\,\cos(2\,\pi\,\phi) + \gamma \nonumber \\
  & = & v_x\,\sin(2\,\pi\,\phi) + v_y\,\cos(2\,\pi\,\phi) + \gamma\,\cdot
\end{eqnarray}
In this expression, $\phi$ is the orbital phase as measured from the time of minimum RV of the Be star (see Fig.\,\ref{RVs}), and $\gamma$ is the systemic velocity. $V_x$ and $V_y$ are the components of the velocity vector of the structure in the orbital plane, with the $x$-axis pointing from the Be star towards its companion, and the $y$-axis pointing in the direction of the orbital velocity vector of the companion star. We further have that $v_x = V_x\,\sin{i}$ and $v_y = V_y\,\sin{i}$ where $i$ is the orbital inclination. In the $(v_x, v_y)$ coordinates, the orbital velocity of the Be star is given by $(0,-K_{\rm Be})$. 

The Doppler map of an emission line is a representation of the line emission region in the $(v_x, v_y)$ plane. For a Keplerian disc, where the orbital velocity of the disc material falls off as $1/\sqrt{r}$, with $r$ the distance from the centre of the star, we expect a ring-like feature in the Doppler map that is seen inside out. Indeed, the highest orbital velocities (hence corresponding to the outer radius of the ring in velocity space) are found at the inner border of the disc, whilst the lowest orbital velocities are found at the outer edge of the disc.

To study the evolution of the disc with epoch, we divided our TIGRE data into three groups, each of which sample a bit more than one full orbital period of $\gamma$~Cas. These three groups correspond to the pre-eruption, eruption and post-eruption epochs, and cover the 2019--2020, 2020--2021 and 2021--2022 observing seasons\footnote{This corresponds to the time intervals (in HJD-2450000) from 8626.97 until 8867.61, from 8992.97 until 9270.56, and from 9418.92 until 9606.57.}. In this exercise, we restricted ourselves to the TIGRE data as they provide the longest homogeneous time series among our optical spectra. We built Doppler maps for the H$\alpha$, H$\beta$, and He\,{\sc i} $\lambda$~5876 emission lines, adopting a velocity step of 5\,km\,s$^{-1}$, which over-samples the $\sim 15$\,km\,s$^{-1}$ resolution element of the HEROS spectrograph. In the calculation of each map, we adopted the value of the systemic velocity $\gamma = -10.5$\,km\,s$^{-1}$, inferred from the orbital solution. The results are illustrated in Figs.\,{\ref{HaDoppmap} and \ref{HbDoppmap}.

\begin{figure*}
  \begin{minipage}{6cm}
    \resizebox{6cm}{!}{\includegraphics[angle=0]{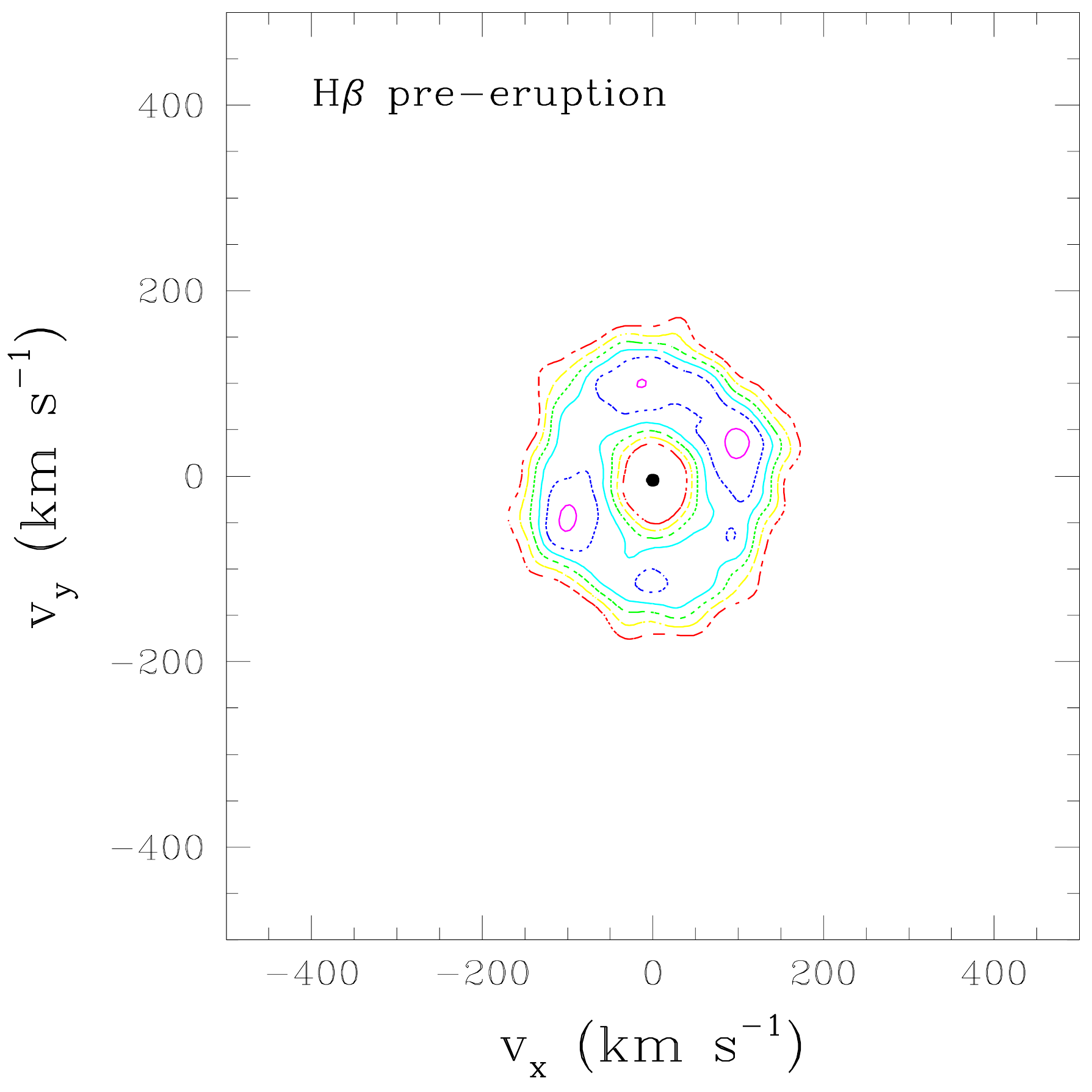}}
%    \hspace*{1.15cm}
%    \resizebox{4.8cm}{!}{\includegraphics[angle=0]{tomogHb1.png}}
  \end{minipage}
  \hfill
  \begin{minipage}{6cm}
    \resizebox{6cm}{!}{\includegraphics[angle=0]{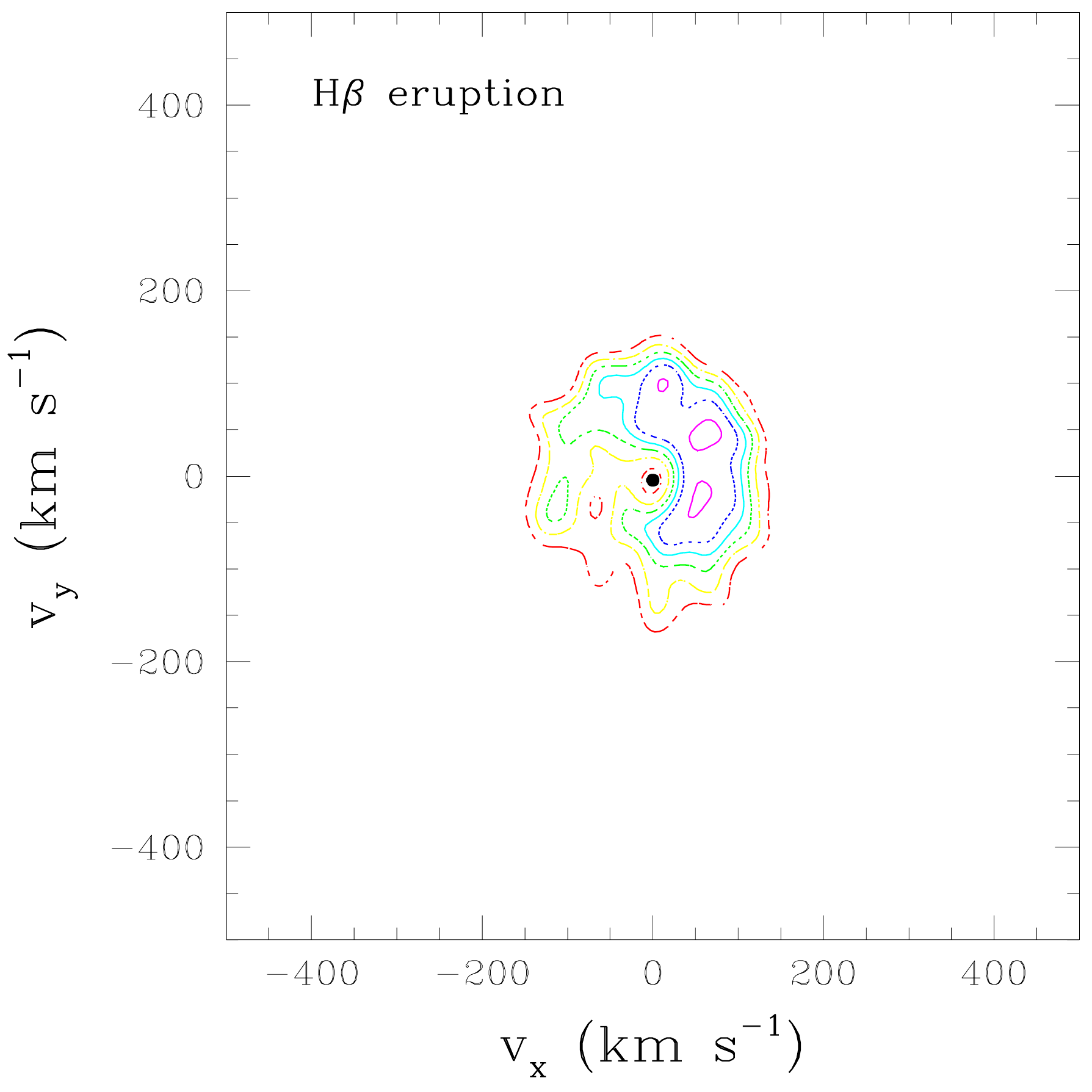}}
%    \hspace*{1.15cm}
%    \resizebox{4.8cm}{!}{\includegraphics[angle=0]{tomogHb2.png}}
  \end{minipage}
  \hfill
  \begin{minipage}{6cm}
    \resizebox{6cm}{!}{\includegraphics[angle=0]{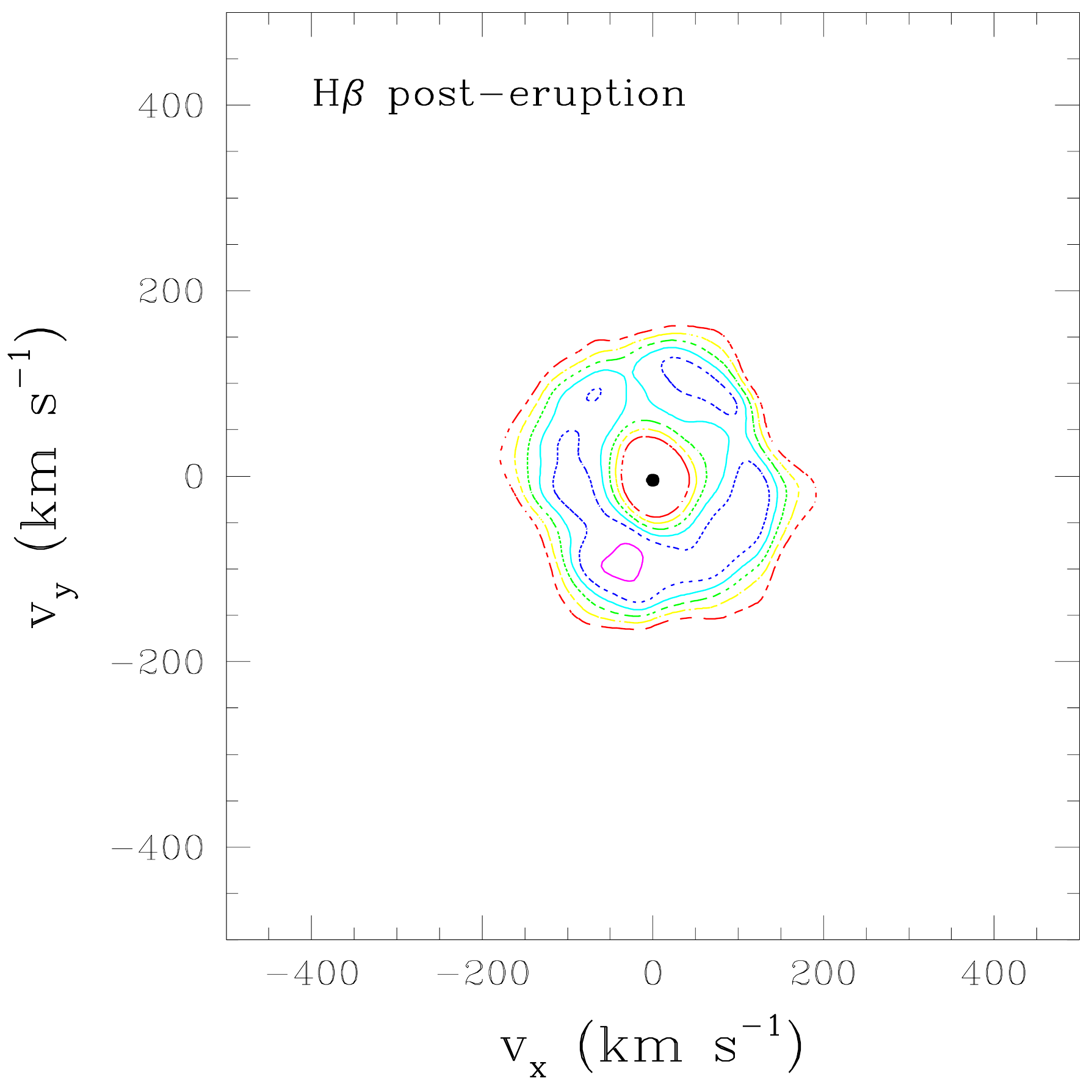}}
%    \hspace*{1.15cm}
%    \resizebox{4.8cm}{!}{\includegraphics[angle=0]{tomogHb3.png}}
  \end{minipage}
  \begin{minipage}{6cm}
    \resizebox{6cm}{!}{\includegraphics[angle=0]{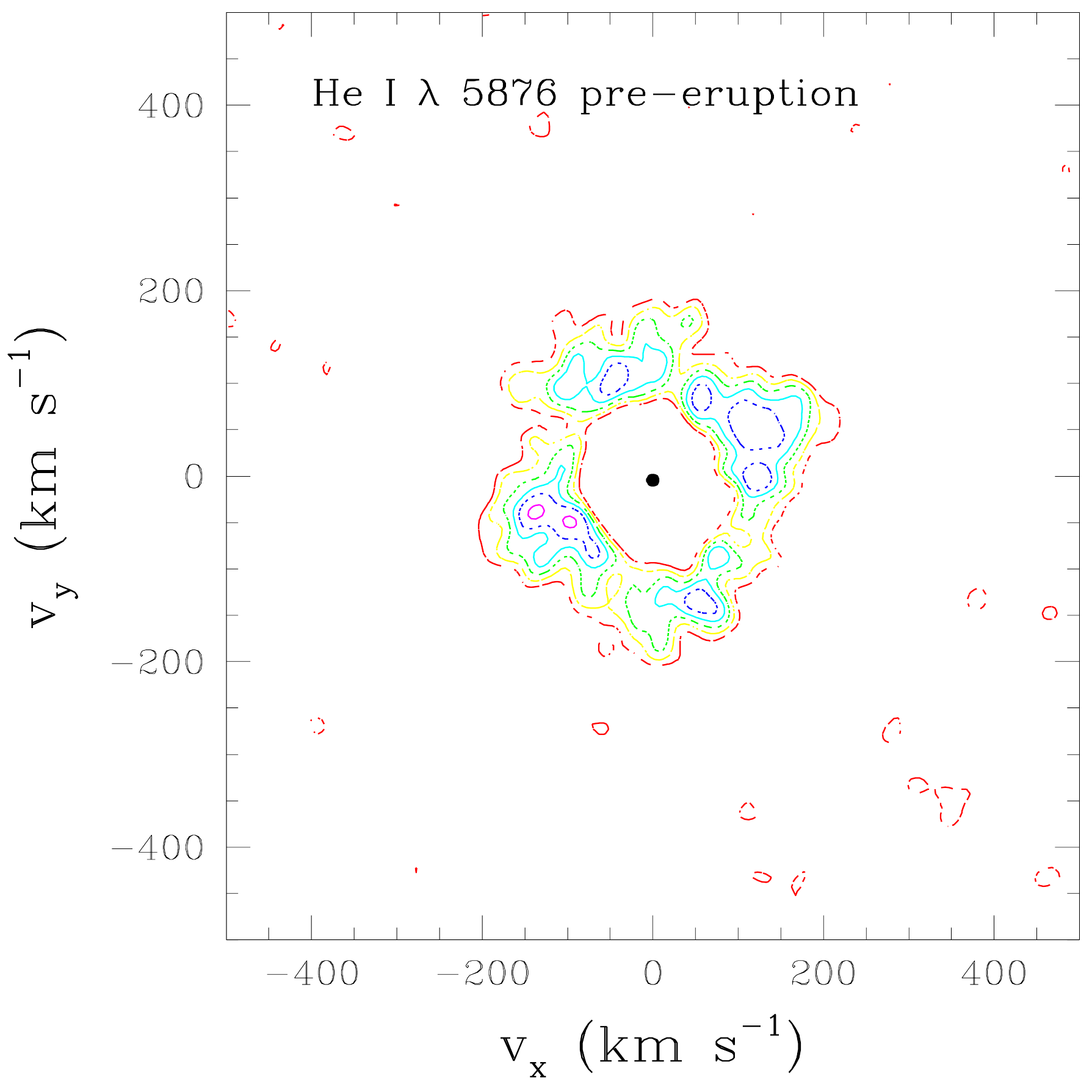}}
  \end{minipage}
  \hfill
  \begin{minipage}{6cm}
    \resizebox{6cm}{!}{\includegraphics[angle=0]{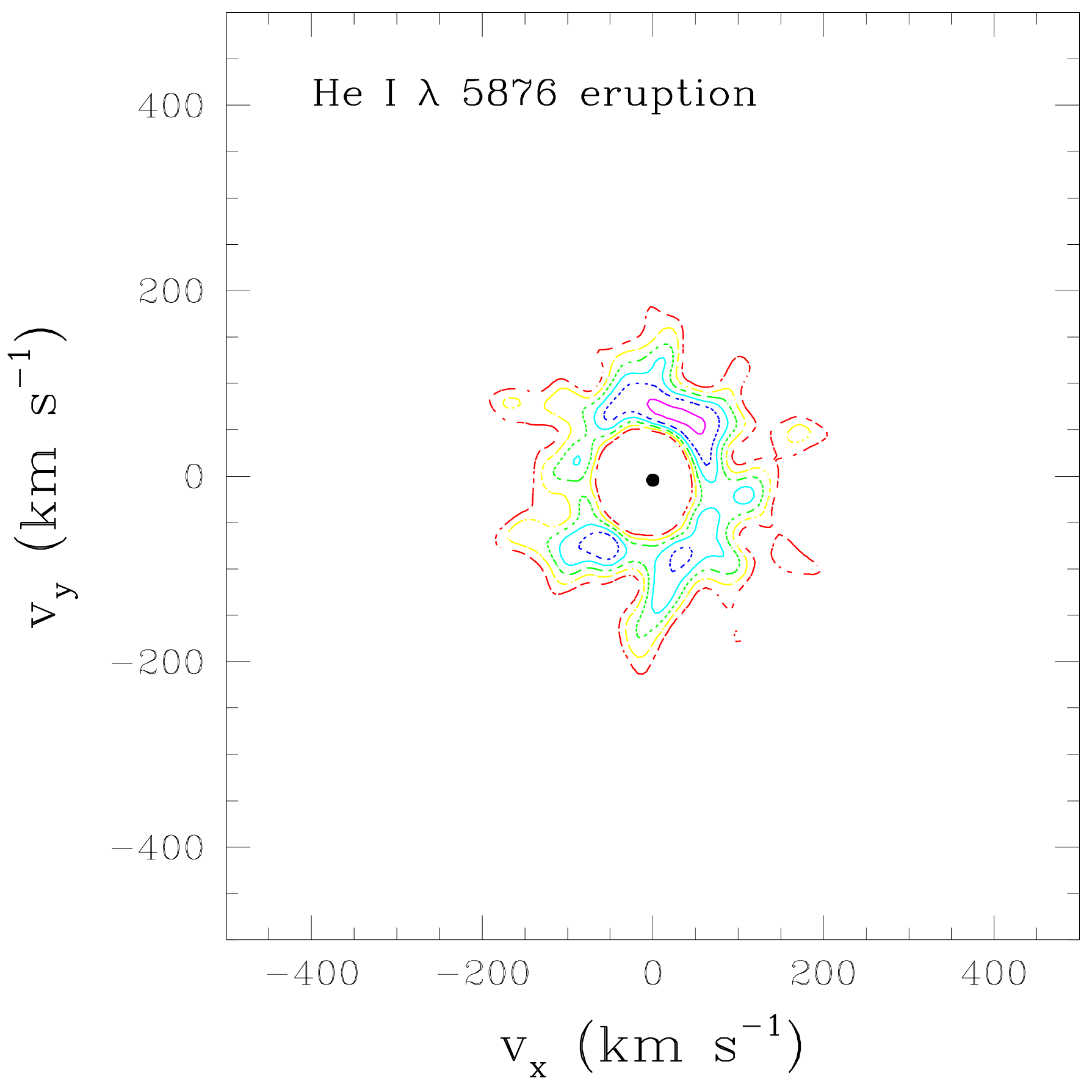}}
  \end{minipage}
  \hfill
  \begin{minipage}{6cm}
    \resizebox{6cm}{!}{\includegraphics[angle=0]{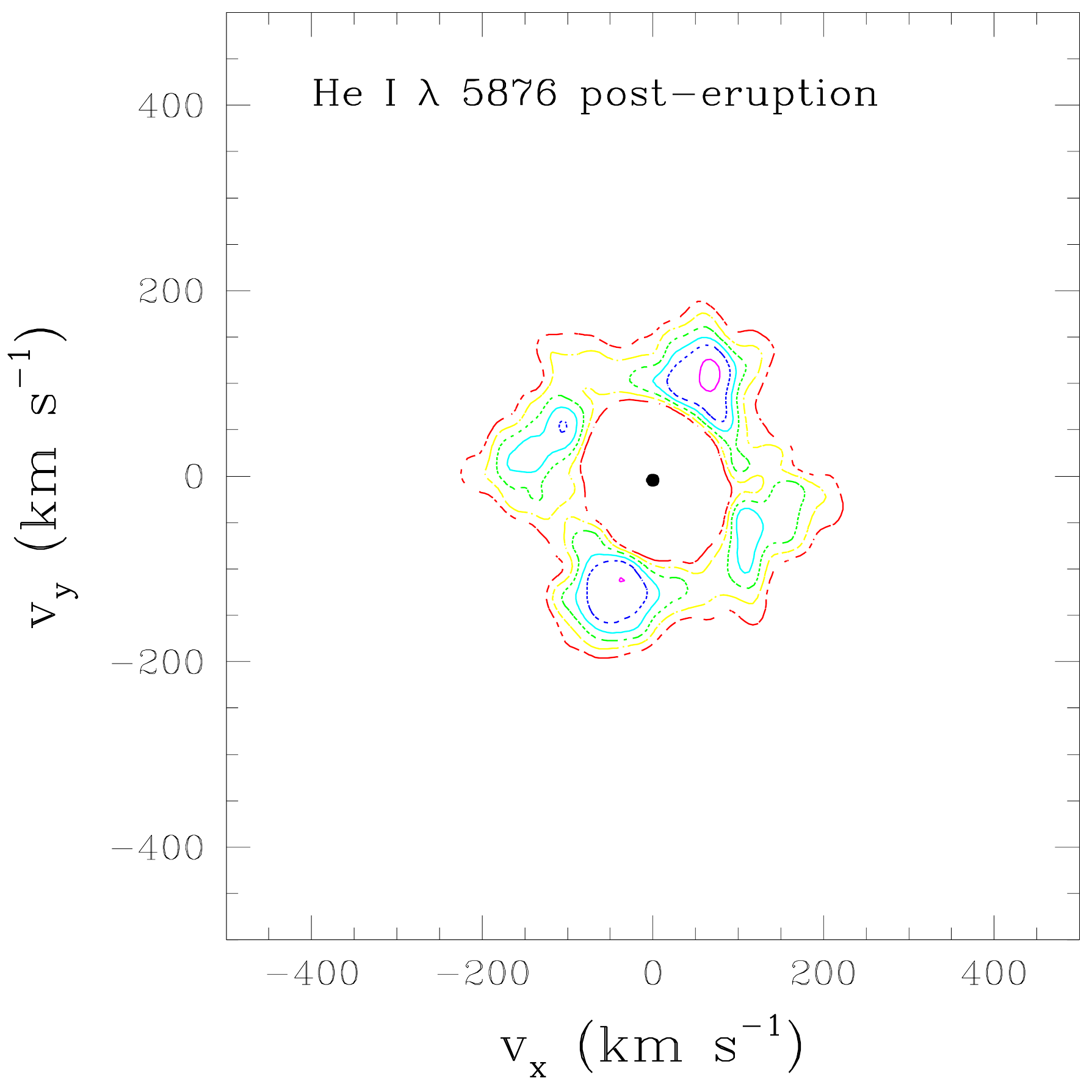}}
  \end{minipage}
  \caption{Doppler maps of the H$\beta$ (top row) and He\,{\sc i} $\lambda$~5876 (bottom row) lines. From left to right the different maps correspond to the pre-eruption, eruption, and post-eruption epochs. The contours correspond to 95\% (magenta), 80\% (blue), 70\% (cyan), 60\% (green), 50\% (yellow), and 40\% (red) of the value of the highest peak in the map. No correction for the photospheric absorption component was applied.\label{HbDoppmap}}
\end{figure*}

As expected, our Doppler maps display ring-like emission regions. Because of the inside-out view in velocity space, the presence or absence of a central hole in these structures reflects the location of the outer edge of the emission regions. The maps of the H$\alpha$ line (Fig.\,\ref{HaDoppmap}) lack a central hole at all epochs, in agreement with the overall picture that this line forms over a large part of the disc, extending to very large radii. By contrast, the map of the He\,{\sc i} $\lambda$~5876 line (see Fig.\,\ref{HbDoppmap}) displays a clear central hole at all epochs, but the radius of the hole is clearly reduced during the eruption, indicating, as expected from the RV separation of the emission peaks, that the formation region of this line has expanded during the eruption. Finally, the Doppler map of the H$\beta$ line displays an intermediate behaviour (Fig.\,\ref{HbDoppmap}). Indeed, whilst a central hole is clearly present before and after the eruption, it has been filled up during the eruption, again indicating an increase in the outer edge of the H$\beta$ emission region.

The Doppler maps can also be used to look for the signature of a phase-locked density perturbation due to the influence of the companion on the Be disc. Such a feature was reported for instance in the case of $\pi$~Aqr \citep{Zha13}, though it disappeared at more recent epochs \citep{Naz19b}. Such a perturbation could stem from tidal interactions of the companion with the Be disc, or, in the specific context of the binary-related interpretation of the hard X-ray emission, it could arise either via accretion by the companion or through the collision of the companion's wind onto the disc. The Doppler maps of the H$\alpha$ line indeed reveal an area of enhanced emission strength in the external regions of the disc (i.e.\ in the innermost part of the disc in velocity space; see the magenta contours in Fig.\,\ref{HaDoppmap}). Yet, our multi-epoch data indicate that this structure is clearly not stationary in the frame of reference of the binary. Indeed, it moves from the upper-right quadrant in the pre-eruption Doppler map, to the lower right in the eruption map and finally to the lower left in the post-eruption map. A similar behaviour is observed for the H$\beta$ line. Therefore, our Doppler maps do not support the existence of an orbitally locked density structure in the disc of $\gamma$~Cas. In other words, the structure responsible for the V/R variations does not seem to be tied to the orbital motion, in line with our conclusion of Sect.\,\ref{lpv}. In this context, it is interesting to recall that long-baseline optical interferometry observations revealed a one-armed oscillation in the disc of $\gamma$~Cas that was found to precess on a timescale near 7\,years \citep{Ber99}, also very different from the orbital period.

Smoothed particle hydrodynamics simulations of Be HMXBs \citep{Mar14,Mar19,Mar21} and equal-mass Be + B binaries \citep{Suf22} indicate that Be discs that are inclined with respect to the orbital plane may get tilted away from the equatorial plane, precess about the orbital axis, and undergo Kozai-Lidov oscillations. These dynamical interactions with the companion act on timescales that are an order of magnitude longer than the orbital period. Whilst these simulations can probably not be applied directly to $\gamma$~Cas, they could offer a possible explanation for some of the long-term variations in the optical spectrum such as the shell phases observed in the 1930s (see Sect.\,\ref{intro}).

These simulations also suggest that dynamical interactions can lead to the re-accretion of disc material by the Be star. However, such a process cannot explain the hard X-ray emission of $\gamma$~Cas and its analogues. In fact, assuming a mass of 16\,M$_{\odot}$ and a radius of 7.9\,R$_{\odot}$ as inferred from the interferometric measurements of \citet{Stee12} and the revised {\it Hipparcos} trigonometric parallax of $\gamma$~Cas of $\varpi = (5.94 \pm 0.12)\,10^{-3}$\,arcsec \citep[][implying a distance of $\sim 168$\,pc]{vLeu07} yields an escape velocity of $v_{\rm esc} = 879$\,km\,s$^{-1}$ for the Be star in $\gamma$~Cas. This is a strict upper limit to the orbital velocity that disc particles on eccentric trajectories might reach. For a pure hydrogen gas, the maximum temperature that could be reached by accretion onto the surface of the Be star is thus given by the condition of complete thermalisation,
  \begin{equation}
    \frac{1}{2}\,(m_e + m_p)\,v_{\rm esc}^2 = (m_e + m_p)\,\frac{G\,M_*}{R_*} = 2\,\frac{3}{2}\,kT \label{thermalT}
  ,\end{equation}
  where the factor $2$ in the right-hand side term stems from the fact that a hydrogen atom consists of two particles (a proton and an electron). This relation leads to an upper limit of $kT = 2.7$\,keV, which is much lower than the $\sim 12.8$\,keV plasma temperature that dominates the observed X-ray emission of $\gamma$~Cas (see Sect.\,\,\ref{Xdata}).      
  Assuming the temperature of the plasma stems from complete thermalisation of the kinetic energy of the gas, Eq.\,(\ref{thermalT}) indicates that velocities of at least 2000\,km\,s$^{-1}$ are required to account for the observed plasma temperature. The only outflow of the Be star that could reach such velocities is the polar wind. However, the polar wind only reaches such velocities at large distances from the star, and, in any case, it is not expected to thermalise by impacting either the star or the inner disc regions.
    
\subsection{Disc truncation\label{trunc}}
In low eccentricity ($e \leq 0.2$) Be binaries, such as $\gamma$~Cas, the Be disc is expected to be truncated by the gravitational interaction with the companion at the 3:1 resonance radius \citep{Oka01}. This radius is given by
\begin{equation}
  R^{3/2}_{3:1} = \frac{\sqrt{G\,M_{\rm Be}}}{2\,\pi}\,P_{\rm orb}\,\frac{1}{3}
\end{equation}\citep[e.g.][]{Zam19}. Adopting again a mass of 16\,M$_{\odot}$ for the primary star, and a radius $R_*$ of 7.9\,R$_{\odot}$, we obtain $R_{3:1} = 22.3\,R_*$.

\citet{Kle17} investigated the issue of disc truncation through an analysis of the spectral energy distribution (SED). From the turndown in the SED between the far-IR and the radio wavelengths, they establish the presence of a truncated disc in several Be stars, including $\gamma$~Cas. For this object, they infer $R_* = 9.24$\,R$_{\odot}$, and $R_{\rm out} = 35 \pm 5\,R_*$. Whilst this outer radius is clearly larger than the 3:1 resonance radius, \citet{Kle17} caution that the actual value of $R_{\rm out}$ could be lower. Indeed, their model does not reproduce the observed SED in the radio domain well, and discarding the cm data from their SED fitting would lead to a reduced disc size. \citet{Suf22} stress that significant amounts of disc material can be present beyond the transition radius. Whilst the dynamics of the material inside this transition radius is dominated by viscosity, the influence of the companion increases significantly beyond this radius. In their simulations, the disc material can thus have eccentric orbits beyond the transition radius, whilst the trajectories are nearly circular within that radius.

In principle, the formation radii of the optical emission lines can be inferred from the data analysed here. For instance, Eq.\,\ref{eqZam} directly connects the velocity separation between the peaks of double-peaked line profiles to the projected rotational velocity of the star and the radius of the line formation region. This relation was used by \citet{Zam19} for the low-activity persistent Be HMXB X~Per. These authors determined the temporal evolution of the outer radius of the H$\alpha$ emission region in X~Per from the $\Delta v$ values of the double-peaked emission line. The corresponding histogram of the inferred H$\alpha$ disc radii displayed its highest peak at the 10:1 resonance, followed by secondary peaks at the 3:1 and 2:1 resonance radii. Applying this method to $\gamma$~Cas is more complicated, not the least because the $v\,\sin{i}$ of $\gamma$~Cas is controversial. From the full width at half maximum of the He\,{\sc i} $\lambda$~4471 and Mg\,{\sc ii} $\lambda$~4481 lines, \citet{Abt02} inferred a $v\,\sin{i}$ of 295\,km\,s$^{-1}$. However, \citet{Har02} showed that values close to 380\,km\,s$^{-1}$ provide a better fit to the photospheric wings of the He\,{\sc i} $\lambda$~6678 line profile than values around 230\,km\,s$^{-1}$ as proposed by \citet{Sle82}. Accounting for gravity darkening effects, \citet{Fre05} determined $v\,\sin{i} = 432 \pm 28$\,km\,s$^{-1}$ from the He\,{\sc i} $\lambda$~4471 and Mg\,{\sc ii} $\lambda$~4481 lines. We thus assume here that $v\,\sin{i}$ must be somewhere in the range 295 to 432\,km\,s$^{-1}$. Adopting the lower value of $v\,\sin{i}$ and applying Eq.\,\ref{eqZam}, we obtain line formation radii during the outburst of $\geq 85\,R_*$ for H$\gamma$, $\sim 26\,R_*$ for the He\,{\sc i} lines, and still $\sim 17\,R_*$ for the Fe\,{\sc ii} $\lambda$~5169 line. Assuming instead the higher value of $v\,\sin{i}$, the radii would be twice larger. These radii are significantly larger than the 3:1 resonance radius and for some lines even exceed the radius of the companion's orbit ($\sim 375$\,R$_{\odot} \simeq 47.5\,R_*$; see Sect.\,\ref{radvel}). This result suggests that Eq.\,\ref{eqZam} cannot be readily applied in the present case. Indeed, if the H$\alpha$ emission region of the disc were that close to the companion, we would expect to observe a phase-locked structure in the corresponding Doppler map, at odds with our conclusion of Sect.\,\ref{tomo}. The most likely explanation for the failure of Eq.\,\ref{eqZam} to provide a good estimate of the disc size is that the disc is not optically thin, at least not in the Balmer H\,{\sc i} lines.   

\citet{Gru06} present an empirical scaling relation between the square root of the absolute value of EW(H$\alpha$) and the radius of the H$\alpha$ emission region calibrated on interferometric measurements of the disc sizes. Assuming this relation holds for a binary system such as $\gamma$~Cas, the EW(H$\alpha$) value of about $-52$\,\AA\ observed in December 2020 -- January 2021 imply $R_{H\alpha}/R_* \simeq$ 10 -- 14, which is much lower than the values inferred above, but similar to the value found by \citet{Stee98}. If this relation yields the correct answer, then the radius of the H$\alpha$ emitting region, even at maximum emission strength, would be well below the 3:1 truncation radius, and would instead be closer to the 9:1 -- 6:1 resonances.
  
\subsection{X-ray data \label{Xdata}}
The new X-ray observations obtained during the 2020 -- 2021 high-emission event sample an emission state of the $\gamma$~Cas Be disc that was never analysed before with {\it Chandra} or {\it XMM-Newton} data. Indeed, optical spectroscopy of $\gamma$~Cas at the epochs of previous X-ray observations, extracted from the BeSS database, shows that the H$\alpha$ line was much stronger in 2020 -- 2021 than for any of these previous observations (Fig.\,\ref{Bess}).
\begin{figure}[h!]
    \resizebox{8cm}{!}{\includegraphics[angle=0]{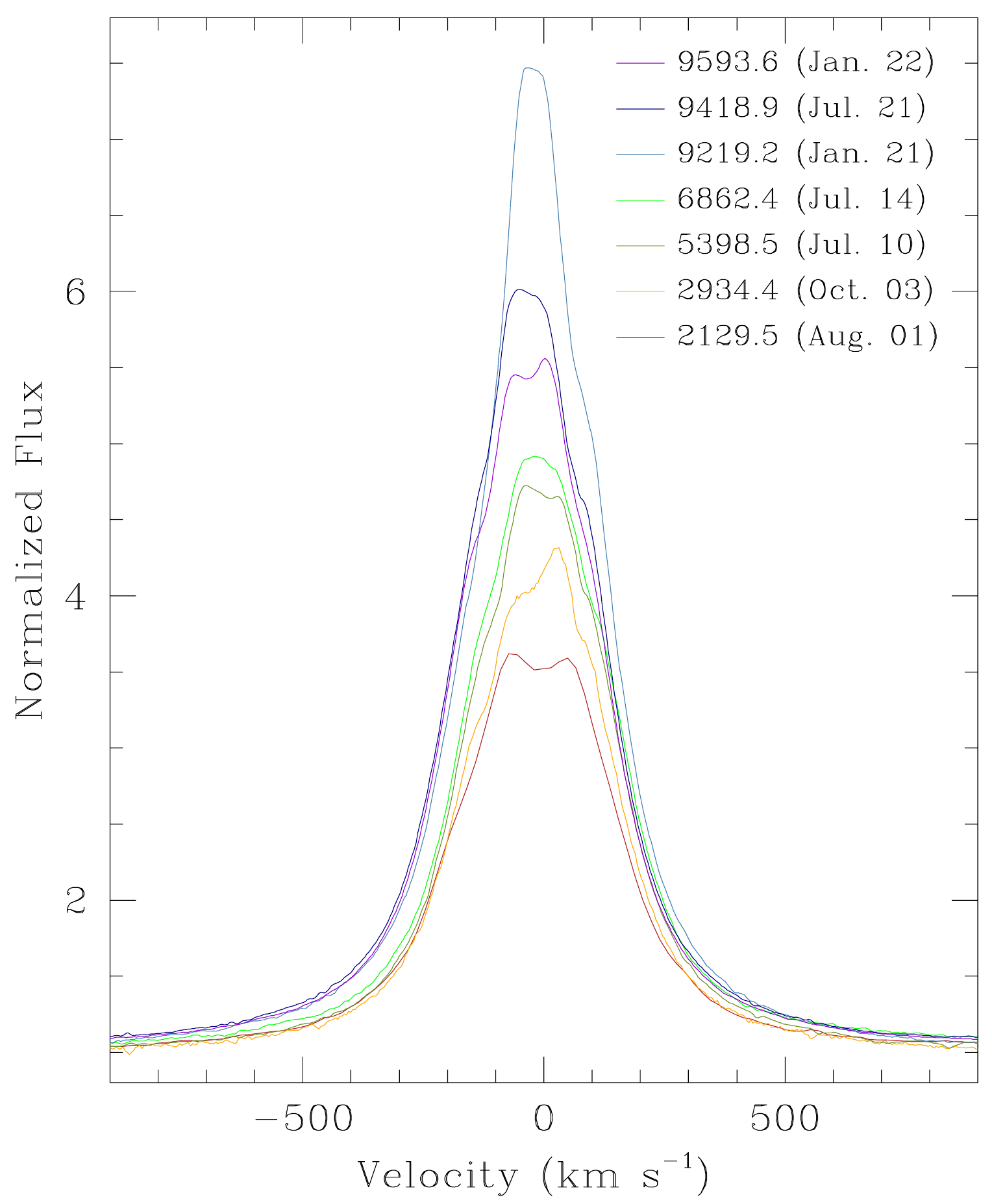}}
    \caption{H$\alpha$ emission line profile of $\gamma$~Cas at the different epochs of the X-ray observations discussed here. The labels indicate the date of the observation in the format HJD$-2\,450\,000$. Except for the July 2021 and January 2022 spectra (TIGRE observations), the data were taken from the BeSS database. \label{Bess}}
\end{figure}

In this section we analyse the new X-ray data along with the previous {\it Chandra} and {\it XMM-Newton} data. Spectral analysis of the data was performed under {\tt xspec} version 12.9.0i with solar abundances set to the values of \citet{Asp09}.
\subsubsection{X-ray light curves \label{Xraylc}}
We extracted EPIC light curves in the 0.5 -- 2.0 (soft, $S$) and 2.0 -- 10.0\,keV (hard, $H$) energy bands with time bins of 10 and 100\,s. The light curves from the two EPIC-MOS and the EPIC-pn instrument are in good agreement. The 10\,s EPIC-pn light curves along with the hardness ratio that we define here as $HR = (H-S)/(H+S)$  are shown in Fig.\,\ref{lc3859_10s} for the January 2021 observation, and in Figs.\,\ref{lc3881_10s} -- \ref{lc4049_10s} for the three other exposures of our monitoring campaign. Using the Fourier periodogram method of \citet{HMM} and \cite{Gos01}, we computed the power spectrum of the light curves with 10\,s bins (Fig.\,\ref{power}). In addition to the Fourier periodogram, we also computed the $Z^2_n$ periodograms \citep{BBB83} to search for stable periodicities in the EPIC-MOS1, EPIC-MOS2, and EPIC-pn event lists down to 5\,s, but none was found.
\begin{figure}[h!]
    \resizebox{8cm}{!}{\includegraphics[angle=0]{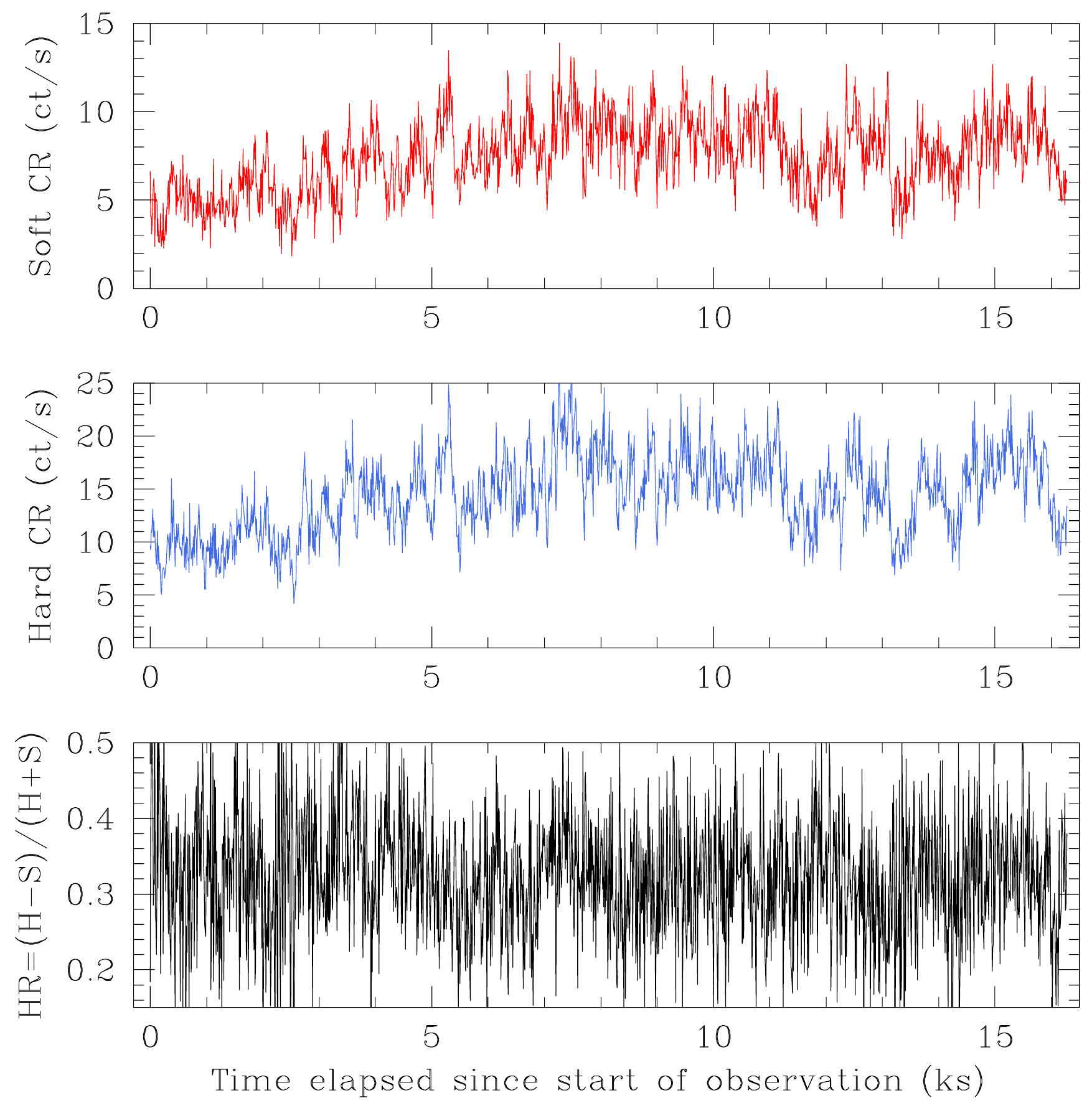}}
    \caption{EPIC-pn soft- (0.5 -- 2.0\,keV, top panel) and hard-band (2.0 -- 10.0\,keV, middle panel) light curves and hardness ratio (bottom panel) of $\gamma$~Cas during revolution 3859 (January\ 2021) with a timestep of 10\,s.\label{lc3859_10s}}
\end{figure}

The general behaviour of the X-ray light curves during our observations as well as the shape of the associated power spectra are very much consistent with previous descriptions \citep[e.g.][]{Par93,Smi98a,Rob00,Shr15}. The light curves exhibit the typical shots in the $S$ and $H$ bands, but they usually produce no significant change in $HR$. \citet{Smi98a} found that the shots occur down to the 4\,s time resolution limit of their RXTE data. Therefore, single bins with a high count rate in our 100\,s light curves, and to a lesser extent also in the 10\,s light curves, likely represent shot aggregates rather than individual events.  

Over the 16\,ks duration of the January 2021 observation, obtained when the H$\alpha$ emission was at its maximum strength (Fig.\,\ref{monit}), the count rate varies by nearly a factor of 4 between its lowest value and its maximum reached during a shot. What is remarkable is the comparatively low value of the count rate during this observation. Whilst this is true for the overall count rate, the difference between the count rates of the January 2021 and February 2021 observations (when the optical line emission was slowly declining) is especially pronounced in the soft band. This affects the mean value of $HR$; it was near 0.3 during the January 2021 observation, whereas it was close to $-0.25$ during the February 2021 observation and near $-0.2$ during the July 2021 observation (when the optical emission had returned to its pre-eruption level). In January 2022, the soft band emission was again strongly diminished ($HR$ around 0.1) though less than in January 2021.  

\citet{Ham16} reported six hardening events during a {\it Suzaku} X-ray Imaging Spectrometer (XIS) observation. These softness dips were seen by comparing the 0.5 -- 1.0 and 1.0 -- 2.0\,keV XIS count rates with the count rates in the 4.0 -- 9.0\,keV band. These dips were most likely due to partial covering absorption ($N_{\rm H} \simeq (2$ -- $8)\,10^{21}$\,cm$^{-2}$) that affects between 40 and 70\% of the 12\,keV plasma component. The longest event that was seen during this {\it Suzaku} observation lasted $\geq 22.5$\,ks. \citet{Smi19} re-analysed the archival {\it XMM-Newton} observations and found a number of such softness dips that differ in amplitude, sharpness and duration. In the 2014 observation, they further noted a hard dip where the flux decreased at high energies, but not in the soft band.

Within each light curve, there are some variations in the hardness ratio, but their amplitudes are generally small compared to the error bars. Overall, the hardness ratio does not display any well-defined trends over the duration of the individual observations. An interesting point is the fact that the longest dip in the {\it Suzaku} light curve actually lasted longer than the duration of our individual {\it XMM-Newton} observations. This opens up the possibility that the unusual hardness of the January 2021 and January 2022 X-ray spectra might be due to such events. Yet, explaining the January 2021 X-ray hardness in this way requires a far more severe reduction in the soft flux than typically observed in the {\it Suzaku} softness dips.

\subsubsection{High-resolution spectra \label{RGSspec}}
\begin{figure*}
  \resizebox{16cm}{!}{\includegraphics[angle=0]{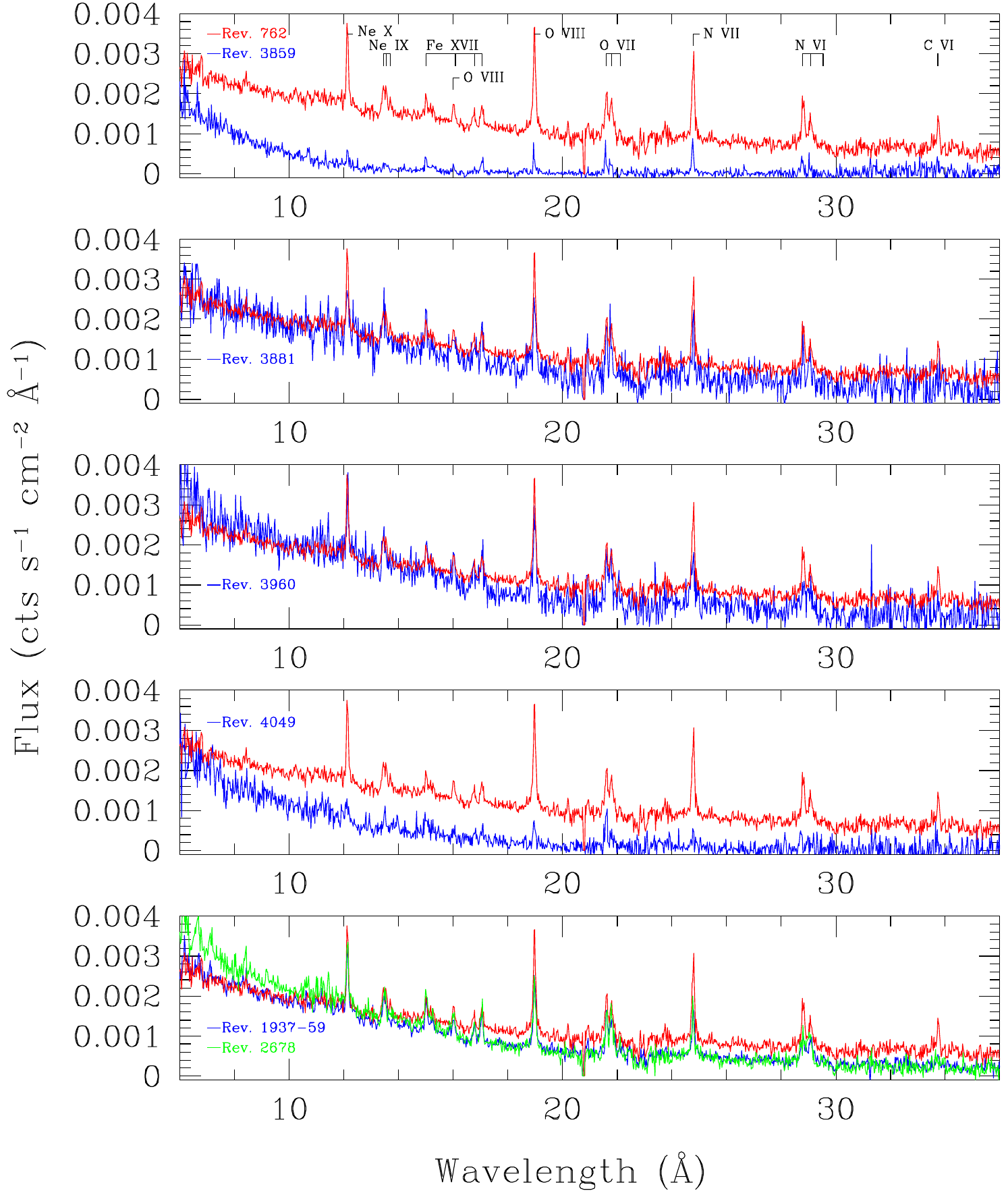}}
  \caption{Variations in the RGS spectrum of $\gamma$~Cas. In each sub-panel, the flux-calibrated RGS spectrum from revolution 762 (\,February 2004) is shown for comparison in red. In the four upper sub-panels, the blue curve corresponds to the data collected during the 2021 -- 2022 campaign, respectively revolutions 3859 (January\ 2021), 3881 (February\ 2021), 3960 (July 2021), and 4049 (January\ 2022) for sub-panels 1, 2, 3, and 4 from the top downwards. For comparison, the blue and green curves in the lowest panel show respectively the combined RGS spectrum from the 2010 campaign (revolutions\ 1937 -- 1959) and the data from July 2014 (revolution\ 2678). The strong suppression of the soft spectrum during revolutions 3859 and 4049 is clearly seen.\label{specRGS}}
\end{figure*}
The RGS spectra illustrated in Fig.\,\ref{specRGS} reveal the presence of a number of emission lines on top of a strong bremsstrahlung continuum. What is striking in this figure is the huge difference between the level of the January 2021 and January 2022 spectra on the one hand, and all other observations on the other hand. This reflects the hardening already pointed out in the previous subsection. The strong fading of the soft-band flux in the January 2021 and January 2022 observations, compared to other epochs, is to a large extent due to the attenuation of the bremsstrahlung continuum of the hottest plasma component, but this is not the only reason. Indeed, as one can see in Fig.\,\ref{specRGS}, the strength of the emission lines (e.g.\ the O\,{\sc viii} Ly$\alpha$ line) is also reduced in January 2021 and January 2022. This indicates that both the hot ($kT \simeq$ 12 -- 14\,keV) spectral component responsible for the bremsstrahlung continuum and the cooler plasma responsible for the lines were attenuated in the January 2021 and January 2022 observations.   

 The most prominent lines in the RGS spectrum are the Ly$\alpha$ transitions of Ne\,{\sc x} (12.13\,\AA), O\,{\sc viii} (18.97\,\AA), N\,{\sc vii} (24.78\,\AA), and C\,{\sc vi} (33.73\,\AA), as well as the He-like {\it f i r} triplets of Ne\,{\sc ix} (13.45 -- 13.70\,\AA), O\,{\sc vii} (21.60 -- 22.10\,\AA), and N\,{\sc vi} (28.78 -- 29.53\,\AA). As already pointed out by \citet{Smi04} and \citet{Smi12}, the forbidden ({\it f}) components of these triplets are strongly suppressed, indicating that the corresponding plasma is either of high density or located close to a strong UV source \citep{Porquet}. Some additional lines are due to Fe\,{\sc xvii} ($\lambda\lambda$\,15.01, 16.78, and 17.06) and possibly Fe\,{\sc xviii} $\lambda$\,16.08, although the latter is blended with the Ly$\beta$ line of O\,{\sc viii}. The lines identified above reach their maximum emissivities for $kT$ values in the range 0.12 -- 0.68\,keV and arise therefore in a much cooler plasma than the bremsstrahlung continuum that is clearly seen in Fig.\,\ref{specRGS}.

Previous investigations of high-resolution HETG or RGS spectra of $\gamma$~Cas concluded that the lines are broadened; they found the lines to be consistent with Gaussian profiles with $\sigma = 478 \pm 50$\,km\,s$^{-1}$ \citep{Smi04,LSM10,Smi12}. Moreover, these authors found that N and possibly Ne are overabundant, whilst Fe seems underabundant. More specifically, the Fe abundance inferred from K-shell lines was found to be significantly sub-solar ($0.12 \pm 0.02$\,Z$_{\odot}$), whereas L-shell lines did not reveal such abundance deviations \citep{Smi04,LSM10}.

To constrain the properties of the cooler plasma components, we attempted fitting the RGS data by a model given by
{\tt phabs(ISM)*phabs*(bvapec(3T)+bvapec(kT$_{\rm hot}$))}, hereafter called the 3T$^+$ model.
The first and second {\tt phabs} photoelectric absorption components account respectively for the interstellar medium and the circumstellar environment. Published values of the interstellar H\,{\sc i} column densities range from about $0.98\,10^{20}$ to $2.70\,10^{20}$\,cm$^{-2}$, with most of the values being close to $1.45\,10^{20}$\,cm$^{-2}$ \citep[see][and references therein]{Gud12}. We thus adopted $1.45\,10^{20}$\,cm$^{-2}$  for the interstellar H\,{\sc i} column density in our spectral fits. The {\tt bvapec} models represent an optically thin thermal plasma model\footnote{We stress that the terminology `optically thin' refers here to the sole hot X-ray emitting plasma, not to the gas responsible for any of the optical emission lines.} with abundances that can be set individually and allowing for velocity-broadened spectral lines \citep{Smi01}. The broadening is assumed to follow a Gaussian profile and to have the same value for all plasma components. Since tests with the Ne abundance left free were not conclusive, we only let the Fe and N abundance vary in our fits, requesting all plasma components to have the same chemical composition. The first three `warm' {\tt bvapec} components account for the lines in the RGS spectra. Their temperatures were left free in the fitting process, but requested to be different from each other. Our choice of three warm {\tt bvapec} components resulted from a series of tests that showed that models with only two components failed to provide a good description of the RGS spectra, whereas a fourth soft component was not needed. The temperature of the fourth, `hot', plasma component was fixed to 12.8\,keV for the fits of the RGS spectra. This value was obtained as the average of $kT_{\rm hot}$ in a preliminary analysis of the EPIC data. In the RGS spectral range, the hot plasma component mostly contributes the bremsstrahlung continuum, and hence the exact value of the hottest temperature has little impact on the fits.

The adjustment was performed independently for all (new and archival) RGS data. In this way, we found that some parameters of the fits vary very little from one epoch to the other. Within their error bars, the nitrogen/hydrogen number abundance ratio of the {\tt bvapec} components, and their velocity broadening were constant and equal to $\frac{n_{\rm N}/n_{\rm H}}{(n_{\rm N}/n_{\rm H})_{\odot}} = 4.6 \pm 0.6$, and $\sigma = 540 \pm 65$\,km\,s$^{-1}$. These parameters were subsequently fixed to these values. The plasma temperatures were also found to be relatively stable, at least for those observations where the soft emission was not strongly reduced. We found $kT_1 = 0.11 \pm 0.01$\,keV, $kT_2 = 0.43 \pm 0.03$\,keV, and $kT_3 = 1.35 \pm 0.10$\,keV. However, we found large variations in the column densities of the circumstellar absorption component. Whilst most observations yielded hydrogen column densities below $2\,10^{21}$\,cm$^{-2}$, the January 2021 and January 2022 yielded column densities of $1.4\,10^{22}$\,cm$^{-2}$ and $5\,10^{21}$\,cm$^{-2}$, respectively. Finally, the Fe abundance was found to be sub-solar, $\frac{n_{\rm Fe}/n_{\rm H}}{(n_{\rm Fe}/n_{\rm H})_{\odot}} = 0.39 \pm 0.11$ but somewhat dependent on the epoch. 

As a next step, we tested a model given by
{\tt phabs(ISM)*phabs*(bvapec(6T) + bvapec(kT$_{\rm hot}$))}, where the temperatures of the first six {\tt bvapec} components were fixed to logarithmically evenly-spaced values of 0.1, 0.2, 0.4, 0.8, 1.6, and 3.2\,keV, to mimic a differential emission measure (DEM) model \citep{Coh21}. The nitrogen abundance and line broadening parameters of these plasma components were set to 4.6 and 540\,km\,s$^{-1}$ as found above. The temperature of the {\tt bvapec(kT$_{\rm hot}$)} component was again set to 12.8\,keV. These DEM models yielded fits of very similar quality to those obtained with the 3T$^+$ model above. Moreover, these DEM-like models confirm that the plasma temperature distribution is not continuous \citep{LSM10,Smi12}. Whilst the components at 0.1, 0.4, and 1.6\,keV always had a significant normalisation parameter (${\rm norm}/\sigma_{\rm norm} \geq 3$, except for the highly attenuated spectra of January 2021 and January 2022), the norm of the components at 0.2 and 3.2\,keV was consistent with, or even equal to, zero. The component at 0.8\,keV had a somewhat intermediate significance. In the January 2021 spectrum, only the 0.1\,keV plasma was detected at ${\rm norm}/\sigma_{\rm norm} \geq 3$, whilst in January 2022, the 0.1 and 0.4\,keV components were detected at this significance. Overall, the DEM-like model thus confirms our choice of the number of plasma components in the 3T$^+$ model as well as the values of their temperatures.

\subsubsection{Broadband spectra \label{broadfits}}
Whilst the properties of the cooler plasma components in the X-ray spectrum of $\gamma$~Cas are best determined by the lines in the RGS spectrum (see the previous subsection), the properties of the hot plasma component are best derived from the continuum and from the Fe\,{\sc xxv} and Fe\,{\sc xxvi} lines around 6.7 -- 6.97\,keV. As a next step, we thus attempted to fit the EPIC spectra (at energies above 0.4\,keV) accounting for the information obtained from the RGS data. The same fitting procedure was applied also to the HETG data to establish the values of the fluxes for the {\it Chandra} observation\footnote{Because uncertainties on the cross-calibration between {\it Chandra} and {\it XMM-Newton} could bias the comparison of the spectral parameters, we do not quote the parameters of the HETG fit in Table\,\ref{tabfits}.}. 

\begin{sidewaystable}
  \tiny
  \caption{Results of the fits of the {\it XMM-Newton} spectra.\label{tabfits}} 
\begin{tabular}{l c c c c c c c c c c c c c c}
\hline
\multicolumn{13}{c}{\tt phabs(ISM)*(phabs$_1$*(bvapce(3T)+bvapec(kT$_{\rm hot}$)+phabs$_2$*(bvapec(kT$_{\rm hot}$+gauss))}\\
\hline
Obs. & N$_{\rm H, 1}$ & kT$_1$ & norm$_1$ & kT$_2$ & norm$_2$ & kT$_3$ & norm$_3$ & kT$_{\rm hot}$ & norm$_{\rm hot}$ & N$_{\rm H, 2}$ & cov$_{\rm hot}$  & [Fe] & $\chi^2$\\
& ($10^{22}$\,cm$^{-2}$) & (keV) & ($10^{-3}$\,cm$^{-5}$) & (keV) & ($10^{-3}$\,cm$^{-5}$) & (keV) & ($10^{-3}$\,cm$^{-5}$) & (keV) & ($10^{-3}$\,cm$^{-5}$) & ($10^{22}$\,cm$^{-2}$)  & (\%) &  &  \\
& & & & & & & & & & & & & \\
\hline
Feb.\ 04 & $0.035^{+.001}_{-.001}$ & $0.14^{+.01}_{-.01}$ & $1.51^{+.11}_{-.11}$ & $0.43^{+.01}_{.01}$ & $2.22^{+.10}_{-.10}$ & $1.06^{+.01}_{-.01}$ & $3.55^{+.10}_{-.10}$ & $12.07^{+.08}_{-.08}$ & $94.6^{+.1}_{-.1}$ & -- & 0 & $0.24^{+.01}_{-.01}$ & 2.65 (2315) \\
\vspace*{-2mm}\\
Jul.\ 10a & $0.09^{+.02}_{-.03}$ & $0.13^{+.02}_{-.01}$ & $2.22^{+1.20}_{-0.78}$ & $0.49^{+.05}_{-.05}$ & $1.75^{+.28}_{-.27}$ & $1.17^{+.06}_{-.13}$ & $3.49^{+.71}_{-.67}$ & $13.09^{+.48}_{-.48}$ & $112.6^{+7.5}_{-7.5}$ & $0.50^{+.16}_{-.12}$ & 44 & $0.48^{+.05}_{-.05}$ & 1.42 (2388)\\
\vspace*{-2mm}\\
Jul.\ 10b & $0.16^{+.01}_{-.01}$ & $0.10^{+.01}_{-.01}$ & $6.85^{+2.35}_{-2.14}$ & $0.35^{+0.04}_{-0.02}$ & $1.86^{+.31}_{-.29}$ & $1.20^{+.09}_{-.13}$ & $2.65^{+.71}_{-.68}$ & $11.53^{+.86}_{-.93}$ & $103.0^{+4.6}_{-4.6}$ & $14.9^{+6.9}_{-5.7}$ & 12 & $0.43^{+.06}_{-.06}$ & 1.24 (2123)\\
\vspace*{-2mm}\\
Aug.\ 10a & $0.11^{+.01}_{-.01}$ & $0.11^{+.01}_{-.01}$ & $5.93^{+1.18}_{-0.97}$ & $0.39^{+0.03}_{-0.03}$ & $2.80^{+.33}_{-.31}$ & $1.23^{+.05}_{-.07}$ & $4.64^{+.76}_{-.74}$ & $11.59^{+.65}_{-.66}$ & $150.8^{+5.6}_{-5.6}$ & $18.7^{+6.5}_{-5.3}$ & 18 & $0.45^{+.05}_{-.05}$ & 1.37 (2876)\\
\vspace*{-2mm}\\
Aug.\ 10b & $0.16^{+.01}_{-.01}$ & $0.11^{+.01}_{-.01}$ & $5.57^{+1.50}_{-1.04}$ & $0.43^{+0.03}_{-0.03}$ & $2.21^{+.31}_{-.29}$ & $1.31^{+.05}_{-.06}$ & $3.78^{+.61}_{-.59}$ & $11.46^{+.71}_{-.81}$ & $104.9^{+4.5}_{-4.5}$ & $24.4^{+10.2}_{-6.9}$ & 16 & $0.41^{+.05}_{-.05}$ & 1.24 (2641)\\
\vspace*{-2mm}\\
Jul.\ 14 & $0.09^{+.01}_{-.01}$ & $0.15^{+.01}_{-.01}$ & $2.03^{+0.34}_{-0.31}$ & $0.66
^{+0.01}_{-0.02}$ & $1.93^{+.16}_{-.15}$ & $1.47^{+.06}_{-.08}$ & $5.75^{+.79}_{-.76}$ & $16.04^{+.46}_{-.46}$ & $151.3^{+8.4}_{-8.4}$ & $0.76^{+.08}_{-.07}$ & 54 & $0.85^{+.05}_{-.05}$ & 1.77 (2472)\\
\vspace*{-2mm}\\
Jan.\ 21 & $0.37^{+.10}_{-.14}$ & $0.13^{+.03}_{-.03}$ & $4.1^{+6.5}_{-2.9}$ & $0.40
^{+0.09}_{-0.03}$ & $1.72^{+.68}_{-.87}$ & $1.49^{+.22}_{-.26}$ & $1.10^{+.68}_{-.49}$ & $12.40^{+.31}_{-.31}$ & $102.6^{+2.8}_{-2.8}$ & $1.38^{+.04}_{-.04}$ & 99 & $0.57^{+.03}_{-.03}$ & 1.68 (1201)\\
\vspace*{-2mm}\\
Feb.\ 21 & $0.05^{+.05}_{-.05}$ & $0.12^{+.03}_{-.02}$ & $1.54^{+1.52}_{-.86}$ & $0.81^{+0.03}_{-0.03}$ & $3.65^{+.37}_{-.74}$ & $1.58^{+.14}_{-.22}$ & $5.40^{+1.64}_{-1.44}$ & $11.42^{+.54}_{-.55}$ & $98.5^{+9.1}_{-9.1}$ & $0.37^{+.03}_{-.03}$ & 57 & $0.58^{+.04}_{-.04}$ & 1.66 (1636)\\
\vspace*{-2mm}\\
Jul.\ 21 & $0.00^{+.06}_{-.00}$ & $0.15^{+.03}_{-.02}$ & $0.93^{+.62}_{-.21}$ & $0.79^{+0.03}_{-0.03}$ & $2.79^{+.58}_{-.32}$ & $1.61^{+.10}_{-.13}$ & $6.24^{+1.17}_{-1.03}$ & $14.25^{+.62}_{-.83}$ & $119.8^{+7.1}_{-7.1}$ & $0.25^{+.07}_{-.03}$ & 94 & $0.68^{+.05}_{-.05}$ & 1.63 (1653)\\
\vspace*{-2mm}\\
Jan.\ 22 & $0.55^{+.02}_{-.02}$ & $\leq 0.08$ & $197^{+43}_{-103}$ & $0.94^{+0.04}_{-0.04}$ & $4.23^{+.74}_{-.72}$ & -- & -- & $11.22^{+.60}_{-.73}$ & $116.0^{+4.1}_{-4.3}$ & $3.60^{+1.08}_{-.79}$ & 20 & $0.57^{+.05}_{-.05}$ & 1.91 (1070)\\
\hline
\end{tabular}
\tablefoot{The ISM neutral hydrogen column density was set to $1.45\,10^{20}$\,cm$^{-2}$, the nitrogen abundance was set to 4.6 times solar and the line broadening to $540$\,km\,s$^{-1}$. The quoted uncertainties on the parameters of the fits correspond to $1\,\sigma$. The normalisation parameters of the {\tt bvapec} components correspond to $\frac{10^{-14}}{4\,\pi\,d^2}\,\int n_e\,n_H\,dV$ where $d$ is the source distance in cm, $n_e$ and $n_H$ are the electron and H densities (in cm$^{-3}$). The norm$_{\rm hot}$ value yields the total normalisation parameter of the hot plasma, that is, it corresponds to the sum of the normalisation factors of both {\tt bvapec(kT$_{\rm hot}$)} components. The column labelled cov$_{\rm hot}$ yields the fraction of norm$_{\rm hot}$ that is covered by the {\tt phabs}$_2$ column density. The Fe abundance is given as $\frac{n_{\rm Fe}/n_{\rm H}}{(n_{\rm Fe}/n_{\rm H})_{\odot}}$.}
\end{sidewaystable}

For this purpose, we used a model given by 
{\tt phabs(ISM)*[phabs$_1$*(bvapec(3T)+bvapec(kT$_{\rm hot}$)) + phabs$_2$*(bvapec(kT$_{\rm hot}$)+gauss)]}.
This model is directly inspired by the results of \citet{LSM10} and \citet{Smi12} who found that some fraction of the hot plasma component is covered by a higher column density \citep[see also][]{Smi04,Tsu18}. Our model thus includes the hot component both with the same absorbing column density as the cooler plasma components, and with a different column. As previously, the ISM column density was fixed to $1.45\,10^{20}$\,cm$^{-2}$, the nitrogen abundance of all plasma components were set to 4.6 and the broadening of the lines was fixed to $540$\,km\,s$^{-1}$. The Fe abundance was allowed to vary in the fits though all plasma components (including kT$_{\rm hot}$) were constrained to have the same iron abundance. Finally, an unresolved Gaussian line was included in the fits with an energy set to 6.4\,keV to represent the fluorescent K$\alpha$ line arising from low-ionisation stages of iron.
\begin{figure}
  \resizebox{9cm}{!}{\includegraphics[angle=0]{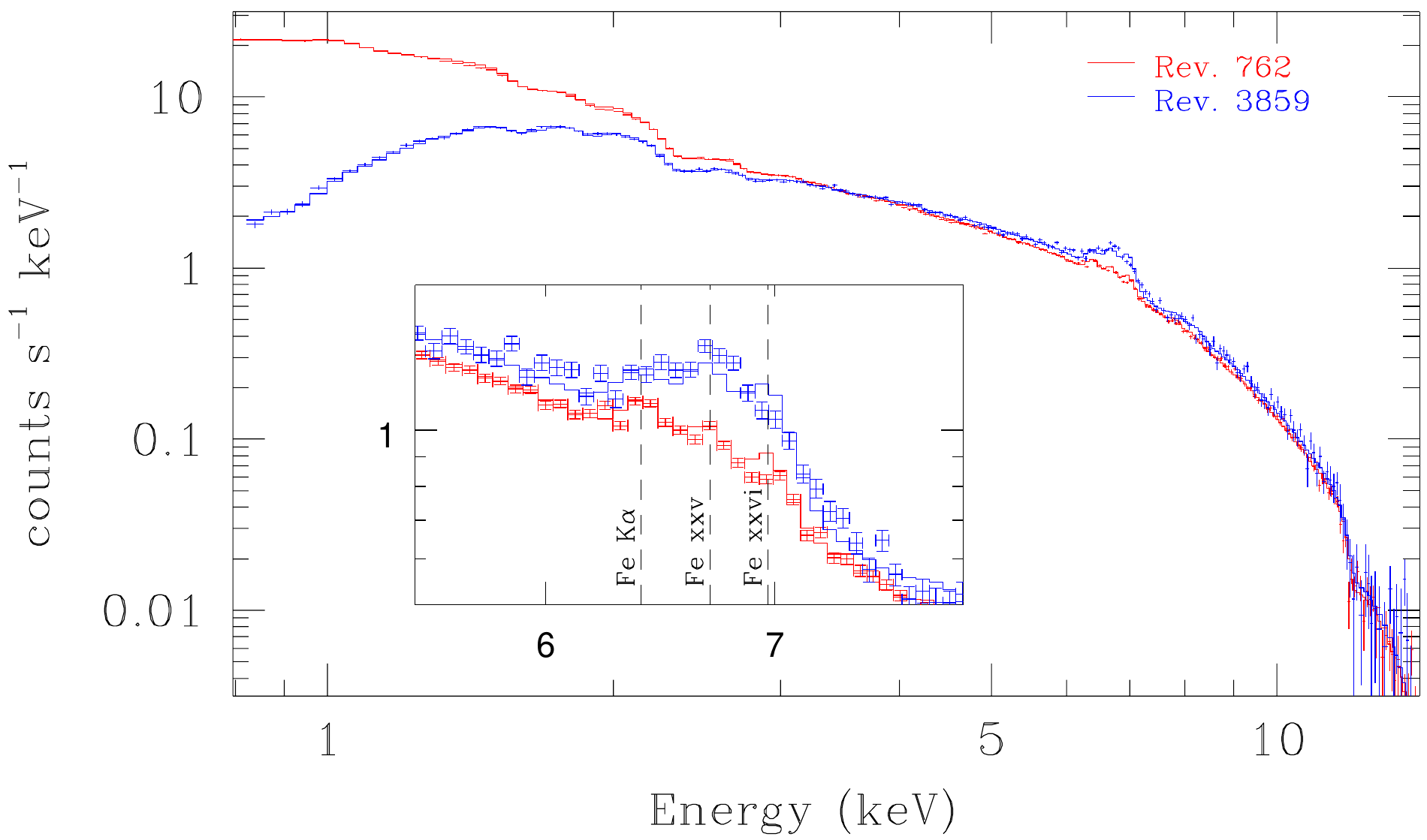}}
  \caption{Comparison between the EPIC-pn data and the best-fit models for revolutions\,0762 (February 2004, red) and 3859 (January\ 2021, blue). Only the EPIC-pn data are shown for clarity, although all EPIC and RGS spectra were fitted simultaneously. Note the strong attenuation of the January 2021 spectrum in the soft energy range that is also seen in the top panel of Fig.\,\ref{specRGS} and discussed in Sects.\,\ref{Xraylc} and \ref{RGSspec}. The insert shows a zoomed-in view of the iron line complex.\label{EPICRev3859}}
\end{figure}
In a first attempt, we fixed the temperatures and norms of the three softer plasma components, as well as the {\tt phabs$_1$} column density to their value inferred from the RGS data, and fitted only the EPIC data. As a second attempt, we then performed a joint fit of the RGS and EPIC data, leaving the temperatures and norms of the cooler plasma components, as well as the circumstellar column density free. This dual approach was mainly aimed at checking the robustness of the parameters, especially the X-ray fluxes and fluorescent line strengths, that we infer from these models. The results of the joint RGS and EPIC fits are listed in Table\,\ref{tabfits}, and the comparison between the EPIC-pn data and the model is illustrated in Fig.\,\ref{EPICRev3859} for the data from revolutions\,0762 (February\ 2004) and 3859 (January\ 2021). Similar plots are provided for the other datasets in Figs.\,\ref{EPICRev3881} -- \ref{EPICRev1959}.
\begin{figure}
  \resizebox{9cm}{!}{\includegraphics[angle=0]{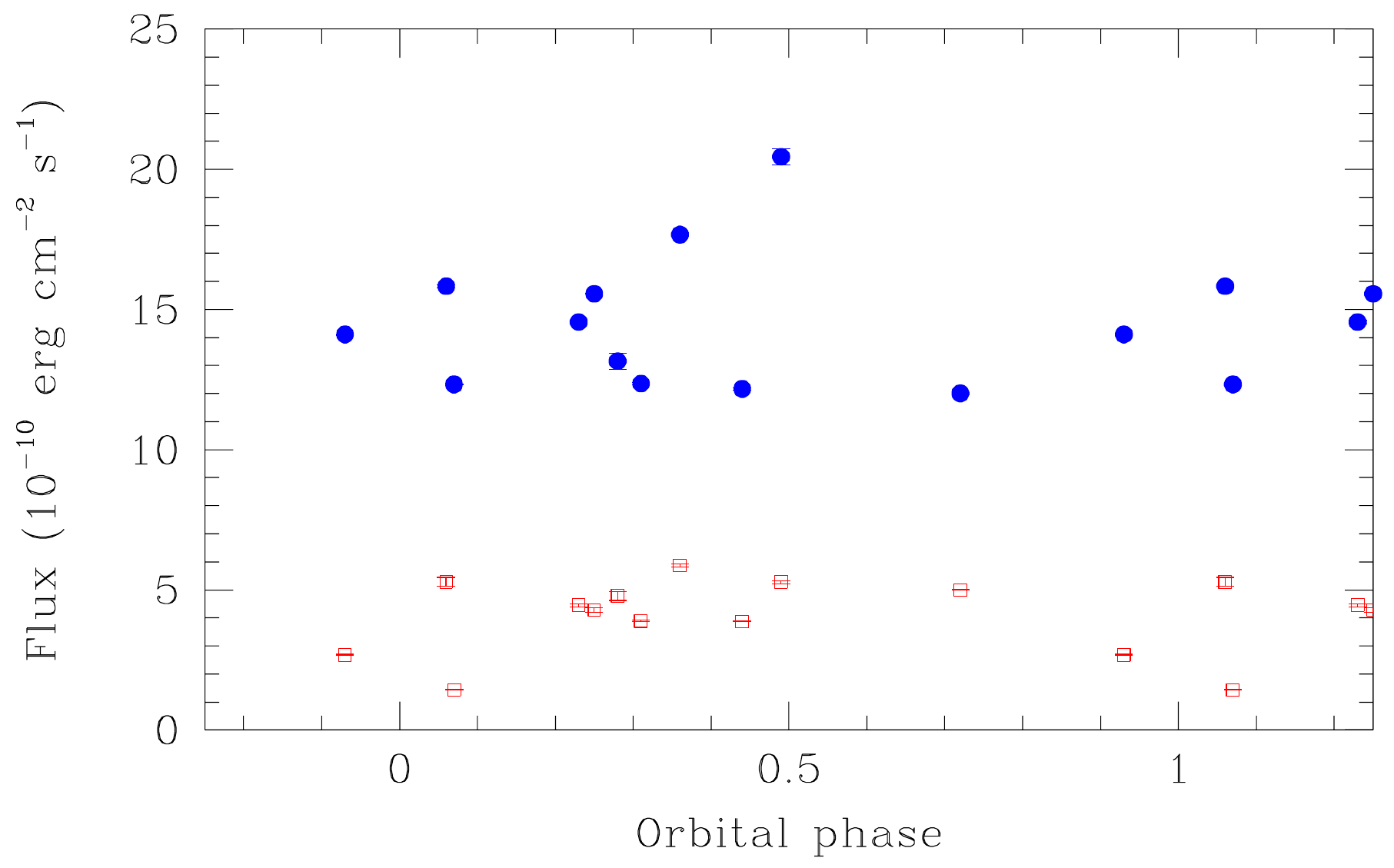}}
  \caption{X-ray fluxes of $\gamma$~Cas as a function of orbital phase. The fluxes in the hard and soft bands are illustrated by the blue filled circles and red open squares, respectively. \label{phase}}
\end{figure}

We caution that these model parameters should not be over-interpreted as our overall description of the emission of $\gamma$~Cas is certainly too simplistic compared to reality. These parameters also depend on the ingredients of the plasma models. Indeed, we tested the models of \citet{Smi12}, which used {\tt mekal} \citep{KaMe00} plasma models with solar abundances from \citet{AG} on the data from the 2010 campaign. Whilst these models provide a good description of the data, with parameters in good agreement with those of \citet{Smi12}, they require a plasma component in the 3.4 -- 4.3\,keV temperature range (in addition to the hot plasma component) to correctly fit the strength of the Fe\,{\sc xxv} triplet \citep[see also][]{Smi04,LSM10}. With the {\tt apec} models adopted here (using up-to-date atomic parameters), such a component is no longer required. Finally, there is certainly a huge degeneracy among different model descriptions that provide comparable fit qualities. Especially, the well-known degeneracy between plasma temperature and column density could lead to non-unique solutions. Therefore, our model parameters should only be seen as representative values reproducing the overall X-ray spectrum with a reasonable accuracy over the full energy domain. These models thus provide a robust estimate of the observed fluxes and Fe K$\alpha$ line EW (see Table\,\ref{FeK}).

The fact that we have several observations at neighbouring binary orbital phases (e.g.\ the January 2021 and July 2021 data) allows us to look for a distinct phase dependence of the fluxes. As one can clearly see in Fig.\,\ref{phase}, no such trend exists in either the soft- or hard-band fluxes. The same conclusion, that is absence of orbital modulation of the X-ray emission, was reached for $\pi$~Aqr \citep{Naz19}.

\begin{table*}
  \caption{Fluorescent Fe K$\alpha$ line strength, X-ray fluxes, and flux hardness ratio. \label{FeK}}
  \begin{tabular}{l c l c c c c c c}
    \hline
    \multicolumn{1}{c}{Obs.} & HJD$-2450000$ & \multicolumn{1}{c}{EW(Fe K$\alpha$)}& \multicolumn{2}{c}{$f_{\rm X,obs}$} & $HR_{\rm flux}$ & \multicolumn{2}{c}{$f_{\rm X,unabs}$}\\
    
    &             & \multicolumn{1}{c}{(eV)}    & \multicolumn{2}{c}{($10^{-11}$\,erg\,cm$^{-2}$\,s$^{-1}$)}  & & \multicolumn{2}{c}{($10^{-11}$\,erg\,cm$^{-2}$\,s$^{-1}$)} \\
    & & & S & H & & S & H \\
    \hline
    Aug.\ 01  & 2132.202 & $34 \pm 14$ & $4.31 \pm 0.09$ & $15.42 \pm 0.10$ & $0.56 \pm 0.01$ & 4.41 & 15.43 \\
    Feb.\ 04  & 3041.662 & $35 \pm  5$ & $5.00 \pm 0.02$ & $12.01 \pm 0.03$ & $0.41 \pm 0.01$ & 5.14 & 12.02 \\
    Jul.\ 10a & 5384.901 & $55 \pm 11$ & $4.46 \pm 0.06$ & $14.55 \pm 0.07$ & $0.53 \pm 0.01$ & 4.56 & 14.56 \\
    Jul.\ 10b & 5401.741 & $57 \pm 14$ & $3.89 \pm 0.03$ & $12.36 \pm 0.04$ & $0.52 \pm 0.01$ & 3.98 & 12.37 \\
    Aug.\ 10a & 5411.182 & $51 \pm 10$ & $5.87 \pm 0.06$ & $17.67 \pm 0.07$ & $0.50 \pm 0.01$ & 6.01 & 17.68 \\
    Aug.\ 10b & 5428.685 & $52 \pm 12$ & $3.88 \pm 0.02$ & $12.17 \pm 0.05$ & $0.52 \pm 0.01$ & 3.97 & 12.18 \\
    Jul.\ 14  & 6863.268 & $59 \pm  7$ & $5.28 \pm 0.06$ & $20.45 \pm 0.29$ & $0.59 \pm 0.01$ & 5.40 & 20.47 \\
    Jan.\ 21  & 9219.190 & $45 \pm  7$ & $1.44 \pm 0.01$ & $12.33 \pm 0.01$ & $0.79 \pm 0.01$ & 1.46 & 12.34 \\
    Feb.\ 21  & 9263.018 & $56 \pm 11$ & $4.79 \pm 0.16$ & $13.16 \pm 0.29$ & $0.47 \pm 0.02$ & 4.89 & 13.17 \\
    Jul.\ 21  & 9420.564 & $40 \pm 10$ & $5.28 \pm 0.16$ & $15.83 \pm 0.05$ & $0.50 \pm 0.01$ & 5.39 & 15.85 \\
    Jan.\ 22  & 9598.080 & $55 \pm 11$ & $2.59 \pm 0.03$ & $14.13 \pm 0.02$ & $0.69 \pm 0.01$ & 2.64 & 14.14 \\
    \hline
  \end{tabular}
  \tablefoot{The fluxes in Cols. 4 and 5, and the inferred flux hardness ratio in Col. 6, are observed values, not corrected for interstellar absorption. The fluxes in the last two columns are corrected for the absorption by the interstellar medium.}
\end{table*}

Several previous studies of $\gamma$~Cas \citep[e.g.][]{Mur86,Smi04,LSM10,Smi12} noted that the strength of the Fe\,{\sc xxv} and {\sc xxvi} lines relative to the continuum required a significantly sub-solar iron abundance for the hot plasma, which might hint at an inverse first ionisation potential (IFIP) effect as seen in many active late-type stars and/or at a non-equilibrium ionisation of the hot plasma (mostly as a result of the flare-like shots that occur on very short timescales). In our analysis, we infer indeed a sub-solar Fe abundance, though our values are generally higher than those of \citet{Smi04} and vary strongly from one epoch to the other. It is worth noting that the variations in the Fe abundance in the overall fit are significantly larger than those found in the fits of the RGS data only, and that the mean Fe abundance is higher than in the RGS fits. This could indicate that the putative variations in the IFIP effect mostly stem from the hot plasma component and thus from the Fe\,{\sc xxv} and {\sc xxvi} lines as previously suggested by \citet{LSM10}. Yet, allowing the Fe abundance of the hot and warm plasma components to differ from each other did not lead to a clearer picture, insofar that, whilst the Fe\,{\sc xxv} and {\sc xxvi} lines were a bit better adjusted, the inferred abundances were spread over a wide range without a significant improvement of the overall fit quality.  

\subsubsection{Fe K$\alpha$ line}
The EWs of the fluorescent Fe K$\alpha$ line were determined in two different ways: directly from the combined EPIC and RGS fits of Sect.\,\ref{broadfits} and by adjusting the EPIC data between 4.0 and 8.0\,keV with a simplified model consisting of a bremsstrahlung continuum and three unresolved Gaussian lines at fixed energies of 6.4, 6.7 and 6.97\,keV. In the second approach, the data from the different EPIC cameras were treated independently to avoid biases due to potential cross-calibration uncertainties. The dispersion of the EWs determined from the different instruments and the relative error on the normalisation of the 6.4\,keV Gaussian were taken as proxies of the error on the EW. The EWs obtained via both methods agree well within the uncertainties. Table\,\ref{FeK} provides the mean of the results obtained with both techniques along with the fluxes and flux hardness ratio, $HR_{\rm flux}$, based on the observed fluxes in the hard and soft bands: $HR_{\rm flux} = \frac{f_{\rm X,H,obs} - f_{\rm X,S,obs}}{f_{\rm X,H,obs} + f_{\rm X,S,obs}}$.

\begin{figure}
  \resizebox{9cm}{!}{\includegraphics[angle=0]{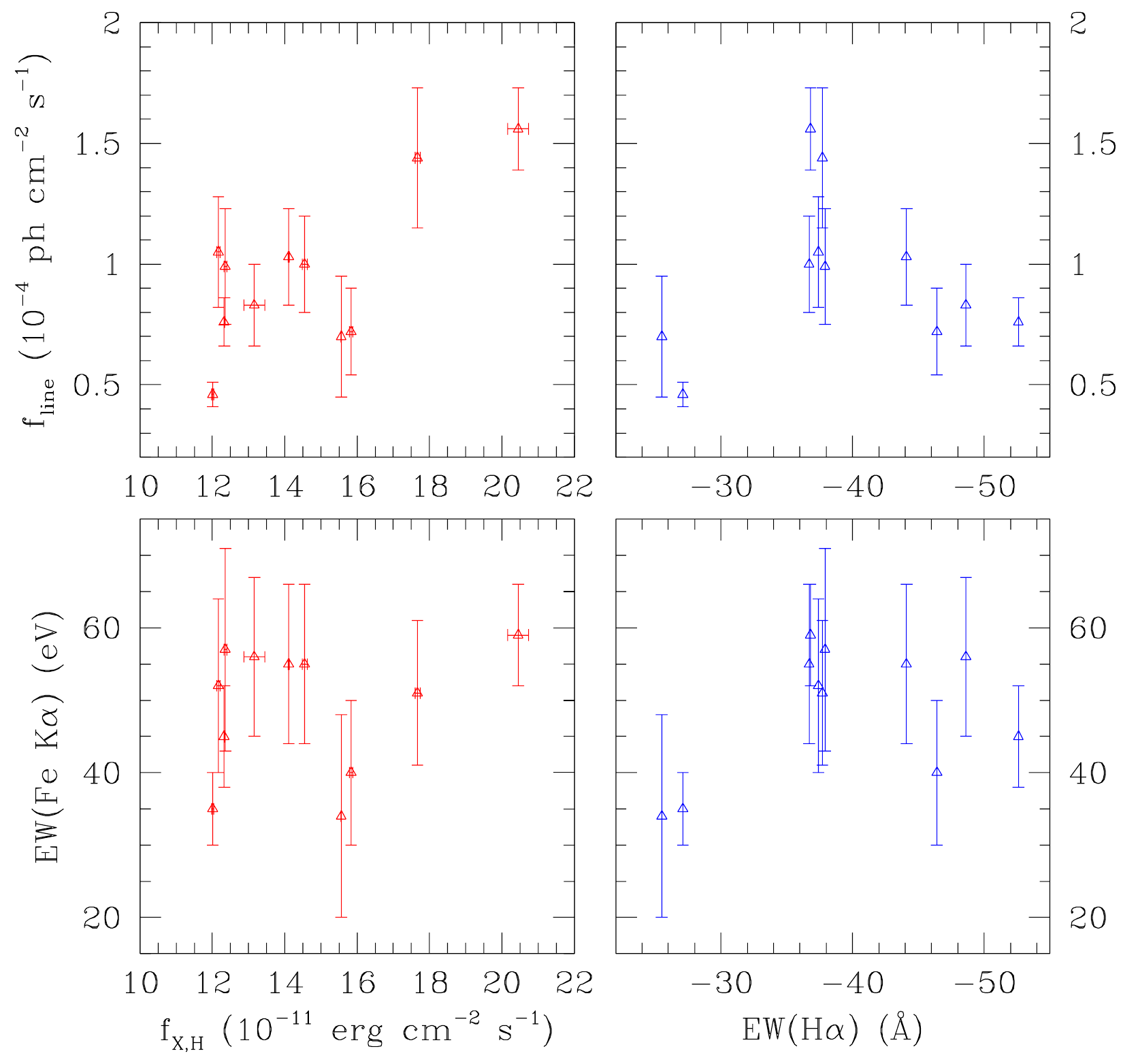}}
  \caption{EW (bottom row) and line flux (top row) of the fluorescent Fe K$\alpha$ line as a function of hard X-ray flux (left column) and EW(H$\alpha$) (right column).\label{Fe}}
\end{figure}

\citet{Tor10} investigate the behaviour of fluorescent Fe K$\alpha$ lines in X-ray binaries observed with {\it Chandra}-HETG. They report a correlation between the line flux and the continuum flux, but an anti-correlation of EW with continuum flux (the so-called Baldwin effect, known in active galactic nuclei). \citet{Gim15} extended this study to {\it XMM-Newton} EPIC-pn data. Whilst they confirmed the presence of an X-ray Baldwin effect in X-ray binaries, they found no such effect for the six $\gamma$~Cas stars with detected Fe K$\alpha$ lines in their sample. Our results
suggest that the Fe K$\alpha$ line flux correlates with the hard X-ray flux (see the upper-left panel of Fig.\,\ref{Fe}), and confirm the absence of a clear dependence of EW(Fe K$\alpha$) on the hard X-ray flux (lower left panel of Fig.\,\ref{Fe}). Finally, the right panels of Fig.\,\ref{Fe} display EW(Fe K$\alpha$) and the line flux versus EW(H$\alpha$), the latter quantity being a proxy of the amount of material in the Be disc. Here, the two lowest values of EW(Fe K$\alpha$) were measured at epochs when the H$\alpha$ emission was relatively weak. However, for EW(H$\alpha$) $\leq -35$\,\AA, the EW of the fluorescent line seems rather insensitive to EW(H$\alpha$). This is expected if the hard X-ray emission, which causes the fluorescence, preferentially illuminates the inner parts of the Be disc, whilst EW(H$\alpha$) essentially reflects changes in the outer parts of the disc. 

\section{Long-term variability \label{longtermvar}}
To examine the long-term behaviour of $\gamma$~Cas, we now compare optical properties ($V$-band photometry and EW(H$\alpha$)) with X-ray data. We recall here that the contribution of the disc to the $V$ band arises from within about 1\,$R_*$, whilst H$\alpha$ forms out to about a dozen stellar radii \citep{Stee98}. Long-term X-ray time series are available from two instruments: the All Sky Monitor \citep[ASM;][]{Lev96} on board RXTE until 2011 and MAXI, on board the ISS since mid-2009 \citep{Mat09}. The treatment of the ASM data is described in \citet{Mot15}. For the MAXI data, several time bins (1, 7, 10, 30, 182, 365, and 500\,d) were considered. The scatter is quite large for the smaller time bins, but the overall behaviours of these light curves agree: there is a broad maximum of the X-ray flux around HJD $\sim 2\,457\,800$ (Fig.\,\ref{VX}). To allow for unbiased comparison with optical photometry, we built a long-term light curve with time steps of one year, but only retaining those dates that fall into the APT optical observation season. Only daily integrations with errors smaller than $0.03$\,ct\,s$^{-1}$ were retained, and the averages were weighted by the inverse of their error. Since we are dealing with X-ray data from two different instruments here, we must pay attention to their cross-calibration. The ASM data cover the energy range from 1.5 to 12\,keV. Yet, to keep contamination of the MAXI data by ISS precession-induced background variations to a minimum, we restricted our analysis to the least affected 2 -- 6\,keV energy band for this instrument. ASM and MAXI count rates were converted into fluxes using conversion factors of respectively $3.469\,10^{-10}$\,erg\,cm$^{-2}$\,count$^{-1}$ and $1.066\,10^{-8}$\,erg\,cm$^{-2}$\,count$^{-1}$. The ASM count-to-flux conversion factor was obtained as described in \citet{Mot15}, whilst the MAXI conversion was inferred from a fit of the MAXI spectrum with an absorbed 1-T thermal plasma model. We then compared the fluxes from ASM and MAXI, converted to the ASM energy band, for the two years of overlapping operations. The MAXI fluxes were found to be lower than the ASM fluxes by about 10\%. These differences most likely reflect cross-calibration uncertainties, and we thus scaled the MAXI fluxes by a factor of 1.109 to best match those of the ASM instrument for the two seasons in common. The APT $V$-band photometry was converted into optical disc fluxes assuming the stellar photospheric emission to correspond to a $V$-band magnitude of 2.8. A correlation of the ASM fluxes against the optical disc fluxes (both in units erg\,cm$^{-2}$\,s$^{-1}$) resulted in a linear regression relation given by
\begin{equation}
  f_{\rm X} \simeq (f_{\rm opt}-1.75\,10^{-7})/131.9\,\cdot
  \label{scale}
\end{equation}
Figure\,\ref{VX} shows the long-term evolution of the X-ray fluxes of $\gamma$~Cas along with the scaled optical disc $V$-band fluxes. This figure, which extends Fig.\,2 of \citet{Mot15}, unveils a strong long-term correlation between the X-ray emission and the optical emission of the disc. We computed the correlation coefficients between the X-ray fluxes and the (unscaled) $V$-band flux of the disc: over the entire 22 years of APT monitoring, the Pearson correlation coefficient amounts to $r = 0.88$, whilst the Spearman rank correlation coefficient is $r_S = 0.90$. We note that the slope of this correlation possibly changed around 2010, although one needs to keep in mind here that this epoch corresponds to the transition between the ASM and MAXI instruments. Nonetheless, the correlation between $V$-band and X-ray flux remains highly significant for the individual instruments: the Spearman rank correlation coefficient amounts to $r_S = 0.75$ for the ASM data taken between 1997 and 2011, and to $r_S = 0.58$ for the MAXI data collected between 2010 and 2019. In contrast with the strong correlation between optical and X-ray fluxes, Fig.\,\ref{historical} shows that the yearly averaged X-ray fluxes are not correlated with EW(H$\alpha$). Indeed, whilst EW(H$\alpha$) continued increasing beyond 2017, culminating in the 2020--2021 eruption, the X-ray fluxes actually follow the trend of the $V$-band fluxes, which underwent an overall decrease since 2017.

\begin{figure}
  \resizebox{9cm}{!}{\includegraphics[angle=0]{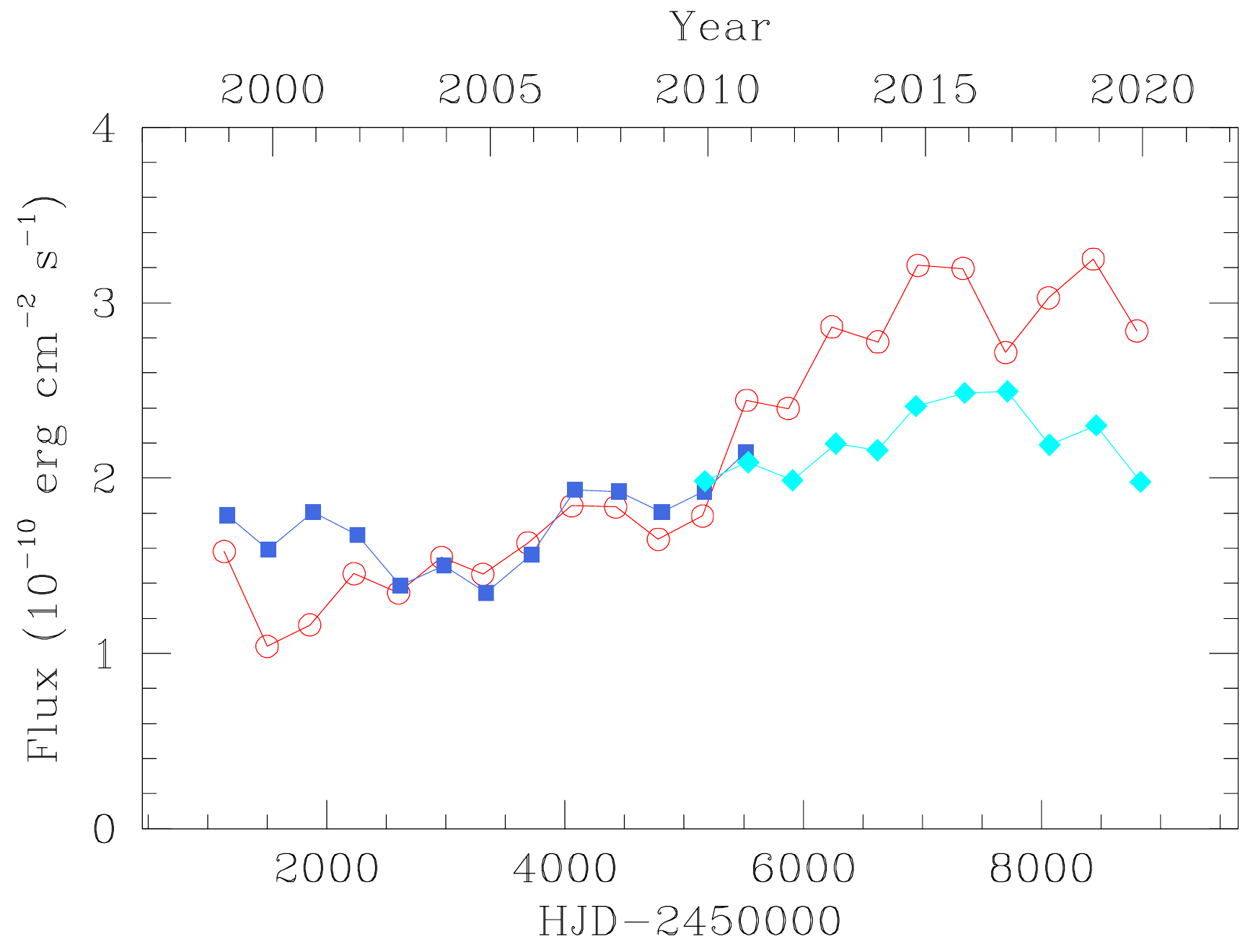}}
  \resizebox{9cm}{!}{\includegraphics[angle=0]{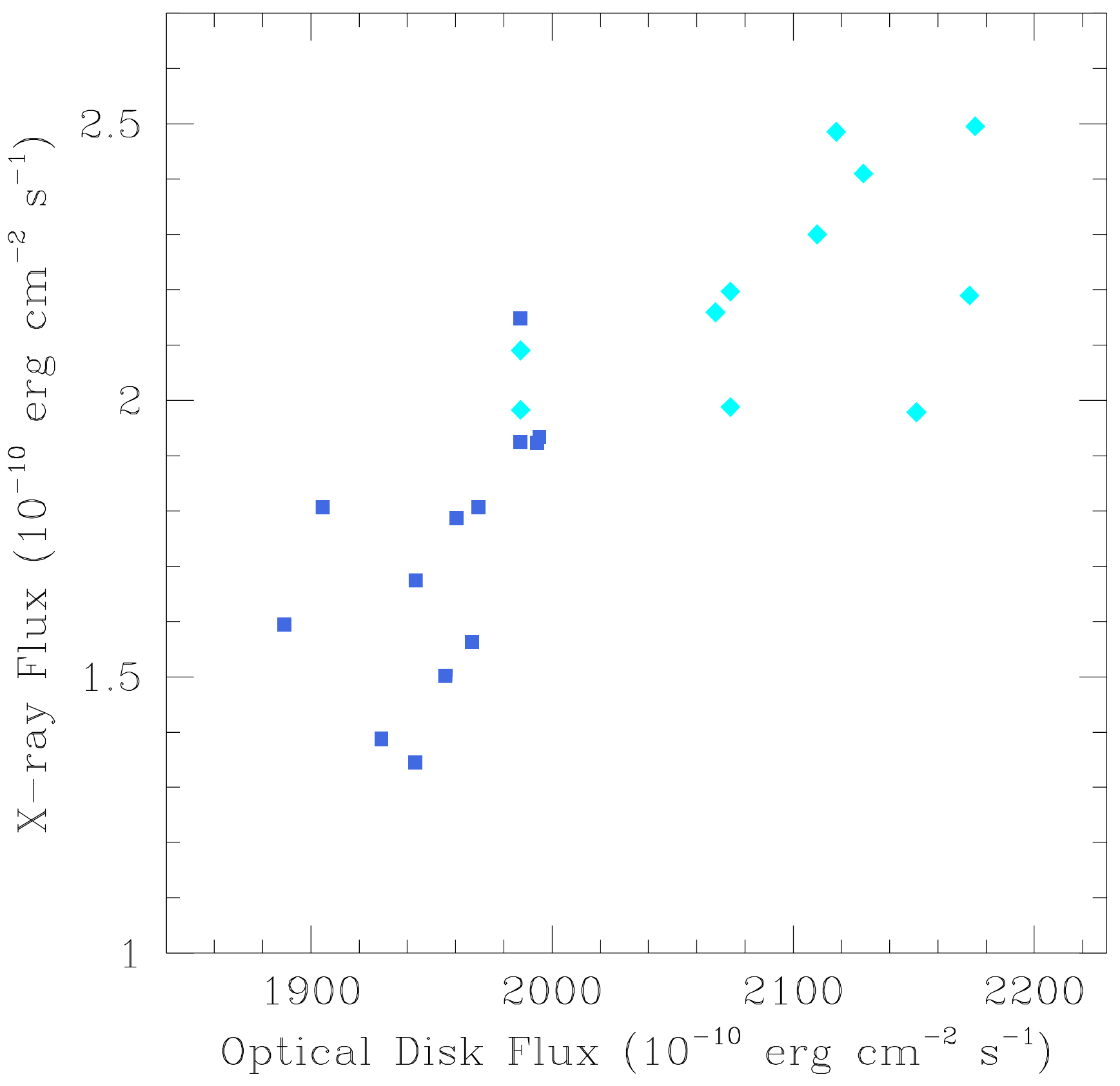}}
  \caption{Long-term evolution of the X-ray fluxes and $V$-band disc fluxes of $\gamma$~Cas. Top panel: evolution of the X-ray flux of $\gamma$~Cas, averaged over the yearly APT observing seasons. The blue squares correspond to the ASM data, and the cyan diamonds correspond to the scaled MAXI fluxes (see the main text). The red open circles indicate the yearly averages of the APT $V$-band flux corrected for the stellar emission and scaled down according to Eq.\,\ref{scale} (see the main text). Bottom panel: Yearly averaged X-ray fluxes as a function of near-contemporaneous APT $V$-band fluxes corrected for the stellar photospheric emission. The symbols have the same meaning as in the top panel.\label{VX}}
\end{figure}

The bottom panel of Fig.\,\ref{historical} illustrates a comparison of the 1.5 -- 12\,keV fluxes measured on the {\it Chandra} and {\it XMM-Newton} spectra with the overall long-term trends inferred from the ASM and MAXI yearly means. The {\it Chandra} and {\it XMM-Newton} fluxes follow the overall trend of the yearly mean fluxes, although with a significant dispersion (that also exists among the daily ASM and MAXI fluxes, but is averaged out in the yearly means).

Finally, we searched the MAXI data for a signature of the orbital period. No significant signal was found at or close to the 203\,day period and its harmonics, independently confirming the lack of such a signal in the fluxes inferred from {\it XMM-Newton} and {\it Chandra} spectra (see Fig.\,\ref{phase}).    
 
\begin{figure}
  \resizebox{9cm}{!}{\includegraphics[angle=0]{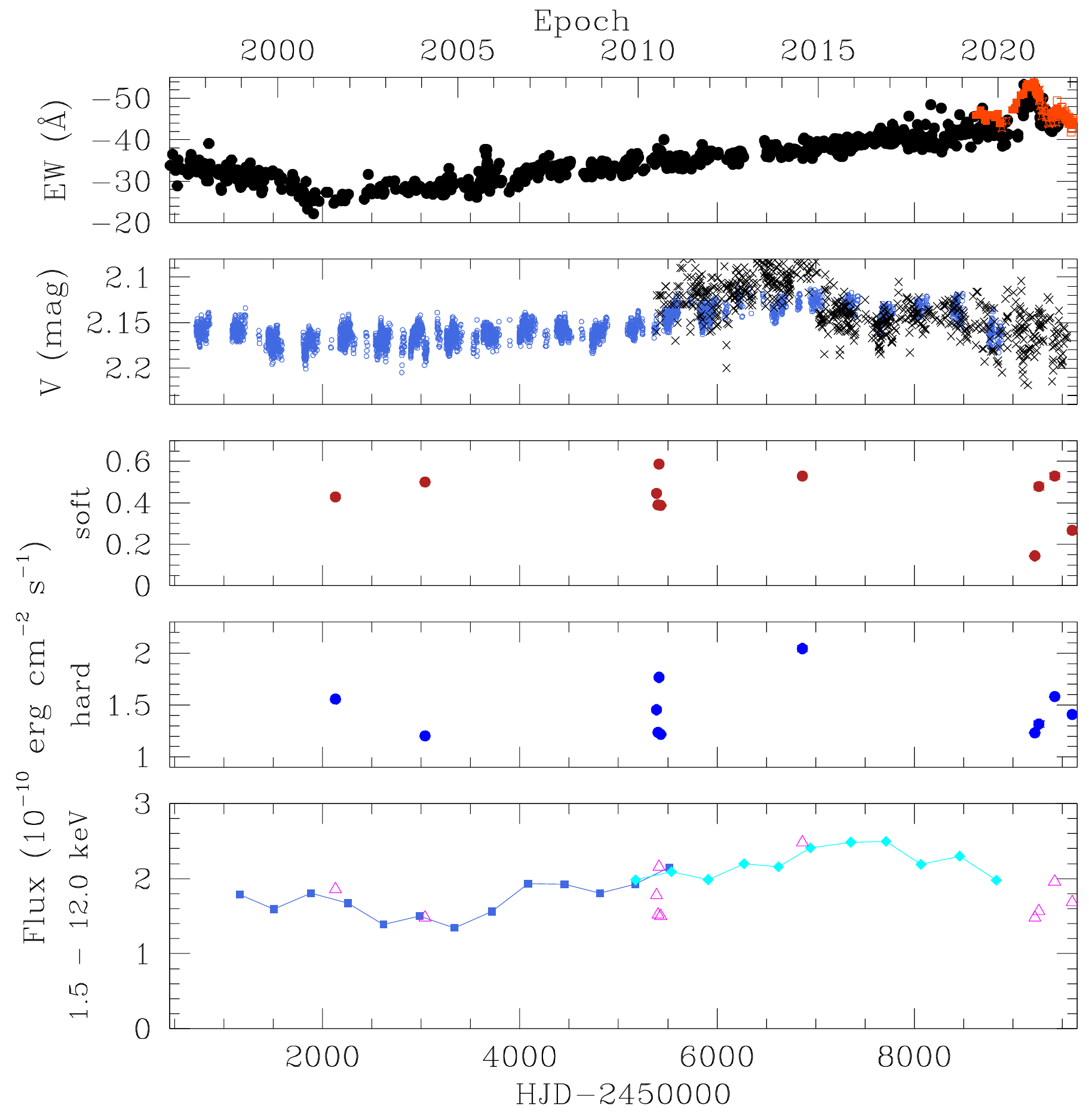}}
  \caption{Long-term view of the variations in $\gamma$~Cas. The top panel illustrates the long-term behaviour of EW(H$\alpha$) with data from \citet{Pol14} in black and data from the present study in orange. The second panel illustrates the APT $V$-band photometry (blue open circles) and the $V$-band photometry by AAVSO observer Wolfgang Vollmann (black crosses). The third and fourth panels show the observed soft and hard X-ray fluxes inferred from {\it Chandra} and {\it XMM-Newton} spectra (see Table\,\ref{FeK}). Finally, the bottom panel yields the evolution of the 1.5 -- 12\,keV X-ray flux of $\gamma$~Cas, averaged over the yearly observing seasons as in Fig.\,\ref{VX}. The fluxes in the same band derived from the {\it Chandra} and {\it XMM-Newton} observations (Table\,\ref{FeK}) are illustrated by the magenta triangles.\label{historical}}
\end{figure}  

\section{Discussion and conclusions \label{discus}}
In this paper we have analysed X-ray observations as well as optical spectroscopy and broadband photometry of $\gamma$~Cas with an emphasis on data collected during an episode of enhanced disc activity that took place around January 2021. We measured the properties of prominent optical emission lines and built epoch-dependent Doppler maps for the H$\alpha$, H$\beta$, and He\,{\sc i} $\lambda$~5876 emission lines in velocity space. We further used archival {\it XMM-Newton}, {\it Chandra}, MAXI, and RXTE-ASM X-ray data, as well as archival $V$-band photometry and H$\alpha$ EWs, to investigate the long-term correlation between optical and X-ray emission.
        
The overall picture that emerges from this study is that there is currently no evidence supporting a scenario where the hard thermal X-ray emission of $\gamma$~Cas (and its analogues) stems from either accretion by a compact companion located at $\sim 47.5$\,R$_*$ from the Be star or the interaction of the wind of an sdO companion with the disc of the Be star. The results of our study imply a hard X-ray emission arising from very near the Be star, involving the star and the inner parts of the disc.

At first sight, this conclusion might look at odds with the lack of response of the X-ray emission of HD~119\,682 to the disappearance of its H$\alpha$ emission in July 2020 \citep{Naz22}. Yet, a close look at Fig.\,1 of \citet{Naz22} shows that, although the EW(H$\alpha$) of this star became positive at several occasions during the monitoring (meaning the line had a net absorption component), the H$\alpha$ line never displayed a purely photospheric profile. Instead, weak variable residual emission was always present in the line, and no change in $V$-band photometry was recorded. This indicates that the disc actually had not completely cleared away. This result is thus fully in line with our conclusion of a hard X-ray emission mechanism that involves the innermost parts of the Be decretion disc. 

The main results of the present study can thus be summarised as follows:
\begin{enumerate}
\item[$\bullet$] During the 2020 -- 2021 observing season, the emission lines in the spectrum of $\gamma$~Cas underwent an eruption, with EW(H$\alpha$) reaching values of $-54$\,\AA. The variations in the optical emission lines during this event are best explained by an episodic increase in the density in the outer parts of the Be disc. Simultaneously, there was no photometric brightening, suggesting that the innermost parts of the disc were not significantly impacted by this event or that a putative brightness increase was compensated for by occultation along our line of sight by a wider and denser circumstellar disc.
\item[$\bullet$] Whilst the signature of the orbital motion of the Be star is clearly present before, during, and after the eruption, no indication of a density structure related to a stable interaction feature of the companion with the outer parts of the Be disc was found. Instead, the density waves in the disc precess on a timescale that differs from the orbital period, unlike what one would expect, for instance, in the case of a collision of the wind of an sdO companion with the disc.
\item[$\bullet$] {\it XMM-Newton} spectra collected during the maximum phase of the eruption and the subsequent decline unveiled no changes in the behaviour of the intra-pointing light curve: the short-term X-ray variability of $\gamma$~Cas remains dominated by rapid shots.
\item[$\bullet$] An X-ray spectrum obtained near maximum H$\alpha$ emission strength, and another one obtained one year later, unveiled a highly diminished soft X-ray emission. These observations possibly reflect extreme versions of the softness dip events that were seen in some previous observations of $\gamma$~Cas. These extreme softness dips were not only due to a reduction in the continuum contributed by the hottest plasma (kT $\sim 12$\,keV), but also to a strong attenuation of the cooler plasma components. Of special interest is the fact that both softness dips occurred near quadrature orbital phases, whilst at the same time another observation taken at the same phase did not display a softness dip. This rules out effects related to the orbital configuration (e.g.\ occultation of or by the companion) as the cause of these dips. Instead, the extreme softness dips observed in January 2021 and January 2022 could be related to the more extreme disc properties during the optical eruption. Indeed, a bigger and denser flared disc \citep{Stee03} or density clumps moving inside such a larger disc \citep{Ber99} could lead to more efficient occultations of the cooler plasma components (which {\it f i r} triplets indicate are located close to the Be star) by the outer parts of the disc. The fact that the hottest plasma component is also affected by these events reinforces the idea that this emission also arises very near the Be star.    
\item[$\bullet$] The hard X-ray flux did not exhibit any conspicuous response (neither an increase nor a decrease) to the increase in the H$\alpha$ emission strength, nor does it show any dependence on the binary orbital phase. However, on timescales of years, the hard X-ray flux appears well correlated with the $V$-band photometry. The $V$-band photometry probes the innermost part of the disc surrounding $\gamma$~Cas (typically out to 1\,R$_*$ above the photosphere). Its good correlation with the hard X-ray data seems to support an X-ray generation close to the star itself, as already pointed out by \citet{Mot15}. The lack of correlation with the H$\alpha$ line, which forms over significantly larger distances of the disc, is another piece of evidence that the hard X-rays arise from regions close to the star, rather than at the disc periphery. 
\end{enumerate}

Of the three existing scenarios to explain the properties of $\gamma$~Cas stars, only the magnetic interaction scenario \citep[][and references therein]{Smi19} predicts a location of the hottest plasma in line with our findings. Alternative mechanisms involving the star and the inner photosphere fail to explain the existence of a plasma at a temperature of about $13$\,keV near the star, as outlined in Sect.\,\ref{tomo}. In the magnetic interaction scenario, the gas is accelerated via magnetic reconnection events taking place between entangled field lines from small-scale stellar magnetic fields \citep{Can11} and fields generated or quenched in the disc via a magnetorotational instability where the small-scale stellar field acts as a seed field \citep{Bal91,Haw91,Krt15}. The material accelerated via the magnetic reconnection and directed towards the star would then have its energy converted into heat. \citet{Rob00} demonstrated that an electron beam of energy 200\,keV impacting the photosphere allows a plasma temperature of nearly $10^8$\,K ($kT = 8.6$\,keV) to be reached. To explain the $\sim 13$\,keV plasma temperature observed here, electron beam energies near 300\,keV would probably be needed. Currently, however, this scenario remains qualitative. Radiative magneto-hydrodynamic simulations similar to those performed by \citet{Fro20} for solar flares but adapted to the configuration of $\gamma$~Cas stars are clearly needed to provide a more quantitative assessment of the interaction scenario.

\section*{Acknowledgements}
We thank Ernst Pollmann for sharing his compilation of H$\alpha$ EWs with us. We deeply acknowledge the variable star observations from the AAVSO International Database contributed by observers worldwide, as well as the amateur spectroscopists who contributed data to the BeSS database. The Li\`ege team thanks the European Space Agency (ESA) and the Belgian Federal Science Policy Office (BELSPO) for their support in the framework of the PRODEX Programme (contract HERMeS). RLO acknowledges financial support from the Brazilian institutions CNPq (PQ-312705/2020-4) and FAPESP (\#2020/00457-4). This work is based on observations collected with {\it XMM-Newton}, an ESA Science Mission with instruments and contributions directly funded by ESA Member States and the USA (NASA). We thank the {\it XMM-Newton}-SOC for their assistance in the scheduling of the ToO observations. This research uses optical spectra collected at the OHP and with the TIGRE telescope (La Luz, Mexico). TIGRE is a collaboration of Hamburger Sternwarte, the Universities of Hamburg, Guanajuato and Li\`ege. This research also made use of MAXI data provided by RIKEN, JAXA and the MAXI team. The ADS, CDS and SIMBAD databases were used in this work.

\begin{appendix}
\section{X-ray light curves of revolutions 3881, 3960, and 4049}
In this section we provide the EPIC-pn light curves of the most recent {\it XMM-Newton} observations for time bins of 10\,s, as well as their associated Fourier power spectra.
\begin{figure}[h]
    \resizebox{8cm}{!}{\includegraphics[angle=0]{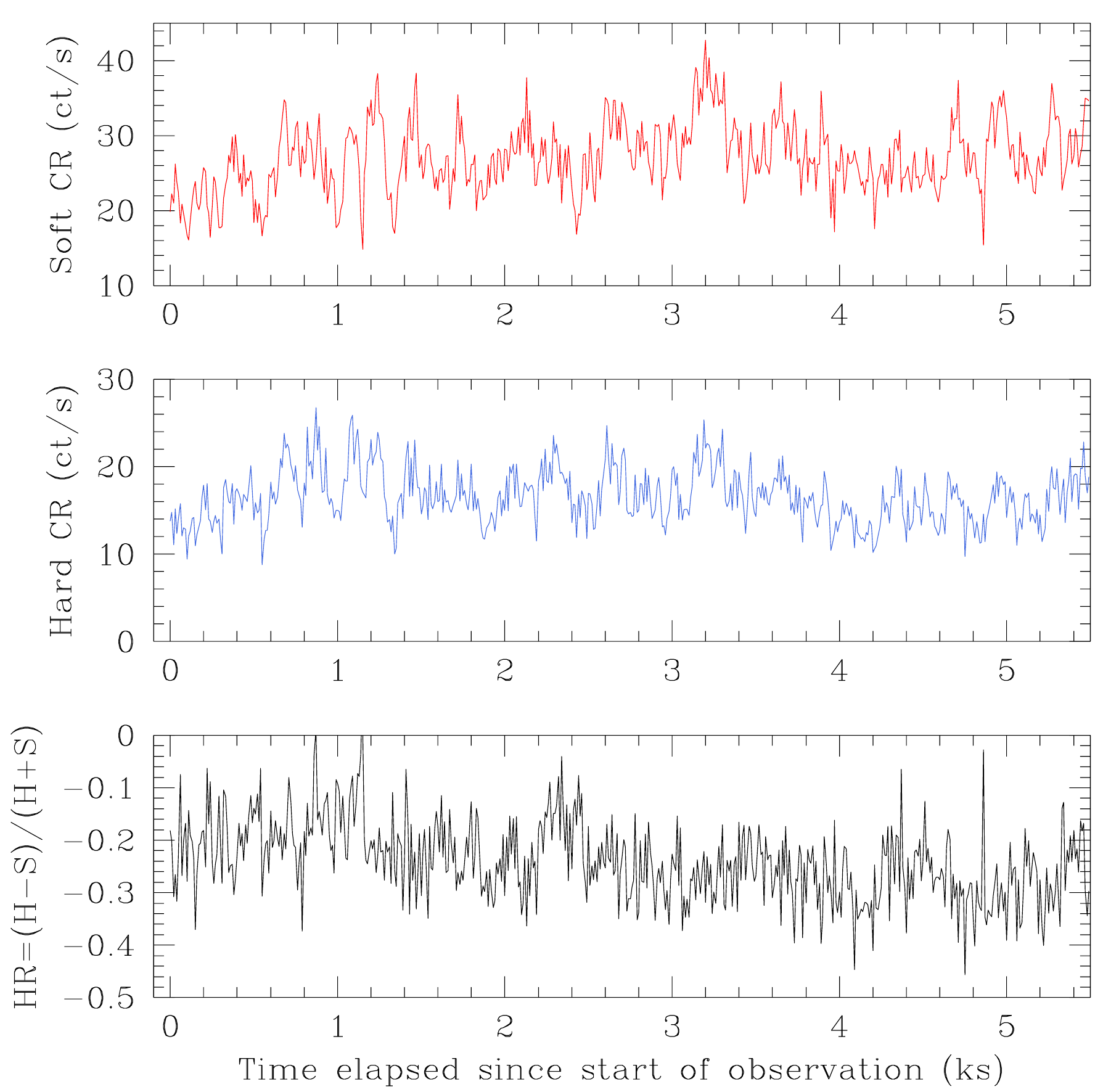}}
    \caption{EPIC-pn soft (0.5 -- 2.0\,keV) and hard (2.0 -- 10.0\,keV) light curves and hardness ratio of $\gamma$~Cas during the low background part of revolution 3881 (February 2021). A timestep of 10\,s was adopted.\label{lc3881_10s}}
\end{figure}
\begin{figure}[h]
    \resizebox{8cm}{!}{\includegraphics[angle=0]{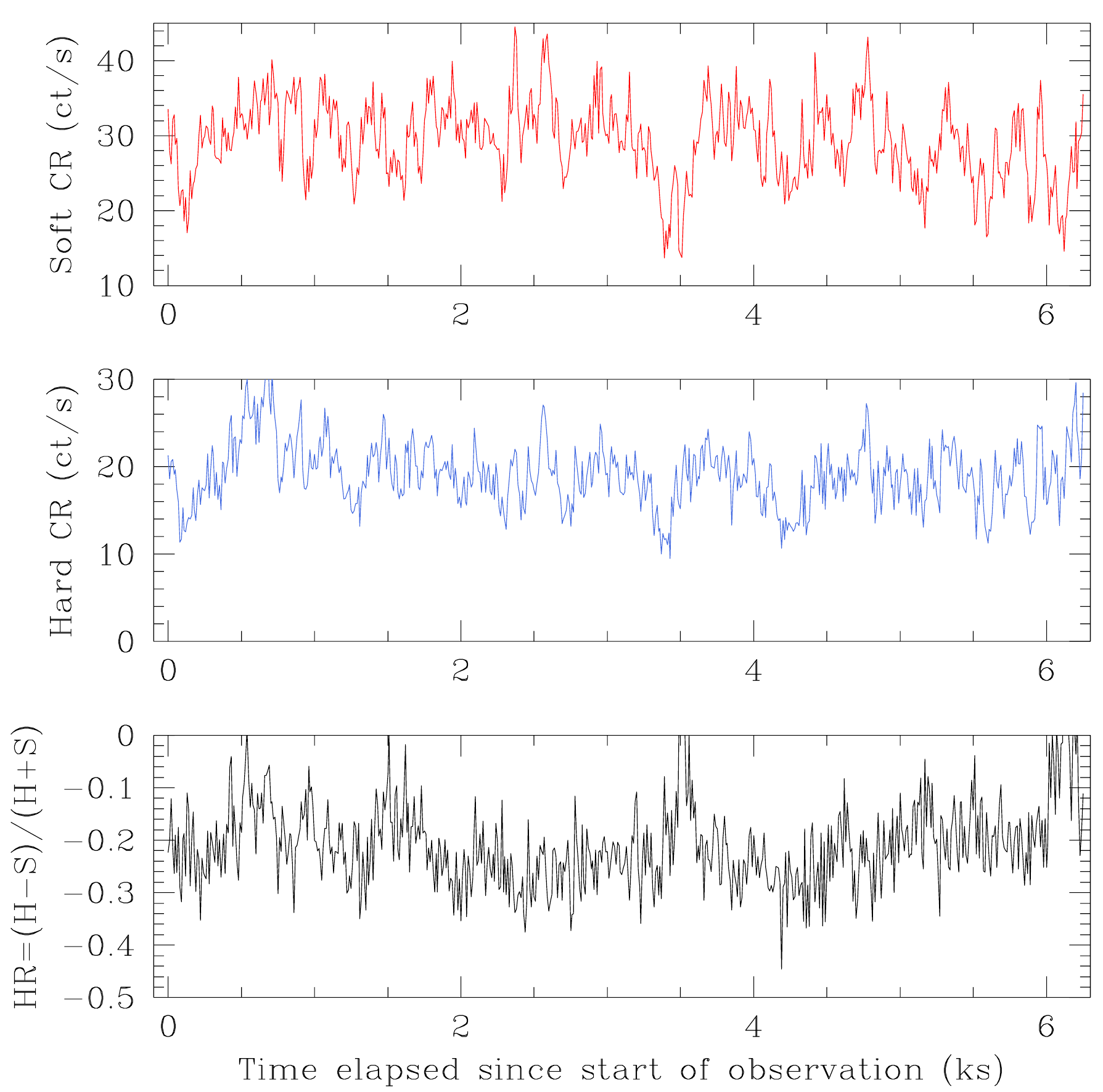}}
    \caption{Same as Fig.\,\ref{lc3881_10s}, but for revolution\ 3960 (July 2021).\label{lc3960_10s}}
\end{figure}
\begin{figure}[h]
  \resizebox{8cm}{!}{\includegraphics[angle=0]{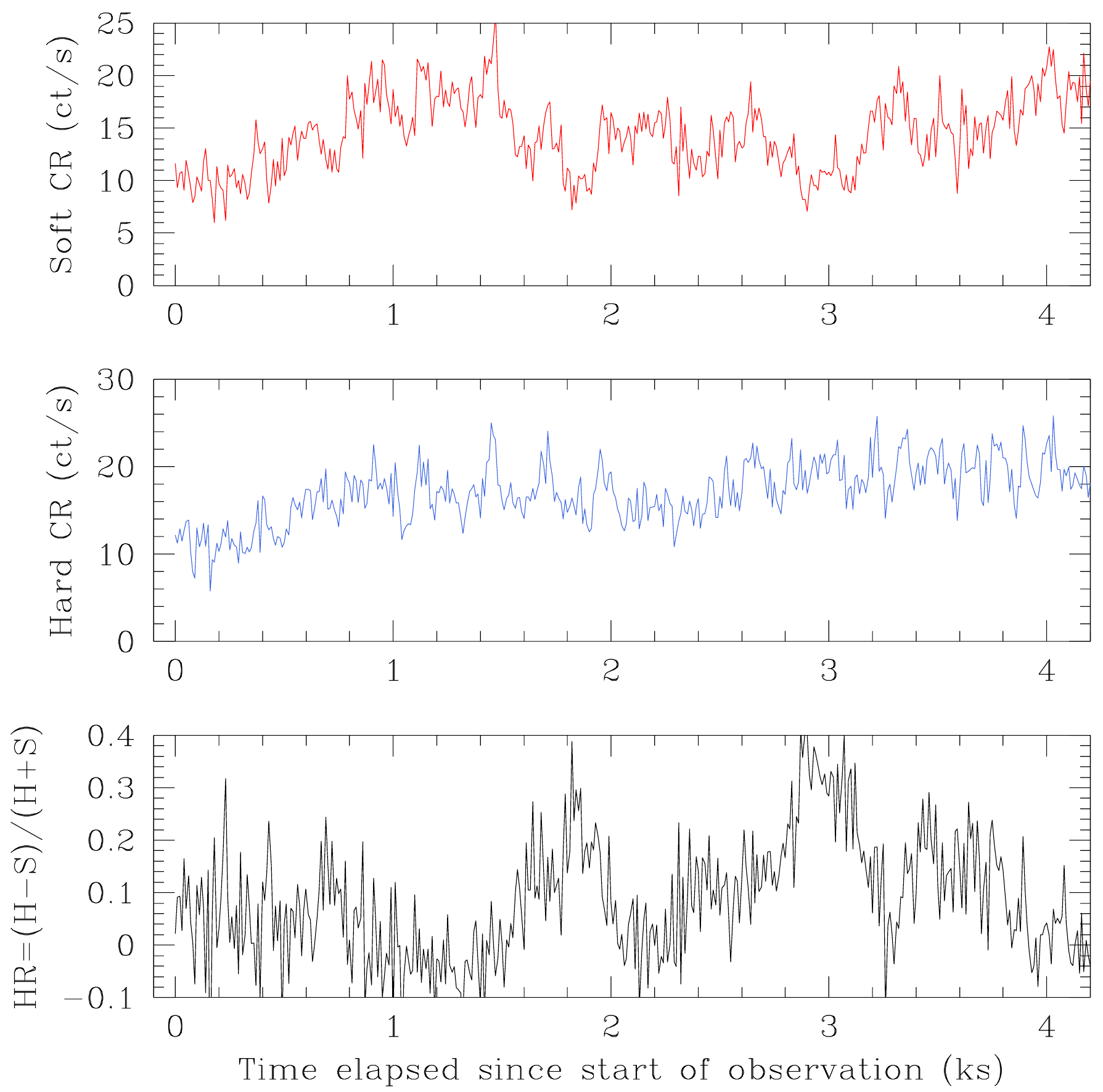}}
    \caption{Same as Fig.\,\ref{lc3881_10s}, but for revolution\ 4049 (January 2022).\label{lc4049_10s}}
\end{figure}
\begin{figure}[h!]
    \resizebox{8cm}{!}{\includegraphics[angle=0]{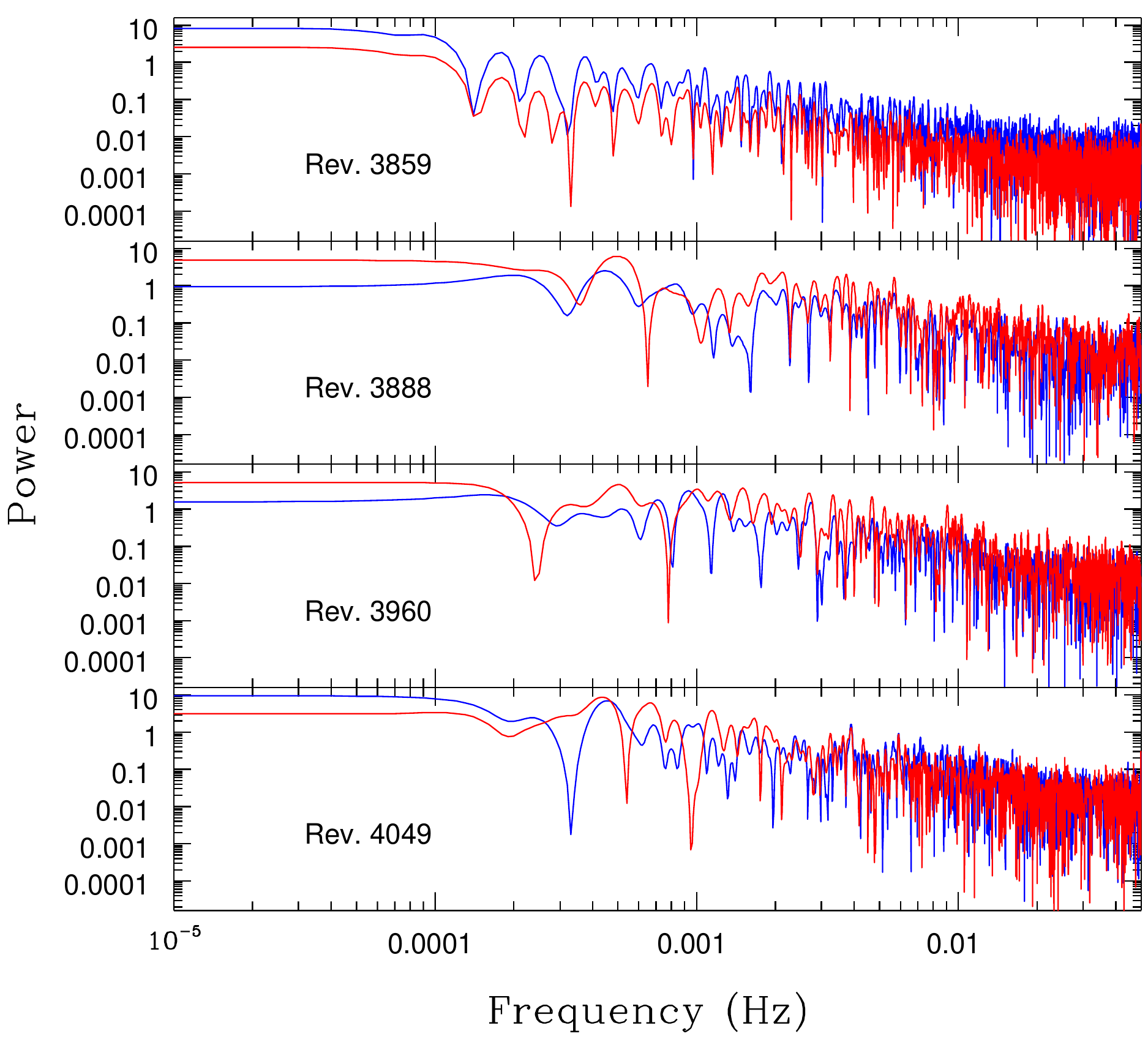}}
    \caption{Fourier power spectra of the EPIC-pn 10\,s light curves recorded during revolutions 3859 (January 2021), 3881 (February 2021), 3960 (July 2021), and 4049 (January 2022), from top to bottom. The red and blue curves in each panel respectively yield the result  for the soft and hard energy bands.\label{power}}
\end{figure}
\pagebreak

\section{EPIC-pn spectra}
The figures in this section illustrate the EPIC-pn spectra along with the best-quality simultaneous adjustment of the EPIC and RGS spectra. Spectra obtained in timing mode are shown in Figs.\,\ref{EPICRev3881} -- \ref{EPICRev4049}, whilst those taken in small window mode are displayed in Figs.\,\ref{EPICRev1937} -- \ref{EPICRev1959}.
\begin{figure}
    \resizebox{8cm}{!}{\includegraphics[angle=0]{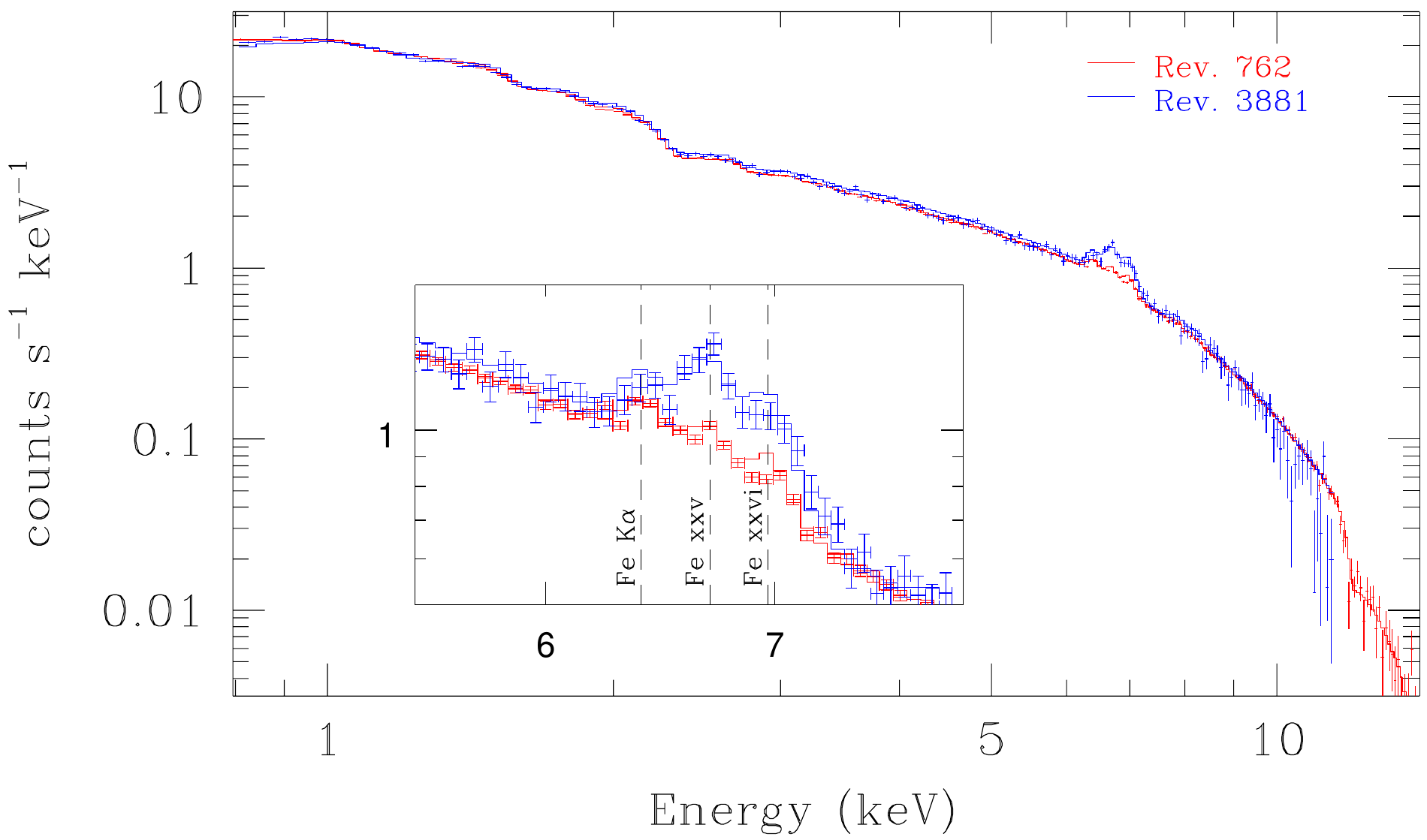}}
    \caption{Comparison between the EPIC-pn data and the best-fit models for revolutions\,0762 (February 2004, red) and 3881 (February 2021, blue). Only the EPIC-pn data are shown, for the sake of clarity, although all EPIC and RGS spectra were fitted simultaneously. The insert shows a zoomed-in view of the iron line complex.\label{EPICRev3881}}
\end{figure}
\begin{figure}
    \resizebox{8cm}{!}{\includegraphics[angle=0]{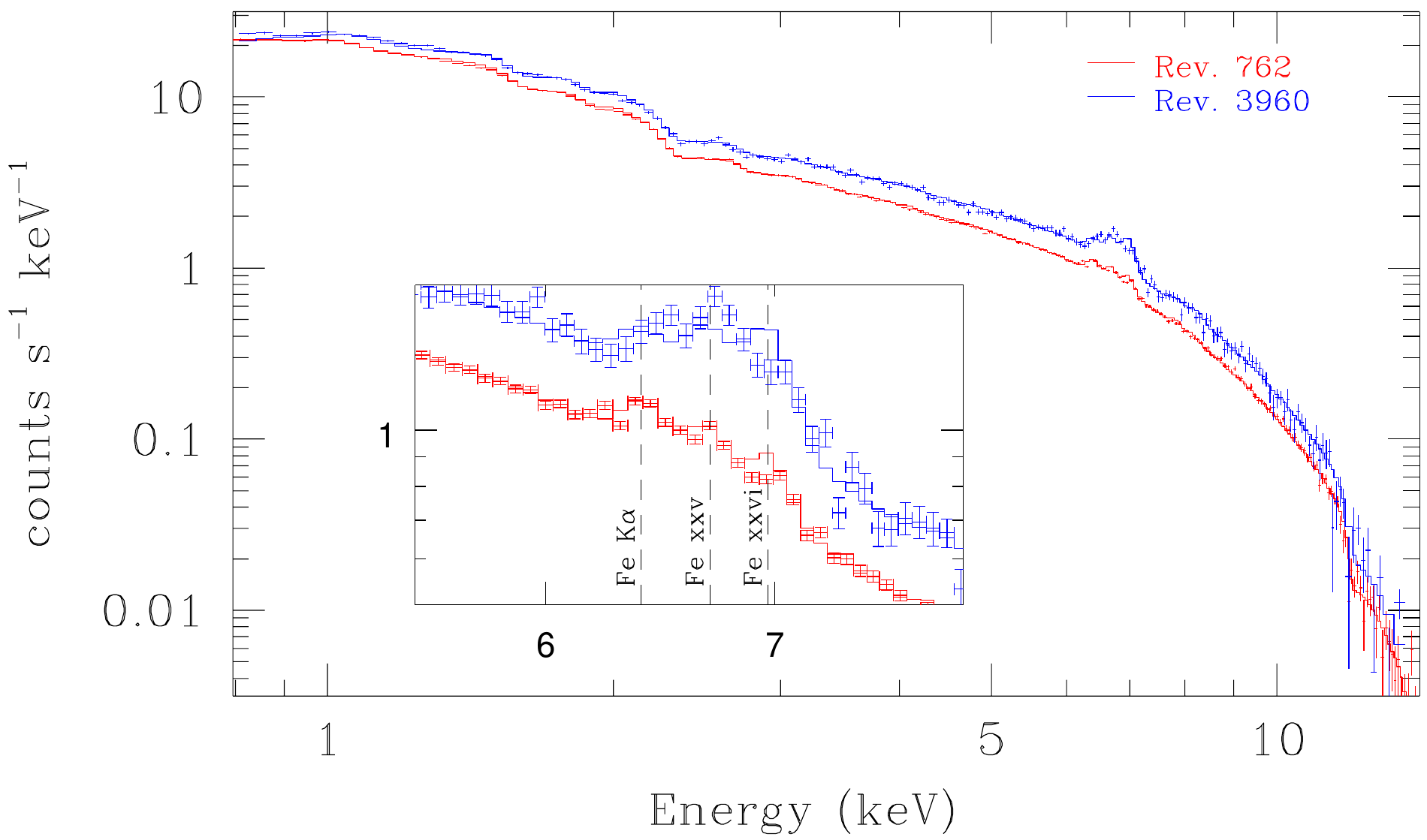}}
    \caption{Same as Fig.\,\ref{EPICRev3881}, but for revolution\ 3960 (July 2021) compared against revolution 0762.\label{EPICRev3960}}
\end{figure}
\begin{figure}
  \resizebox{8cm}{!}{\includegraphics[angle=0]{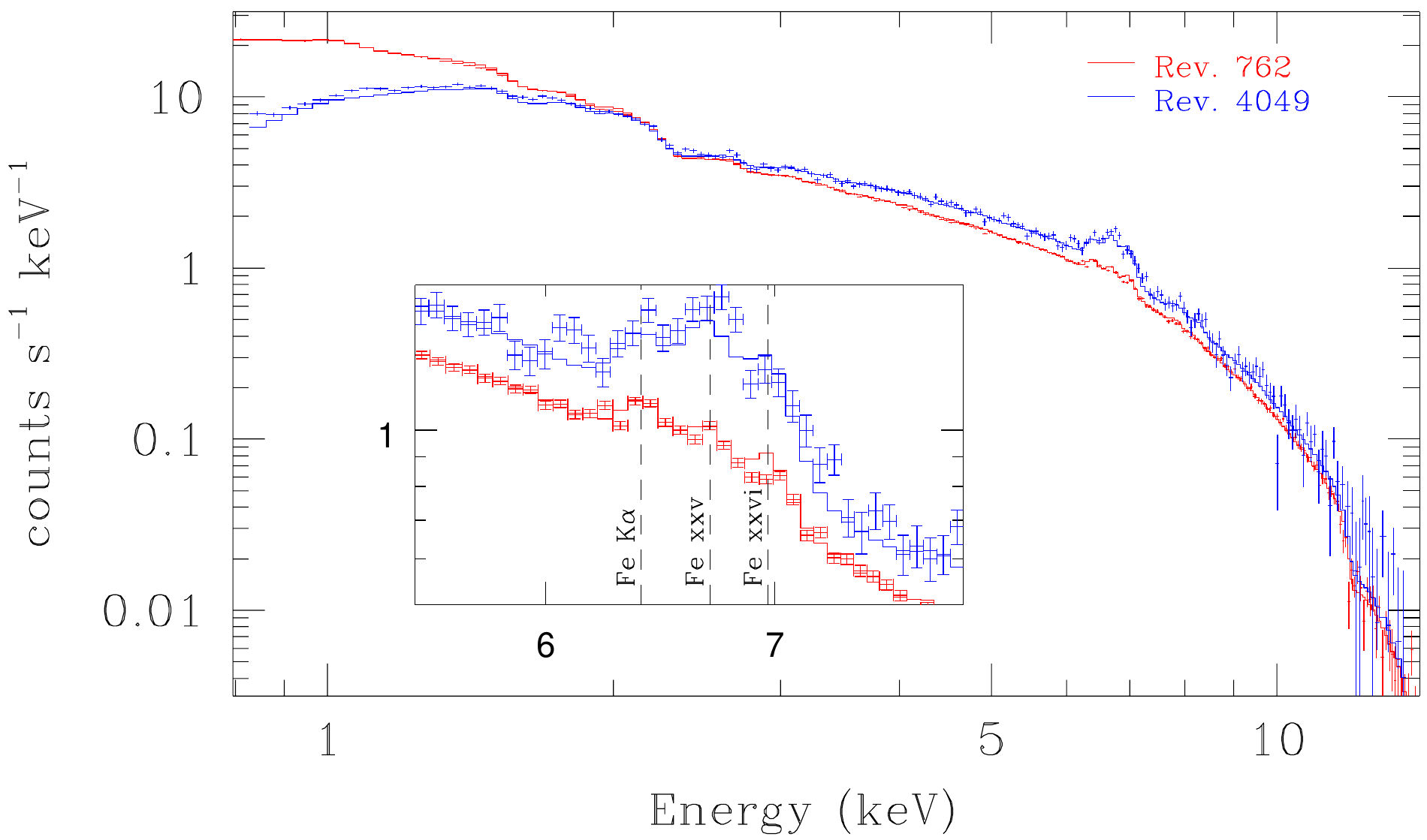}}
  \caption{Same as Fig.\,\ref{EPICRev3881}, but for revolution\ 4049 (January 2022) compared against revolution\,0762.\label{EPICRev4049}}
\end{figure}

\begin{figure}
    \resizebox{8cm}{!}{\includegraphics[angle=0]{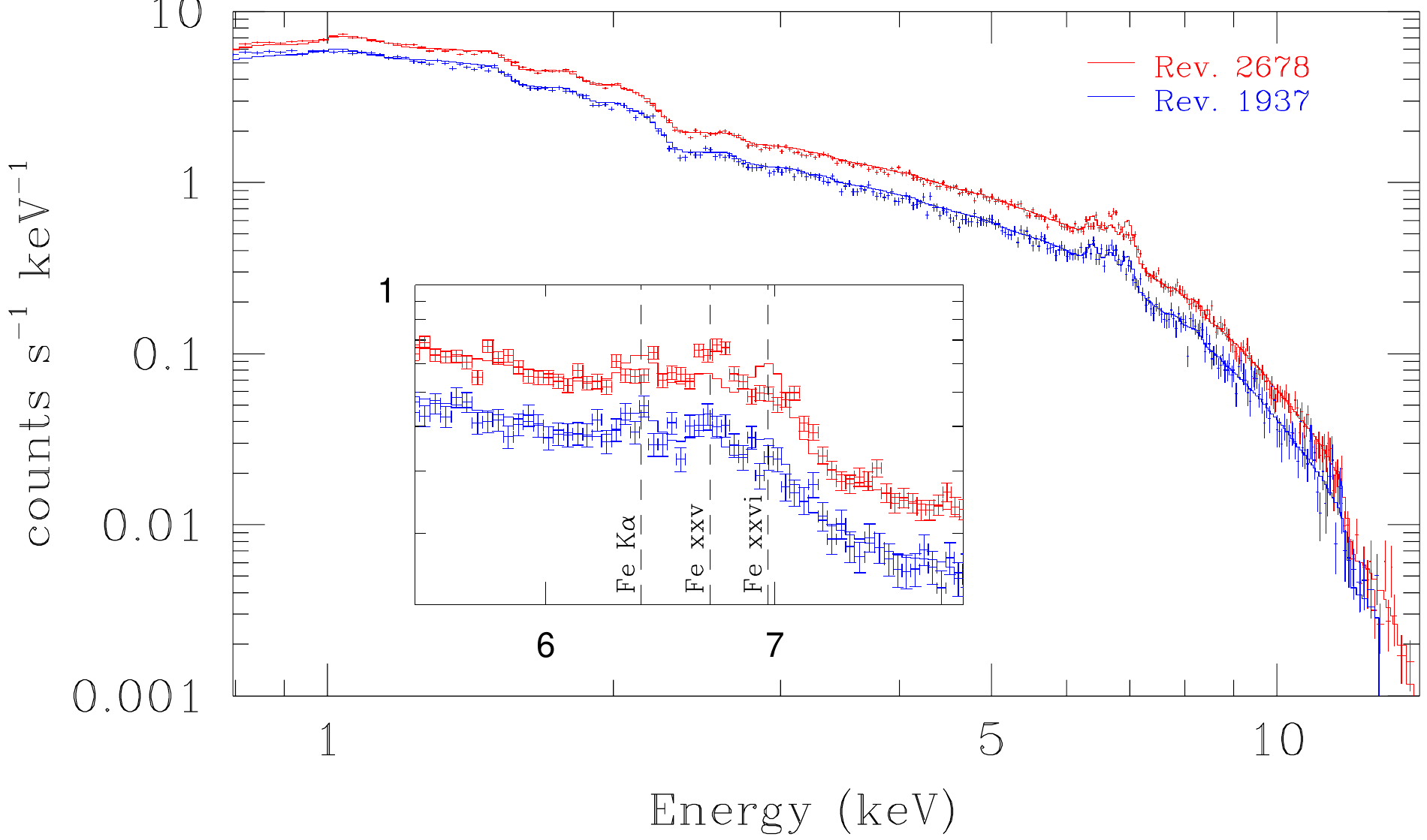}}
    \caption{Comparison between the EPIC-pn data and the best-fit models for revolutions\,1937 (July 2010a, blue) and 2678 (July 2014, red). Only the EPIC-pn data are shown, for the sake of clarity, although all EPIC and RGS spectra were fitted simultaneously. The insert shows a zoomed-in view of the iron line complex.\label{EPICRev1937}}
\end{figure}
\begin{figure}
    \resizebox{8cm}{!}{\includegraphics[angle=0]{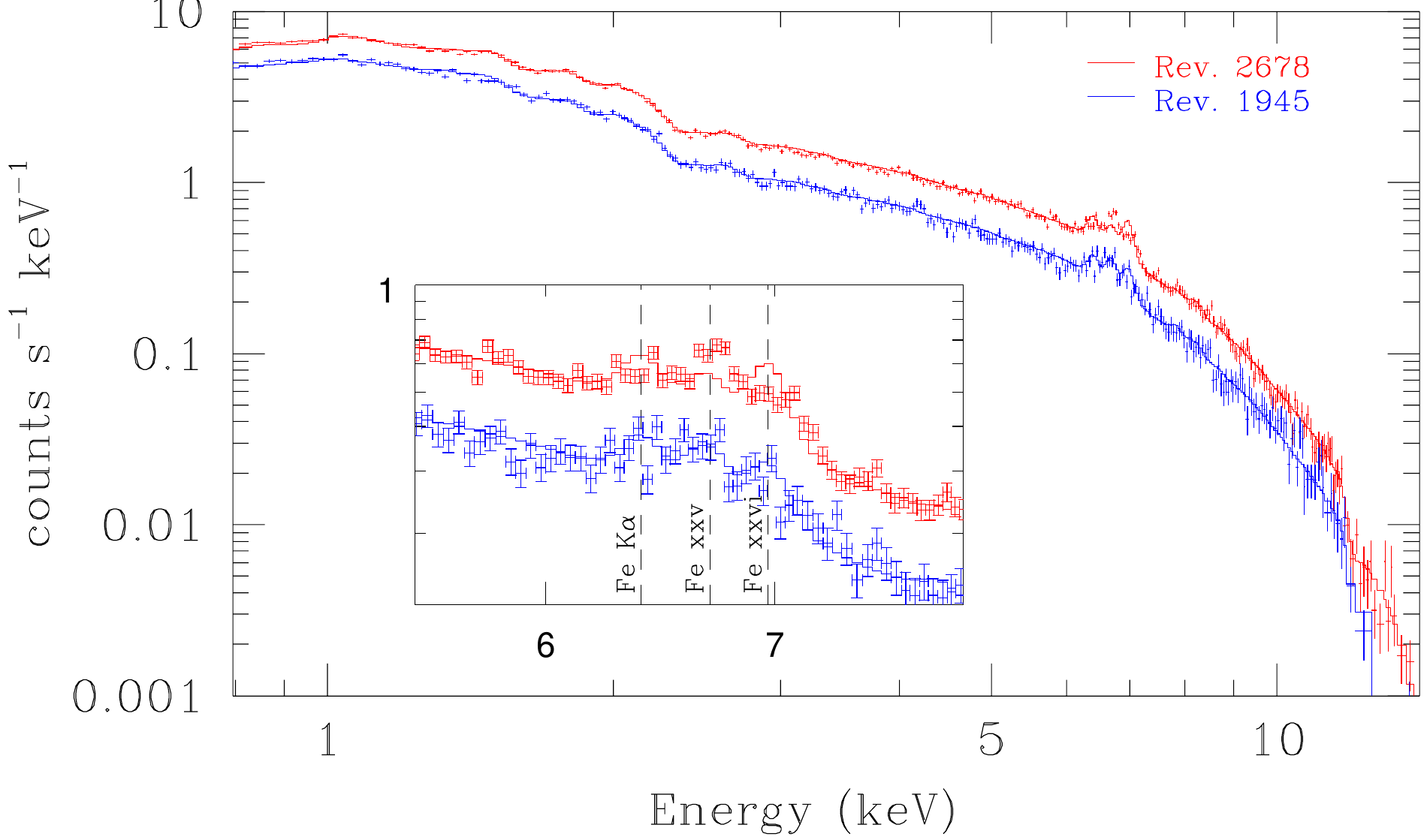}}
    \caption{Same as Fig.\,\ref{EPICRev1937}, but for revolution\ 1945 (July 2010b) compared to revolution\,2678.\label{EPICRev1945}}
\end{figure}
\begin{figure}
    \resizebox{8cm}{!}{\includegraphics[angle=0]{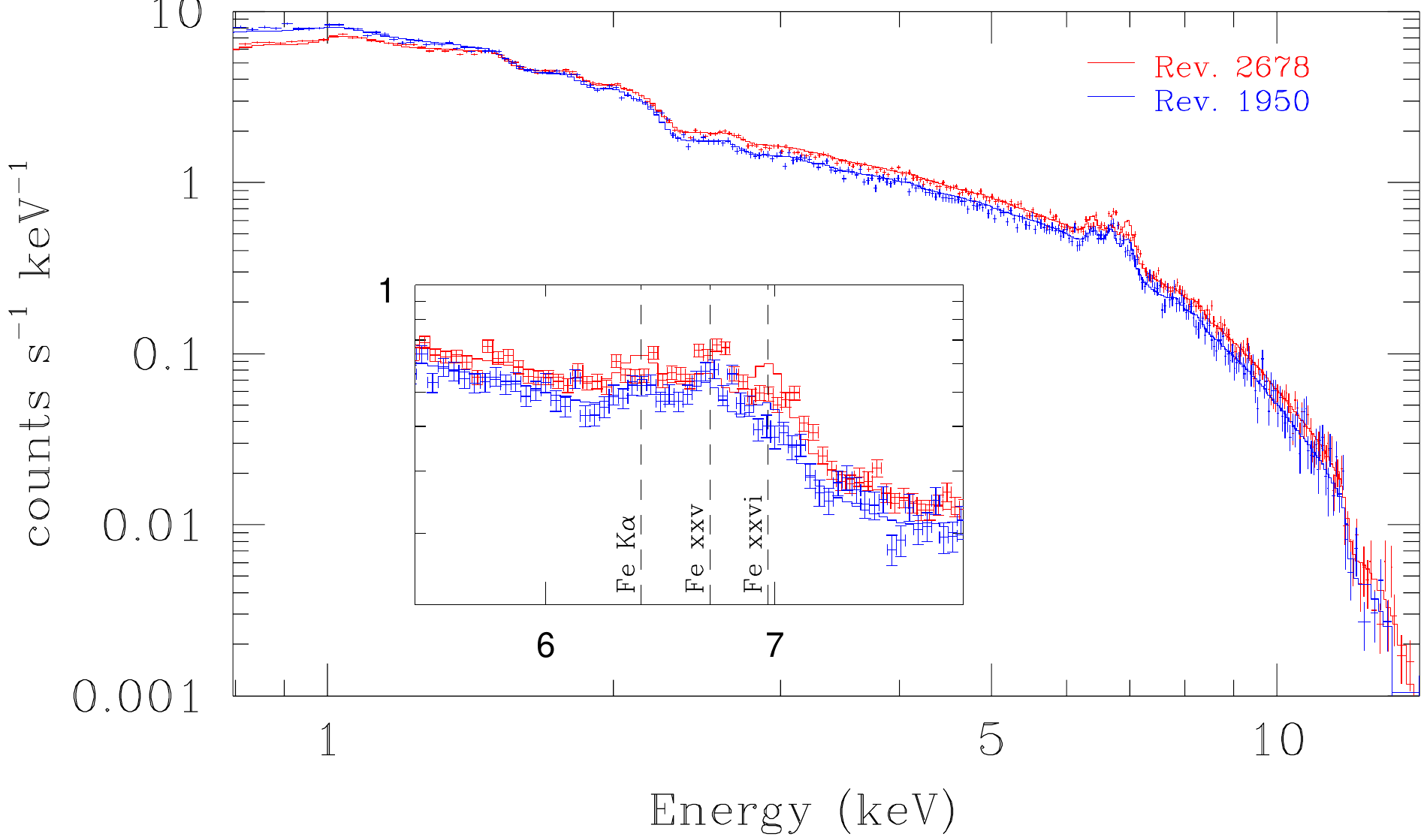}}
    \caption{Same as Fig.\,\ref{EPICRev1937}, but for revolution\ 1950 (August\ 2010a) compared to revolution\,2678.\label{EPICRev1950}}
\end{figure}
\begin{figure}
    \resizebox{8cm}{!}{\includegraphics[angle=0]{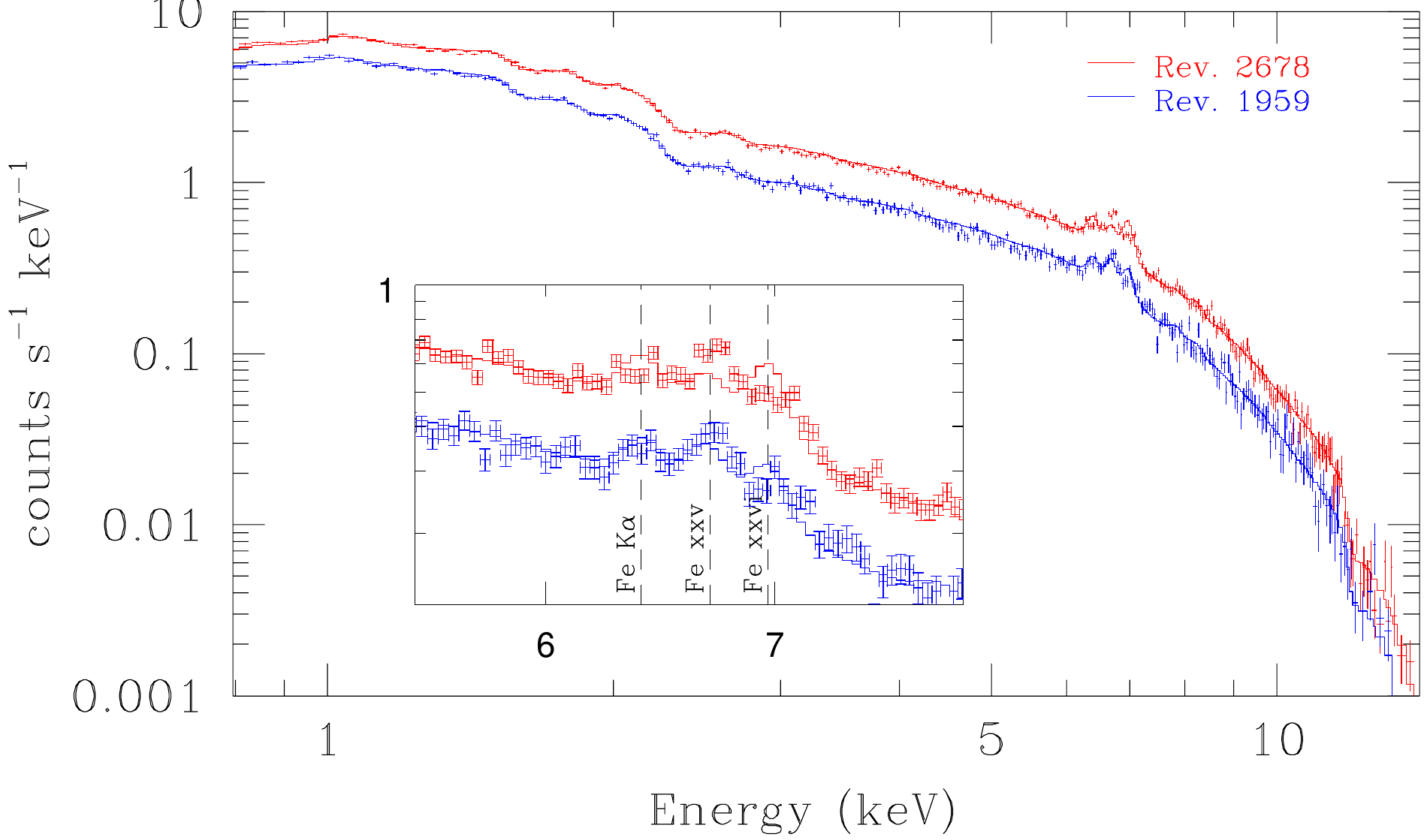}}
    \caption{Same as Fig.\,\ref{EPICRev1937}, but for revolution\ 1959 (August\ 2010b) compared to revolution\,2678.\label{EPICRev1959}}
\end{figure}
\end{appendix}
\end{document}